\RequirePackage{snapshot}

\documentclass[sigconf]{acmart}

\usepackage{soul}

\usepackage{graphicx}
\usepackage{balance}

\usepackage{tikz}
 
\newcommand*{\pieempty}{
\begin{tikzpicture}[baseline={([yshift=-.5ex]current bounding box.center)}]
\draw (0,0) circle (1ex);
\end{tikzpicture}
}

\newcommand*{\piehalf}{
\begin{tikzpicture}[baseline={([yshift=-.5ex]current bounding box.center)}]
\draw (0,0) circle (1ex);\fill (0,-1ex) arc (-90:90:1ex) -- (0,0) -- cycle;
\end{tikzpicture}
}

\newcommand*{\piefull}{
\begin{tikzpicture}[baseline={([yshift=-.5ex]current bounding box.center)}]
\filldraw (0,0) circle (1ex);
\end{tikzpicture}
}

\newcommand*{\piehalfred}{
\begin{tikzpicture}[baseline={([yshift=-.5ex]current bounding box.center)}]
\draw (0,0) circle (1ex);\fill[red] (180:1ex) arc (180:0:1ex) -- cycle;
\end{tikzpicture}
}

\newcommand*{\piered}{
\begin{tikzpicture}[baseline={([yshift=-.5ex]current bounding box.center)}]
\draw[black,fill=red] (0,0) circle (1ex);
\end{tikzpicture}
} \usepackage{booktabs}   
\usepackage{subcaption} 

\usepackage{listings}
\usepackage{todonotes}

\usepackage{geometry}
\usepackage{tikz}
\usepackage{pgfmath}
\usepackage[ruled,vlined,linesnumbered]{algorithm2e}
\usepackage{multirow}
\usepackage{colortbl}
\usepackage{stmaryrd}
\usepackage{verbatim}

\usepackage{relsize}

\usepackage{url}

\usepackage{paralist}

\usepackage{booktabs}

\usepackage{longtable}

\usepackage{amsmath}
\usepackage{mathtools}
\usepackage{breqn}
\usepackage{stmaryrd}
\usepackage[inline]{enumitem}
\usepackage{empheq}
\usepackage{cleveref}

\usepackage{pdflscape}

\DeclareMathOperator{\defas}{:=}

\newcommand{\MH}[1]{ \ifbool{WithComments}{ {\color{magenta} \textbf{Melanie says:} #1}} {}}
\newcommand{\BG}[1]{ \ifbool{WithComments}{ {\color{red} \textbf{Boris says:} #1}} {}}
\newcommand{\RD}[1]{ \ifbool{WithComments}{ {\color{blue} \textbf{Ralf says:} #1}} {}}
\newcommand{\SL}[1]{ \ifbool{WithComments}{ {\color{orange} \textbf{Seokki says:} #1}} {}}
\newcommand{\RS}[1]{ \ifbool{WithComments}{ {\color{cyan} \textbf{Ralf and Seokky say:} #1}} {}}

\newcommand{\hlfancy}[2]{\sethlcolor{#1}\hl{#2}}
\newcommand{\copied}[1]{ \ifbool{WithComments}{ \hlfancy{yellow}{#1}} {#1}}

\definecolor{black}{rgb}{0,0,0}
\definecolor{grey}{rgb}{0.8,0.8,0.8}
\definecolor{red}{rgb}{1,0,0}
\definecolor{green}{rgb}{0,1,0}
\definecolor{darkgreen}{rgb}{0,0.5,0}
\definecolor{darkpurple}{rgb}{0.5,0,0.5}
\definecolor{darkdarkpurple}{rgb}{0.3,0,0.3}
\definecolor{blue}{rgb}{0,0,1}
\definecolor{shadegreen}{rgb}{0.95,1,0.95}
\definecolor{shadeblue}{rgb}{0.95,0.95,1}
\definecolor{shadered}{rgb}{1,0.85,0.85}
\definecolor{shadegrey}{rgb}{0.85,0.85,0.85}
\definecolor{oddRowGrey}{rgb}{0.80,0.80,0.80}
\definecolor{evenRowGrey}{rgb}{0.85,0.85,0.85}
\definecolor{lightpurple}{rgb}{0.88,1.0,1.0}

\definecolor{crblue}{rgb}{0.1,0.1,0.9}
\definecolor{tabbg}{rgb}{ 0.8,0.8,0.9}
\newcommand{\CR}[1]{#1}

\definecolor{revac}{HTML}{785EF0}
\definecolor{revbc}{HTML}{DC267F}
\definecolor{revcc}{HTML}{FE6100}
\definecolor{revcm}{HTML}{0000A0}

\definecolor{revgreen}{rgb}{0,0.5,0}
\newrobustcmd{\reva}[1]{\textcolor{black}{{#1}}}
\newrobustcmd{\revb}[1]{\textcolor{black}{{#1}}}
\newrobustcmd{\revc}[1]{\textcolor{black}{{#1}}}
\newrobustcmd{\revm}[1]{\textcolor{black}{{#1}}}

\newcommand{\nrcbag}[1]{\{\!\{#1\}\!\}}
\newcommand{\nrcset}[1]{\{#1\}}
\newcommand{\nrctuple}[1]{\langle #1 \rangle}

\newcommand{\attrType}[2]{{#1}:{#2}}
\newcommand{\intType}{\textsc{int}}
\newcommand{\strType}{\textsc{str}}
\newcommand{\boolType}{\textsc{bool}}
\newcommand{\dateType}{\textsc{date}}

\newcommand{\typeOf}[1]{\mathbf{type}(#1)}
\newcommand{\attrLabels}{\mathbb{L}}
\newcommand{\nType}{\tau}
\newcommand{\pType}{\mathcal{P}}
\newcommand{\tType}{\mathcal{T}}
\newcommand{\rType}{\mathcal{R}}
\newcommand{\sType}{\mathcal{S}}
\newcommand{\dbType}{\mathcal{D}}
\newcommand{\bagType}[1]{\nrcbag{#1}}
\newcommand{\tupType}[1]{\nrctuple{#1}}

\newcommand{\anyType}{\mathcal{A}}
\newcommand{\schemaOf}[1]{\textsc{sch}({#1})}
\newcommand{\makeType}[1]{\mathcal{#1}}

\newcommand{\nInst}{I}
\newcommand{\primInst}{\mathbb{P}}

\newcommand{\aTup}{t}

\newcommand{\aRel}{R}
\newcommand{\rel}{\aRel}
\newcommand{\bt}[2]{{#1}^{#2}}

\newcommand{\aval}[2]{{#1}:\,{#2}}
\newcommand{\db}{D}
\newcommand{\multOf}[2]{\textsc{mult}(#1,#2)}
\newcommand{\nullVal}{\bot}

\newcommand{\query}{Q}
\newcommand{\concat}{\circ}
\newcommand{\qEval}[1]{\llbracket{#1}\rrbracket}
\newcommand{\valPlaceholder}{?\xspace}
\newcommand{\multPlaceholder}{*\xspace}

\newcommand{\tree}{T}
\newcommand{\ted}{d}

\newcommand{\valid}{valid}
\newcommand{\survivor}{retained}
\newcommand{\compatible}{consistent}
\newcommand{\sbatts}[2]{\frac{\color{blue}{#1}}{#2}}

\newcommand{\selection}{\sigma}
\newcommand{\projection}{\pi}
\newcommand{\join}{\Join}
\newcommand{\union}{\cup}

\newcommand{\rename}{\rho}
\newcommand{\aggregation}[2]{\gamma_{{#1} \to {#2}}}
\newcommand{\crossprod}{\times}

\newcommand{\nestTupSimp}{\mathcal{N}^T}
\newcommand{\nestTup}[2]{\nestTupSimp_{{#1} \to {#2}}}
\newcommand{\nestRelSimp}{\mathcal{N}^R}

\newcommand{\nestRel}[2]{\nestRelSimp_{{#1} \to {#2}}}

\newcommand{\flatten}{F}

\newcommand{\RA}{\mathcal{RA}}

\newcommand{\somejoin}{\diamond}
\def\ojoin{\setbox0=\hbox{$\Join$}
  \rule[.07ex]{.25em}{.52pt}\llap{\rule[0.97ex]{.25em}{.5pt}}}
\def\leftouterjoin{\mathbin{\ojoin\mkern-7.3mu\Join}}
\def\rightouterjoin{\mathbin{\Join\mkern-7.3mu\ojoin}}
\def\fullouterjoin{\mathbin{\ojoin\mkern-7.3mu\Join\mkern-7.3mu\ojoin}}

\newcommand{\whynot}{\Phi}
\newcommand{\aWhynot}{\langle\query, \db, \aTup\rangle}
\newcommand{\expl}{\mathcal{E}}

\newcommand{\paramOf}{param}

\newcommand{\reparams}[1]{\textsc{Re}({#1})}
\newcommand{\SR}[1]{\textsc{SR}({#1})}
\newcommand{\MSR}[1]{\textsc{MSR}({#1})}
\newcommand{\leqMSR}{\preceq_{\whynot}}

\newcommand{\changed}[2]{\ensuremath\Delta({#1},{#2})}

\newcommand{\anOpDS}{op}
\newcommand{\aSchemaOpt}{S}
\newcommand{\aSchemaOptCollection}{\mathcal{\aSchemaOpt}}
\newcommand{\aProvenanceAnnotation}{A}
\newcommand{\aProvenanceAnnotationCollection}{\mathcal{\aProvenanceAnnotation}}

\newcommand{\matches}{\simeq}

\newcommand{\matching}{{\mathcal M}}

\newcolumntype{H}{>{\columncolor{black}\color{white}}c}

\makeatletter
\newcommand{\inlineitem}[1][]{
\ifnum\enit@type=\tw@
    {\descriptionlabel{#1}}
  \hspace{\labelsep}
\else
  \ifnum\enit@type=\z@
       \refstepcounter{\@listctr}\fi
    \quad\@itemlabel\hspace{\labelsep}
\fi}

\newcommand{\mypar}[1]{\noindent\textbf{{#1}.}}

\newcommand{\card}[1]{\vert{#1}\vert}

\newcommand{\mathtext}[1]{\thickspace\text{#1}\thickspace}

\DeclareMathAlphabet{\mathbbold}{U}{bbold}{m}{n}

\newcommand{\ptime}{\texttt{PTIME}\xspace}

\newcommand{\nphard}{NP-hard\xspace}

\newtheorem{Theorem}{Theorem}
\newtheorem{definition}{Definition}

\newtheorem{example}{Example}

\newcommand{\myproofpar}[1]{\smallskip\noindent\underline{{#1}:}\,}

\newcommand{\sbtMap}{{\mathcal M}_{sbt}}

\newcommand{\termNRAB}{nested relational algebra for bags\xspace}
\newcommand{\abbrNRAB}{\ensuremath{\mathcal{NRAB}}\xspace}
\newcommand{\baseNRAB}{\ensuremath{\mathcal{NRAB}^0}\xspace}
\newcommand{\spcNRAB}{\ensuremath{\mathcal{SPC}}\xspace}
\newcommand{\spcuNRAB}{\ensuremath{\mathcal{SPC}^+}\xspace}
\newcommand{\abbrNIP}{NIP\xspace}
\newcommand{\abbrNIPs}{NIPs\xspace}
\newcommand{\rpnosa}{RPnoS\xspace}
\newcommand{\rp}{RP\xspace}

\newcommand{\trimfigspace}[1][-4mm]{\vspace{#1}}

\newcommand*\squeezespaces[1]{
  \thickmuskip=\scalemuskip{\thickmuskip}{#1}
  \medmuskip=\scalemuskip{\medmuskip}{#1}
  \thinmuskip=\scalemuskip{\thinmuskip}{#1}
  \nulldelimiterspace=#1\nulldelimiterspace
  \scriptspace=#1\scriptspace
}
\newcommand*\scalemuskip[2]{
  \muexpr #1*\numexpr\dimexpr#2pt\relax\relax/65536\relax
}

\newif\ifswitch
\let\switchon\switchtrue
\let\switchoff\switchfalse

\def\rowswitch#1\\{
\ifswitch
  #1\\
\fi
}

\makeatletter
\newcommand{\removelatexerror}{\let\@latex@error\@gobble}
\makeatother

\newbool{Techreport}
\newrobustcmd{\ifnottechreport}[1]{\ifbool{Techreport}{}{#1}}
\newrobustcmd{\iftechreport}[1]{\ifbool{Techreport}{#1}{}}

\newbool{WithComments}
\boolfalse{WithComments}

\booltrue{Techreport}

\begin{document}

\title{To \emph{not} miss the forest for the trees - A holistic approach for explaining missing answers over nested data (extended version)}

\author{Ralf Diestelk\"amper}
\affiliation{
  \institution{University of Stuttgart - IPVS, Germany}
}
\email{ralf.diestelkaemper@ipvs.uni-stuttgart.de}

\author{Seokki Lee}
\affiliation{
  \institution{University of Cincinnati, USA}
}
\email{lee5sk@ucmail.uc.edu}

\author{Melanie Herschel}
\affiliation{
  \institution{University of Stuttgart - IPVS, Germany}
}
\email{melanie.herschel@ipvs.uni-stuttgart.de}

\author{Boris Glavic}
\affiliation{
  \institution{Illinois Institute of Technology, USA}
}
\email{bglavic@iit.edu}

\begin{abstract}
Query-based explanations for missing answers identify which operators of a query are responsible for the failure to return a missing answer of interest. This type of explanations has proven to be useful in a variety of contexts including debugging of complex analytical queries. Such queries are frequent in big data systems such as Apache Spark. 
We present a novel approach for producing query-based explanations. 
Our approach is the first to support nested data and to consider operators that modify the schema and structure of the data (e.g., nesting and projections) as potential causes of missing answers. To efficiently compute explanations,
we propose a heuristic algorithm that applies two novel techniques: (i) reasoning about multiple \emph{schema alternatives} for a query and (ii) re-validating at each step whether an intermediate result can contribute to the missing answer.
Using an implementation of our approach on Spark, we demonstrate that it is the first to scale to large datasets and that it often finds explanations 
that existing techniques fail to identify.

\end{abstract}

 \maketitle

\section{Introduction}\label{sec:intro}

\revm{Debugging analytical queries
in data-intensive scalable computing~(DISC) systems such as Apache Spark or Flink is a tedious process. Query-based explanations can aid users in this process by narrowing down the debugging task to parts of the query that are responsible for the failure to compute an expected answer.} 
In this work, we present an approach for producing query-based explanations and implement this approach on Spark.
We represent data in the nested relational model and queries in the nested relational algebra for bags~\cite{grumbach:pods93}. This allows us to cover a large variety of practical queries expressible in big data systems, like in~\cite{amsterdamer:pvldb11}. 

In general, missing answers approaches have three inputs: a \emph{why-not question} specifying which missing results are of interest, a query, and an input data. Three categories of explanations have been considered~\cite{herschel:vldbj17}: (i)~\textit{instance-based} explanations attribute missing answers to missing input data; (ii)~\textit{query-based} explanations pinpoint which parts of the query, typically at the granularity of individual operators, cause the derivation of the expected results to fail; and (iii) \textit{refinement-based} explanations produce a rewritten query that returns the missing answer. Our approach 
returns query-based explanations  that consist of \emph{a set of operators}. Each explanation 
indicates a set of operators that should be fixed for the missing answers to be returned.

\begin{figure}[t]
 \begin{minipage}{.23\textwidth}
\includegraphics[width=0.9\textwidth]{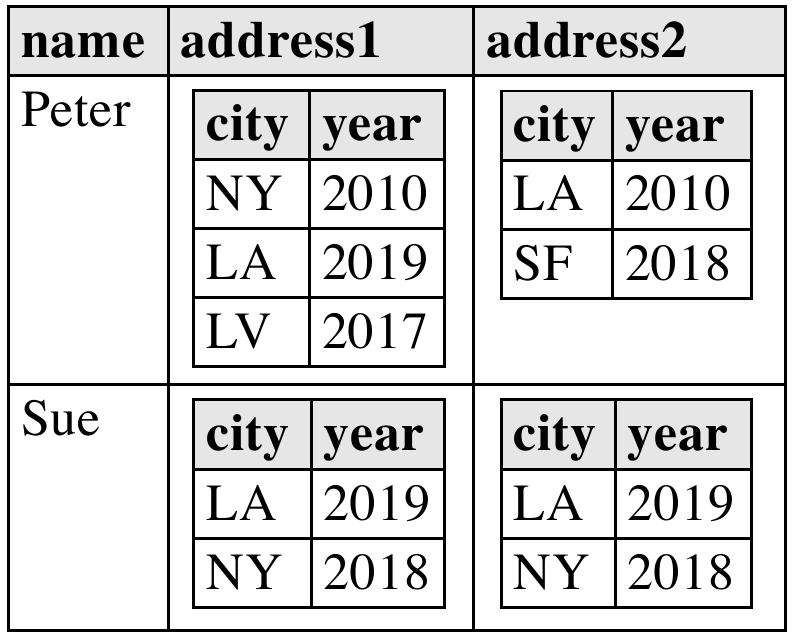}\\[-5mm]
\subcaption{Sample input data}
\label{tab:nested-data}
\end{minipage}
 \begin{minipage}{.23\textwidth}
 \centering
 \includegraphics[width=0.4\textwidth]{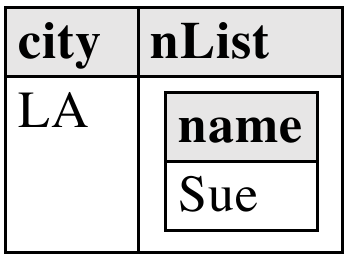}
 \subcaption{Sample output}
 \label{tab:output}
 \end{minipage}

 \begin{minipage}{0.5\textwidth}
  \centering
\begin{minipage}{\textwidth}
\centering
  \includegraphics[width=0.9\textwidth]{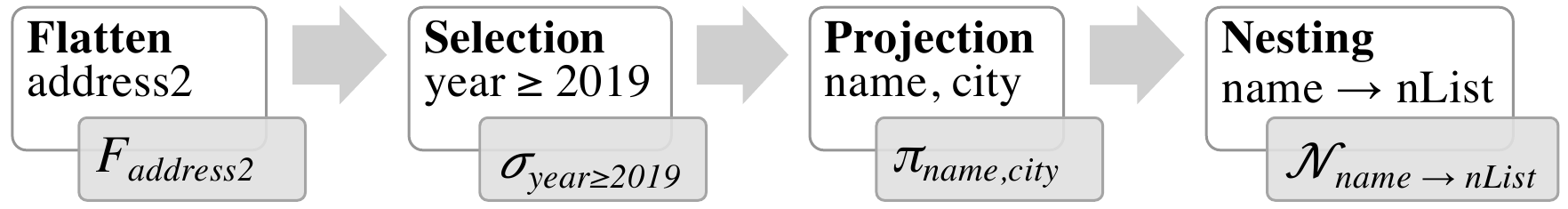}
  \subcaption{Operator pipeline for sample program}
  \label{fig:sample-pipelines}
  \end{minipage}
\end{minipage}
\trimfigspace
 \caption{\CR{Given \texttt{person} input data (a), we obtain a list of cities with associated persons (b) when running a Spark program that corresponds to the operator pipeline shown in (c). \label{fig:running-example}}}
\vspace{-0.3cm}
\end{figure}

\begin{example}\label{ex:running-example}

  Consider the person table shown in \Cref{tab:nested-data}. Each person tuple contains two nested address relations (cities with associated years). These may correspond to work and home addresses. 
   \Cref{fig:sample-pipelines} shows a query that 
  returns cities that are the workplace of at least one person since 2019. For each such city, the query 
  returns the list of persons that work in this city. The query is composed of four operators (explained further below). 
  The query's result over the person table consists of a single nested tuple (\Cref{tab:output}). An analyst may wonder why NY is not in the result and pose this concern as a  why-not question.
  Multiple query-based explanations exist. For instance, the selection~\texttt{year $\geq$ 2019} prevents the tuple (NY, \{(Sue)\}) that matches the why-not question to appear in the result. 
 This results in an explanation pinpointing this single selection operator as problematic. Another possibility is that the analyst assumed attribute \texttt{address2} stores work addresses, while in fact, \texttt{address1} does. 
  However, given the data in \texttt{address1}, this is not sufficient for explaining the missing answer, as no tuple featuring NY has a sufficiently recent year. Thus, an explanation involving a ``misconfigured'' flattening operation also requires adjusting the selection, which results in 
  an explanation that includes  both the flatten and selection operator. 
\end{example}

The idea of providing operators as query-based explanations for a missing answer is at the core of lineage-based approaches~\cite{DBLP:conf/sigmod/ChapmanJ09,DBLP:conf/edbt/BidoitHT14,herschel:jdiq15,deutsch:edbt20}.
They identify \emph{compatible} tuples in the input data that contain the values necessary to produce the missing answers and trace them through the query to determine \emph{picky operators}.  These operators filter \emph{successors} of compatible tuples. \revm{The rationale is that it may be possible to change the parameters of a picky operator such that it no longer filters the successors of compatible tuples.}

\begin{example}
Applying the lineage-based explanation approach to our example for the why-not question asking for NY, we identify tuple (NY, 2018) nested in the \texttt{address2} attribute of \texttt{Sue} as the only compatible tuple. When tracing this tuple through the query's operators, we observe that it is in the lineage of the flatten operator's intermediate result. In other words, its successor passes this first operator and is in the input of the subsequent selection. The selection's result does not include any successor of this compatible. Thus, we would  identify the selection as a picky operator 
  and return it as an explanation. 
\label{ex:lineage}
\end{example}

 \Cref{ex:lineage} already makes non-trivial adaptations to state-of-the-art solutions for relational data. It extends the set of supported operators with flatten and nesting and assumes tracing support for nested tuples. Straightforward extensions of existing solutions would trace top-level tuples only and, thus, return no result at all. More importantly, a purely lineage-based formulation of the problem fails to find all query-based explanations from \Cref{ex:running-example}.

In this paper, we propose a novel formalization of query-based why-not explanations fitting both flat and nested data models. We further present a practical algorithm to compute such explanations, which is implemented and evaluated in Apache Spark. 

\mypar{Why-not explanations for flat and nested data based on repa-rametrizations} Alternative approaches to lineage-based why-not explanations 
have been investigated recently~\cite{bidoit:cikm15,deutsch:pvldb18,diestelkaemper:tapp19}. However, their practical use is limited since they only support conjunctive queries over relational data or lack an efficient or effective algorithm or implementation. Inspired by~\cite{diestelkaemper:tapp19}, our formalization is based on \emph{reparameterizations} of query operators. These are changes to the parameters of one or more operators that ``repair'' the query such that the missing answer is returned. We define an explanation to be the set of operators modified by a \emph{minimal successful reparameterizations (MSRs)}, which is a reparameterization that is minimal wrt. to a partial order based on the number of operators that are modified (we do not want to modify operators unless needed) and the side effects of the reparameterization (``repairs'' should avoid changes to the original query result). Our formalization has two advantages over past work: (i) it guarantees that neither false negatives (operators not returned that have to be changed) nor false positives are returned (operators part of explanations that do not have to be changed); 
and (ii) explanations may include operators such as projections and nesting (not supported by past work). Such richer explanations require reasoning about the effect that changes to the \emph{schema and (nesting) structure} of intermediate results have on the final query result.  
\revc{However, this precision and expressiveness come at a price:  
computing MSRs is \nphard and even restricted cases that are in \ptime require further optimizations to be practical.} 

\mypar{A scalable heuristic algorithm leveraging schema alternatives and revalidation} In light of this result, we explore a heuristic algorithm that approximates explanations. \revb{Given the corners we cut to be efficient, e.g., disregarding reparameterizations of equi-joins to theta-joins that rely on cross products and are of little practical interest in DISC systems, our algorithm may miss certain operators and corresponding MSRs in its returned explanations. }
Even though our algorithm is heuristic in nature and, like past approaches, uses lineage and forward tracing of compatibles,  
it often finds explanations they cannot produce. This is due to two novel technical contributions: (i) Our algorithm reasons about multiple \textit{schema alternatives}. It traces 
changes of the schema and (nesting) structure of intermediate results caused by possible reparameterizations of operators, e.g., flattening \texttt{address1} instead of \texttt{address2} in our example.
(ii) Like previous approaches, it uses
\emph{compatibles}
to find missing answers.
In contrast to them, 
it \emph{revalidates} compatibility of successors of compatible tuples to avoid false positives (tuples are incorrectly identified as compatible). All past lineage-based approaches are subject to this issue that is exacerbated by considering nested data.  
For instance, in our example, the complete second input tuple is initially flagged as compatible. After flattening, only one of its two successors is compatible.

\mypar{Implementation and evaluation} We implement our algorithm in Apache Spark. However, the algorithm itself is system-independent. We highlight 
design choices that make our approach the first to scale to large datasets (we evaluate on datasets several orders of magnitude larger than previous work) and to offer the most expressive query-based explanations to date for both relational and nested data models. We have validated these claims experimentally.

We review related work in \Cref{sec:related} and introduce preliminaries in \Cref{sec:prelim}. Our why-not explanations are covered in \Cref{sec:why-not}. We present our heuristic algorithm in \Cref{sec:compute-msr}, our  
implementation and evaluation 
in \Cref{sec:experi}, and 
conclude in \Cref{sec:conclusions}.

 \vspace{-0.3cm}
\section{Related Work}
\label{sec:related}

\mypar{Why-not explanations}
Most closely related to our work are query-based (e.g.,~\cite{DBLP:conf/sigmod/ChapmanJ09, DBLP:conf/edbt/BidoitHT14, bidoit:cikm15,deutsch:pvldb18,deutsch:edbt20}) and refinement-based approaches for explaining missing answers (e.g.,~\cite{tran:sigmod10}). All these approaches target flat relational data and, except for~\cite{belhajjame:edbt18, DBLP:conf/sigmod/ChapmanJ09}, which target workflows, support queries limited to subclasses of relational algebra plus aggregation.
As we have seen in the introduction, these approaches do not trivially extend to handling nested data with a richer set of operators and would return fewer explanations than one may expect. 
The only work we are aware of that considers nested data is~\cite{diestelkaemper:tapp19}. The formalization of why-not explanations presented in this paper extends~\cite{diestelkaemper:tapp19} by  defining admissible reparameterizations for a wide set of operators and by utilizing the tree edit distance~\cite{BP05} to quantify the impact a reparametrization may have on the query result. We further present an algorithm matching our formalization. 

\mypar{Query-by-example (QBE) and query reverse engineering (QRE)}
Query-based explanations for missing answers are also closely related to QBE~\cite{DA16a,Z77,DG19c} and QRE techniques~\cite{barcelo-19-tvrenpdq,KL18,tan-17-renagq,tran-14-qren}, which generate a query from a set of input-output examples provided by the user. In contrast to QBE, our explanations start from a given database, query, and output. Opposed to some QRE approaches, which return a query equivalent to an unknown query $Q$, our explanations apply on a given input query that is assumed to be erroneous. Furthermore, in contrast to QBE, QRE, and refinement-based approaches, our approach points out which operators need to be  modified rather than returning a complete query. 

\mypar{Query refinement and the empty answer problem}
Query refinement is also related to our approach~\cite{Mishra:2009,Mishra:2008,Mottin:2016}.
Query refinements come in two forms: relax queries to return more results or contract queries to return fewer results.
The former addresses the empty answer problem where a query fails to produce any result, and the latter deals with queries that return too many answers.
Both address quantitative constraints on the query result: the rewritten query should return fewer or more answers, but we do not care what these answers are. In contrast, our work addresses qualitative constraints: the query should return answers with a certain structure and/or content.

\mypar{Provenance in DISC systems}
DISC systems natively support nested data formats such as JSON, XML, Parquet, or Protocol Buffers. Provenance capture for DISC systems has been studied in, e.g., \cite{amsterdamer:pvldb11, interlandi:vldbj18,logothetis:spcc13, ikeda:cidr11, Zheng:2019, Diestelkaemper:2020}. Why-not explanations  are practically relevant in these systems. However, we are not aware of any scalable solution
that computes
why-not explanations.  

\mypar{Provenance for nested data}
Since why-not explanations typically build on the provenance of existing results, our work also relates to work on provenance models for nested data.
Like~\cite{foster:pods08,amsterdamer:pvldb11,Diestelkaemper:2020}, we use a nested data model and query language (a nested relational algebra for bags inspired by~\cite{grumbach:pods93} in our case).

\section{Preliminaries and Notation}\label{sec:prelim}

\subsection{Nested Relational Types and Instances}

Nested relations are bags (denoted as $\bagType{\cdot}$) of tuples where the attributes of a tuple are either of a primitive type (e.g., booleans or integers), 
tuples themselves, or nested relations. \revm{This follows existing models for nested relations~\cite{grumbach:pods93,libkin:jcss97}.} 

\begin{definition}[Nested Relation Schema]\label{def:nested-relation}
Let $\attrLabels$ be an infinite set of names. A \emph{nested type} $\nType$ is an element conforming to the grammar shown below, where each $A_i \in \attrLabels$. A type $\rType$ is called a \emph{nested relation schema}. A nested database schema $\dbType$ is a set of $\rType$ types. 

\begin{center}
\begin{tabular}{rlrl}
$\pType$&$\defas \intType \mid \strType  \mid \boolType \mid \ldots $ & $\rType$&$\defas \bagType{\tType}$ \\
$\tType$&$\defas \tupType{ \attrType{A_1}{\anyType}, \ldots, \attrType{A_n}{\anyType}}$ & $\anyType$&$\defas \pType \mid \tType \mid \rType$ \\
\end{tabular}
\end{center}
\end{definition}

\begin{definition}[Nested Relation Instance]\label{def:nci}
\revm{Let $\primInst$ denote the domain of  primitive type $P$. }
We assume the existence of a special value $\nullVal$  (null) which is a valid value for any nested type. 
We use $\typeOf{\nInst}$ to denote the type of an instance  $\nInst$. The instances $\nInst$ of type $\nType$ are defined recursively based on the following 
rules for primitive types, homogeneous bags, and tuples:  $\frac{\nInst \in \primInst}{\typeOf{\nInst} = P}$,   $\frac{\typeOf{\nInst_1} = \nType, \ldots, \typeOf{\nInst_n} = \nType}{\typeOf{\nrcbag{\nInst_1, \ldots, \nInst_n}} = \bagType{\nType}}$,  \\$\frac{\typeOf{\nInst_1} = \nType_1, \ldots, \typeOf{\nInst_n} = \nType_n}{\typeOf{\nrctuple{\attrType{A_1}{\nInst_1}, \ldots, \attrType{A_n}{\nInst_n}}} = \tupType{\attrType{A_1}{\nType_1}, \ldots, \attrType{A_n}{\nType_n}}}$.

\end{definition}
\begin{example}
All tuples of the nested relation shown in \Cref{tab:nested-data} are of type
$\nrctuple{name: \strType, \:  address1: \nType_{r}, \: address2 : \nType_{r}}$,
where $\nType_{r}$ is a nested relation of type $\nrcbag{\nrctuple{city: \strType, \: year: \dateType}}$.
\end{example}

\subsection{Nested Relational Algebra}
\label{subsec:nra}

 \ifbool{Techreport}{
	\begin{table*}[t]
\centering \scriptsize
\begin{tabular}{| p{1.8cm} l l |} \hline
  \cellcolor{tabbg} \textbf{Operator} & \cellcolor{tabbg} \textbf{Semantics} & \cellcolor{tabbg} \textbf{Output type} $\typeOf{\cdot}$\\ \hline
  Table access & $ \qEval{R} = \nrcbag{ \bt{t}{n} \mid \bt{t}{n} \in \rel}$& $ \rType$ \\ \hline
  Projection &  $\qEval{\projection_L(R)} =  \nrcbag{ \bt{t}{l} |  l = \sum_{t': t'.L = t} \multOf{R}{t'} }$ & $\bagType{\tupType{\attrType{A_{i_1}}{\nType_{i_1}}, \ldots, \attrType{A_{i_m}}{\nType_{i_m}}}}$  for $L = \{A_{i_1}, \ldots, A_{i_m} \}$\\ \hline
  Renaming &  $\qEval{\rho_{B_1 \gets A_1, \ldots, B_n \gets A_n}(R)} = \nrcbag{ \bt{t}{l} | \bt{t'}{l} \in R \wedge  t = \nrctuple{B_{1}: t'.A_{1}, \ldots, B_{n}:t'.A_{n}} }$ & $ \bagType{\tupType{\attrType{B_1}{\nType_1}, \ldots, \attrType{B_n}{\nType_n}}}$ for $\typeOf{A_i} = \nType_i$
  \\
       \hline
  Selection & $\qEval{\selection_\theta(R)} = \nrcbag{\bt{t}{l} | \bt{t}{l} \in R \wedge t \models \theta} $ &
   $\rType$\\ \hline
  Inner join & $ \qEval{R \Join_\theta S} = \nrcbag{ \bt{(t \concat t')}{k \cdot l} | \bt{t}{k} \in R \wedge  \bt{t'}{l} \in S \wedge t \concat t' \models \theta }$ & $\rType \concat \sType$ 
  \\ \hline
  Left outer join & $   \qEval{R \leftouterjoin_\theta S} = R \Join_\theta S \: \cup$  &  $\rType \concat \sType$ 
  \\
  & \hspace{1cm} $\nrcbag{ \bt{(t \concat t_{\nullVal})}{k} | \bt{t}{k} \in R \wedge t \notin \projection_{\schemaOf{R}}(R \Join_\theta S) \wedge t_{\nullVal} = \nrctuple{B_1:\nullVal, \ldots, B_m:\nullVal} }$ & \\
                                        \hline
  Right outer join & $ \qEval{R \rightouterjoin_\theta S} = R \Join_\theta S \: \cup$   & $\rType \concat \sType$ 
  \\
  & \hspace{1cm} $\nrcbag{ \bt{(t'_{\nullVal} \concat s)}{l} | \bt{t'}{l} \in S \wedge t' \notin \projection_{\schemaOf{S}}(R \Join_\theta S) \wedge t'_{\nullVal} = \nrctuple{A_1:\nullVal, \ldots, A_n:\nullVal} }$ & \\
  \hline
  Full outer join & $ \qEval{R \fullouterjoin_\theta S} = (R \leftouterjoin_\theta S \cup R \rightouterjoin_\theta S) - (R \Join_\theta S)$ & $\rType \concat \sType$ 
  \\ \hline
  Tuple flatten & $\qEval{\flatten^{T}_{A}(R)} = \nrcbag{ \bt{(t \concat t.A)}{k} | \bt{t}{k} \in R}$ 
                &
$\makeType{R} \concat \nrcbag{\nType}$ where $\typeOf{\projection_A(R)} = \bagType{\nType}$
  \\ \hline
  Relation inner flatten & $ \qEval{\flatten^{I}_{A}(R)} =  \nrcbag{ \bt{(t \concat u)}{k \cdot l} | \bt{t}{k} \in R \wedge \bt{u}{l} \in t.A }$ 
                                         &
$\makeType{R} \concat \nType$ where $\typeOf{\projection_{A}(R)} = \nType$
  \\ \hline
  Relation outer flatten & $\qEval{\flatten^{O}_{A}(R)} = \flatten^{I}_{A}(R) \: \cup \nrcbag{ \bt{(t \concat u_{\nullVal})}{k} | \bt{t}{k} \in R \wedge t.A = \emptyset \wedge u_{\nullVal} = \nrctuple{B_1:\nullVal, \ldots, B_m:\nullVal} }$
                                         & $\makeType{R} \concat \nType$ where $\typeOf{\projection_{A}(R)} = \nType$
  \\ \hline
  Tuple nesting & $ \qEval{\nestTup{A}{C}(R)} = \nrcbag{ \bt{(t.M \concat \nrctuple{ \attrType{C}{t.A} })}{k} \mid \bt{t}{k} \in R }$& $ \bagType{\nType_{M} \concat \tupType{ \attrType{C}{\nType}}}$ where  $M = \schemaOf{R} - \{A\}$ and $\typeOf{\projection_{A}(R)} = \bagType{\nType}$
  \\ \hline
Relation  nesting &
            $  \qEval{\nestRel{A}{C}(R)} = \nrcbag{ \bt{(t.M \concat ns(R,M,A,C,t))}{1} | t \in gr(R,M) }$ & $ \bagType{\nType_{M} \concat \tupType{ \attrType{C}{\nrcbag{\nType_{A}}}}}$
                                                                                                           where  $M = \schemaOf{R} - \{A\}$ and $\typeOf{\projection_{A}(R)} = \nType$

\\
         & $ gr(R,M) = \{ t.M \mid \bt{t}{n} \in R \}$, $ns(R,M,A,C,t) = \nrctuple{\attrType{C}{\qEval{\projection_{A}(\selection_{M=t.M}(R))}}}$ & \\ \hline
  Aggregation & $ \qEval{\aggregation{f(A)}{B}(R)} = \nrcbag{ \bt{(t \concat \nrctuple{ \attrType{B}{f(t.A)}})}{k}  | \bt{t}{k} \in R }$ &  $ \rType \concat \bagType{\tupType{\attrType{B}{\typeOf{f(A)}}}}$\\ \hline
  Union & $ \qEval{R \cup S} = \nrcbag{ \bt{t}{k+l} \mid \bt{t}{k} \in R \wedge \bt{t}{l} \in S}$& $\rType$ \\ \hline
  Deduplication &   $\delta(R) = \nrcbag{ \bt{t}{1} \mid \bt{t}{k} \in R }$ & $  \rType$ \\ \hline
\end{tabular}
\caption{Evaluation semantics and output types for the operators of our \termNRAB \abbrNRAB. 
} 
 \MH{Projection: do we need a $t' \in R$ somewhere? Also, text does not match table}
\label{tab:ql}

\end{table*}

 }
{
}

Data of the above 
model is manipulated through a \termNRAB (\abbrNRAB). \revm{
  We define  
  \abbrNRAB based on the
  algebra from~\cite{grumbach:pods93,libkin:jcss97}, which we denoted as \baseNRAB. Let $R$ and $S$ denote relations. \baseNRAB includes operators with bag semantics for selection $\selection_\theta(R)$, restructuring $map_f(R)$, cartesian product $R \times S$, additive union $R \cup S$, difference $R - S$, duplicate elimination  $\epsilon(R)$, and bag-destroy~$\delta(R)$. We further define \spcNRAB as the subset of $\baseNRAB$ sufficient to express select-project-join queries, and \spcuNRAB the algebra that additionally includes additive union to express select-project-join-union queries. These less expressive fragments of \baseNRAB represent the operators commonly supported by lineage-based missing-answers approaches. We use them later for a comparative discussion. }

\revm{Similarly to~\cite{rodriguez:cikm16,amsterdamer:pvldb11}, we propose  additional operators based on query constructs supported by big data systems. Together with the operators of \baseNRAB, they form our algebra \abbrNRAB. Our additional operators include 
  attribute renaming $\rho_{B_1 \gets A_1, \ldots, B_n \gets A_n}(R)$ that renames each attribute $A_i$ of $R$ into $B_i$, the projection $\projection_{A_1, \cdots, A_n}(R)$ and join variants (i.e., $R \Join_\theta S$, $R \leftouterjoin_\theta S$, $R \rightouterjoin_\theta S$, and $R \fullouterjoin_\theta S$), as well as aggregation and variants of nesting and flattening. }
\revm{We introduce these operators to achieve a close correspondence between big data programs and the algebra since we aim at explanations that aid users in debugging their programs. Similarly to~\cite{rodriguez:cikm16}, we can derive these operators
from \baseNRAB operators. Before discussing selected operators of our algebra in more detail, we introduce some notational conventions.}

\mypar{Notation} We denote tuples as $t, t', t_1, \ldots$, nested relations as $R, S, T, \ldots$, and nested databases as $D, D', \ldots$. $\rType$ and $\dbType$ denote the type of a nested relation $R$ and database $D$, respectively. Furthermore, $t.A$ denotes the projection of tuple $t$ on  a set of attributes or single attribute~$A$. $\schemaOf{R}$ is the list of attribute names of relation $R$. Operator~$\concat$  concatenates tuples and tuple types. We also apply $\concat$ to relation types, 
e.g., $\nrcbag{\nrctuple{\attrType{A}{\nType_1}}} \concat \nrcbag{\nrctuple{\attrType{B}{\nType_2}}} = \nrcbag{\nrctuple{\attrType{A}{\nType_1}, \attrType{B}{\nType_2}}}$.
We use 
$\bt{t}{n} \in \rel$ to denote that tuple $t$ appears in relation $\rel$ with multiplicity $n$ and we use arithmetic operations on multiplicities, e.g., $\bt{t}{2+3}$ means that tuple $t$ appears $5$ times.  $\multOf{R}{t}$ denotes the multiplicity of tuple~$t$ in relation~$\rel$.
We use $\qEval{\query}_{\db}$ to denote $\query$'s evaluation result over $\db$. We omit $\db$ if it is clear from the context.
Finally, $\typeOf{\query}$ denotes the result type of $\qEval{\query}$.

Now, we define selected operators \revm{with ambiguous bag semantics or without well-known semantics}. We assume $R$ is an n-ary input relation of type $\rType = \nrcbag{\tupType{\attrType{A_{1}}{\nType_{A_1}}, \ldots, \attrType{A_{n}}{\nType_{A_n}}}}$.

 \iftechreport{
\mypar{Table access} If $R$ is a relation with type $\rType$, then the table access operator for $R$ is denoted as $R$.
\begin{align*}
  \qEval{R} &= \nrcbag{ \bt{t}{n} \mid \bt{t}{n} \in \rel}\\
\typeOf{R} &= \rType
\end{align*}

\mypar{Projection} Let $L =\{A_{1}, \ldots, A_{l}\}$ where $A_i \in \schemaOf{R}$ for $i \in \{1,\ldots, l\}$. Furthermore, we require that for each $j, k \in \{1, \ldots, m\}$ \MH{$m$ or $l$ as subscript? Also, why do we need the $C_i$'s here? It was about the order of ``output'' attributes being different from ``input attributes'', right? Then I guess $m$ subscript is OK but $j$ overloaded?} we have $A_j \neq A_k$ when $j \neq k$. 
Defining $\nType_{out}$ as $\tupType{\attrType{A_{1}}{\nType_{A_1}}, \ldots, \attrType{A_{l}}{\nType_{A_l}}}$ where $\nType_{A_i} = \nType_{A_j}$ for $A_i = A_j$,
the projection $\projection_L(R)$ of relation $\rel$ on $L$ is defined as:
 \begin{align*}
   \qEval{\projection_L(R)} &=  \nrcbag{ \bt{t}{l} | \typeOf{t} = \nType_{out} \wedge l = \sum_{t': t'.L = t} \multOf{R}{t'} }\\
\typeOf{\projection_L(R)} &= \bagType{\tupType{\attrType{A_{1}}{\nType_{A_1}}, \ldots, \attrType{A_{l}}{\nType_{A_l}}}}
 \end{align*}

 \mypar{Renaming}\BG{Removed the requirement that the function is injective}
 Let $f$ be a injective function $\schemaOf{R} \to \attrLabels$ (recall that $\attrLabels$ is the set of all allowable identifiers). We write $f$ as a list of elements $B_i \gets A_i$ which each represents one input output pair, 
 e.g., we rename attribute $A_i$ as $B_i$. Renaming $\rho$ renames the attributes of relation $\rel$ using function $f$.
 \begin{align*}
   \qEval{\rho_{B_1 \gets A_1, \ldots, B_n \gets A_n}(R)} &= \\ \nrcbag{ \bt{t}{l} | \bt{t'}{l} \in R \wedge t &= \nrctuple{B_{1}: t'.A_{1}, \ldots, B_{n}:t'.A_{n}} }\\
   \typeOf{\rho_{B_1 \gets A_1, \ldots, B_n \gets A_n}(R)} &=  \bagType{\tupType{\attrType{B_1}{\nType_1}, \ldots, \attrType{B_n}{\nType_n}}}
 \end{align*}

 \mypar{Selection} Let $\theta$ be a condition consisting of comparisons between attributes from relation $R$ and constants, and logical connectives. Selection $\selection_\theta(R)$ filters out all tuples that do not fulfill condition $\theta$ (denoted as $t \models \theta$).
 \begin{align*}
   \qEval{\selection_\theta(R)} &= \nrcbag{\bt{t}{l} | \bt{t}{l} \in R \wedge t \models \theta}\\
   \typeOf{\selection_\theta(R)} &= \rType
 \end{align*}

 \mypar{Join} Let $\theta$ be a condition over the attributes of relations $R$ and $S$. The (inner, left, right, and full outer) joins are defined as follow:
 \begin{itemize}
 \item Inner join
 \begin{align*}
   \qEval{R \Join_\theta S} &= \nrcbag{ \bt{(t \concat t')}{k \cdot l} | \bt{t}{k} \in R
      \wedge  \bt{t'}{l} \in S \wedge t \concat t' \models \theta }\\
   \typeOf{R \Join_\theta S} &= \rType \concat \sType
 \end{align*}
\item Left outer join
 \begin{align*}
   \qEval{R \leftouterjoin_\theta S} &= R \Join_\theta S \: \cup \nrcbag{ \bt{(t_R \concat t_{\nullVal})}{k} | \bt{t}{k} \in R \\
   & \hspace{-10mm} \wedge t \notin (R \Join_\theta S)  \wedge t_{\nullVal} = \nrctuple{B_1:\nullVal, \ldots, B_m:\nullVal} }\\
   \typeOf{R \leftouterjoin_\theta S} &= \rType \concat \sType
 \end{align*}
\item Right outer join
  \begin{align*}
    \qEval{R \rightouterjoin_\theta S} &= R \Join_\theta S \: \cup \nrcbag{ \bt{(t'_{\nullVal} \concat t')}{l} | \bt{t'}{l} \in S \\
    &  \hspace{-10mm} \wedge t' \notin (R \Join_\theta S) \wedge t'_{\nullVal} = \nrctuple{A_1:\nullVal, \ldots, A_n:\nullVal} } \\
   \typeOf{R \rightouterjoin_\theta S} &= \rType \concat \sType
 \end{align*}
\item Outer join
  \begin{align*}
   \qEval{R \fullouterjoin_\theta S} &= (R \leftouterjoin_\theta S \cup R \rightouterjoin_\theta S) - (R \Join_\theta S) \\
   \typeOf{R \fullouterjoin_\theta S} &= \rType \concat \sType
 \end{align*}
\end{itemize}
}

\mypar{Flatten}
The flatten operator unnests the values of an attribute $A \in \schemaOf{R}$ 
which must be of a tuple or relation type.
If $A$ is of a tuple type
$\nType = \tupType{\ldots}$,
then  the \textbf{tuple flatten} operator returns a tuple $\bt{(t \concat t.A)}{k}$ for each $\bt{t}{k}$ in $R$: $\qEval{\flatten^{T}_{A}(R)} = \nrcbag{ \bt{(t \concat t.A)}{k} | \bt{t}{k} \in R}$. Its result type is the concatenation of  $\makeType{R}$ and $\nType$: $ \typeOf{\flatten^{T}_{A}(R)} = \makeType{R} \concat \nrcbag{\nType}$.

If $A$ is of a nested relation type $\nType = \bagType{\tupType{\attrType{B_1}{\nType_1'}, \ldots, \attrType{B_m}{\nType_m'}}}$, then \textbf{inner relation flatten} returns each tuple $\bt{u}{l}$ in the nested relation concatenated with the tuple $ \bt{t}{k}$ it was initially nested in:  $  \qEval{\flatten^{I}_{A}(R)} = \nrcbag{ \bt{(t \concat u)}{k \cdot l} | \bt{t}{k} \in R \wedge \bt{u}{l} \in t.A }$ and  $\typeOf{\flatten^{I}_{A}(R)} = \makeType{R} \concat \nType$. We require that none of the attribute names $B_i$ already exist in $\makeType{R}$.

An \textbf{outer relation flatten} behaves similarly to inner relation flatten but additionally returns tuples of $R$ padded with null values where the value of the flattened attribute is the empty relation. That is, using $u_{\nullVal} = \nrctuple{B_1:\nullVal, \ldots, B_m:\nullVal}$, we define $\qEval{\flatten^{O}_{A}(R)} = \flatten^{I}_{A}(R)\,\,\, \cup \nrcbag{ \bt{(t \concat u_{\nullVal})}{k} \mid \bt{t}{k} \in R
   \wedge t.A = \emptyset} $. 

\mypar{Nesting}
Analogously to the flatten operators, we define two nesting operators: tuple nesting and relation nesting.

Given an attribute set $\squeezespaces{0.5} A \subseteq \schemaOf{\rel}$, \textbf{tuple nesting} removes  attribute(s) $A$ from each tuple  
$\squeezespaces{0.5} t \in R$  and adds new attribute $C$ of type $\nType_{A}$ (the tuple type in relation type $\typeOf{\projection_A(R)}$) storing 
$t.A$.
Using  $M = \schemaOf{R} - A$ and $\nType_{M}$ to denote the tuple type of $\typeOf{\projection_{M}(R)}$, we define
$\qEval{\nestTup{A}{C}(R)} = \nrcbag{ \bt{(t.M \concat \nrctuple{ \attrType{C}{t.A} })}{k} \mid \bt{t}{k} \in R }$. Accordingly, $\typeOf{\nestTup{A}{C}(R)} = \bagType{\nType_{M} \concat \tupType{ \attrType{C}{\nType_{A}}}}$.

{\sloppy

  \textbf{Relation nesting} $\nestRel{A}{C}(R)$ groups $R$ on 
  $M$. For each group in $gr(R,M) = \{ t.M \mid \bt{t}{n} \in R\}$, the operator returns a tuple with the group-by values ($t.M$) and a fresh attribute $C$ of relation type $\nType_{A} = \typeOf{\projection_{A}(R)}$ that stores the projection of all tuples from the group on $A$ as a nested relation $ns(R,M,A,C,t) = \nrctuple{\attrType{C}{\qEval{\projection_{A}(\selection_{M=t.M}(R))}}}$. Overall, the result of relation nesting is
  $$\qEval{\nestRel{A}{C}(R)} = \nrcbag{ \bt{(t.M \concat ns(R,M,A,C,t))}{1} | t \in gr(R,M) }$$
  with associated type
  $\typeOf{\nestRel{A}{C}(R)} = \bagType{\nType_{M} \concat \tupType{ \attrType{C}{\nrcbag{\nType_{A}}}}}$.
}

\mypar{Aggregation}
Consider an aggregation function $f$ of type $\bagType{\tupType{\attrType{C}{\nType}}} \to \nType_{out}$ and let $\nType_{in} = \bagType{\tupType{\attrType{C}{\nType}}}$.  The  \textbf{aggregation} operator applies $f$ to the set of values of unary tuples in the results of $\projection_A(R)$ and stores the result in a new attribute $B$ that is of type $\nType_{out}$. Attribute $A$  has to be of type $\nType_{in}$.  Thus, $ \qEval{\aggregation{f(A)}{B}(R)} = \nrcbag{ \bt{(t \concat \nrctuple{ \attrType{B}{f(t.A)}})}{k}  | \bt{t}{k} \in R }$ and its output type is  $\typeOf{\aggregation{f(A)}{B}(R)} = \rType \concat \bagType{\tupType{\attrType{B}{\nType_{out}}}}$.

\iftechreport{
\mypar{Union}
Let $R$ be a relation with schema $\rType$ 
$ =  \bagType{\tupType{\attrType{A_1}{\nType_1}, \ldots, \attrType{A_n}{\nType_n}}}$ and $S$ be a relation with schema $\sType$ 
$ = \bagType{\tupType{\attrType{B_1}{\nType_1}, \ldots, \attrType{B_n}{\nType_n}}}$.
\begin{align*}
  \qEval{R \cup S} &= \nrcbag{ \bt{t}{k+l} \mid \bt{t}{k} \in R \wedge \bt{t}{l} \in S}\\
  \typeOf{R \cup S} &= \rType
\end{align*}
 Recall that we employ the convention  that $\bt{t}{0} \in R$ is true if the tuple $t$ is not part of relation $R$.

 \mypar{Duplicate Elimination} Operator $\delta$ eliminates duplicates.
 \begin{align*}
   \delta(R) &= \nrcbag{ \bt{t}{1} \mid \bt{t}{k} \in R }\\
   \typeOf{\delta(R)} &= \rType
 \end{align*}
}

\begin{example}\label{ex:group-by-aggregation} \sloppy
  The operator pipeline of \Cref{fig:sample-pipelines} corresponds to \revm{the following expression in \abbrNRAB:}
  \[\nestRel{name}{nList} \left( \projection_{name, city}\left(\selection_{year \geq 2019}\left( \flatten^I_{address2}\left(\texttt{person}\right) \right) \right) \right)\]

\end{example}

 \section{Why-Not Explanations}  
\label{sec:why-not}
We are now ready to  
formalize the problem of computing why-not explanations for nested (and flat) data.

\subsection{Why-Not Questions}\label{sec:why-not-questions}

A why-not question describes a (set of) missing, yet expected (nested) tuple(s) in a query's result $\qEval{Q}_\db$.
We let users specify why-not questions as \emph{nested instances with placeholders} (\abbrNIPs). Intuitively, a \abbrNIP incorporates placeholders to represent \emph{a set of} missing answers, any of which is acceptable to the user. We introduce the \emph{instance placeholder} $\valPlaceholder$ that can stand in for any value of a type and the \emph{multiplicity placeholder}  $\multPlaceholder$,  which can only be used as element of a nested relation type and represents $0$ or more tuples of a nested relation's tuple type. \BG{If we need space, remove:} Note that for finite domains, the expressive power of why-not questions with placeholders is not larger than why-not questions based on fully specified tuples. But efficiently supporting the former avoids the exponential blow-up incurred when naively translating them to the latter representation.

\begin{definition}[Instances with Placeholders]\label{def:instances-with-place}
  Let $\nType$ be a nested type. The rules to construct nested instances with placeholders~(\abbrNIPs) of type $\nType$ are: If $~\typeOf{\nInst} = \nType$ or $\nInst = \valPlaceholder$, then $\nInst$ is a \abbrNIP of type $\nType$. Furthermore, if $\nType = \tupType{\attrType{A_1}{\nType_1}, \ldots,\attrType{A_n}{\nType_n}}$, then $\nrctuple{\nInst_1, \ldots, \nInst_n}$ is a \abbrNIP of type $\nType$ if each $\nInst_i$ is a \abbrNIP of type $\nType_i$.
Finally, if $\nType = \bagType{\nType_{tup}}$, then $\nrcbag{\nInst_1, \ldots, \nInst_n}$ is a \abbrNIP of type $\nType$ if (i) $\forall$ $I_i$ either $\typeOf{I_i} = \nType_{tup}$, $\nInst_i = \valPlaceholder$, or $\nInst_i = \multPlaceholder$ and (ii) $\not \exists$ $i \neq j \in \{1, \ldots, n\}$ such that $\nInst_i = \nInst_j = \multPlaceholder$.
\end{definition}

\begin{example}
\label{ex:nip}
\revc{
  A \abbrNIP that conforms to the output schema of our running example is $t_{ex} = \nrctuple{city: NY, nList : \nrcbag{\valPlaceholder, \multPlaceholder}}$. It stands for all tuples with  city equal to  NY and at least one 
  name in nList.}
\end{example}

 Next, we define the set of nested instances that match a \abbrNIP.

\begin{definition}[Matching \abbrNIPs]\label{def:tuple-matching-whynot}
  An instance $\nInst$ of type $\nType$ matches a \abbrNIP $\nInst'$ of type $\nType$, written as $\nInst \matches \nInst'$ if one of these conditions holds:
  \begin{enumerate}
  \item $\nInst' =~\valPlaceholder$
  \item $\nInst = \nInst'$
  \item $\typeOf{\nInst} = \tupType{\attrType{A_1}{\nType_1}, \ldots, \attrType{A_n}{\nType_n}}$ and $\forall i \in [1,n]$, $\nInst.A_i \matches \nInst'.A_i$
  \item $\typeOf{\nInst} = \bagType{\nType_{tup}}$ and there exists an assignment $\matching \subseteq \nInst \times \nInst' \to \mathbb{N}$ such that all conditions below hold:
  \begin{enumerate}
  \item for all $\aTup \in \nInst$ and $\aTup' \in \nInst'$, \revc{if} $\matching(\aTup, \aTup') > 0$ 
  \revc{then either}
    $\aTup = \aTup'$, $\aTup' =~\valPlaceholder$, or $\aTup' = \multPlaceholder$
    \item for all $\aTup \in \nInst$, $\sum_{\aTup' \in \nInst'} \matching(\aTup,\aTup') = \multOf{\nInst}{\aTup}$
    \item for all $\aTup' \in \nInst'$, either $\sum_{\aTup \in \nInst} \matching(\aTup, \aTup') = \multOf{\nInst'}{\aTup'}$ or $\aTup' = \multPlaceholder$
    \end{enumerate}
  \end{enumerate}

\end{definition}

Condition~(4) ensures \revc{that multiplicies are taken into account.} 

\begin{example}\label{ex:matching-nix}
\revc{Consider \abbrNIP $\aTup_{ex}$ from \Cref{ex:nip} as well as \abbrNIP $\aTup_{ex}' =  \nrctuple{city: NY, nList : \nrcbag{\valPlaceholder, \valPlaceholder}}$. Only the former matches the tuple $\aTup = \nrctuple{city: NY, nList : \nrcbag{\nrctuple{name: Sue}^2, \nrctuple{name: Peter}}}$.
Since $\aTup$ is of a tuple type, condition (3) in \Cref{def:tuple-matching-whynot} must hold.
While both satisfy condition (2) because $\aTup_{ex}.city \matches \aTup.city$  and $\aTup_{ex}'.city \matches \aTup.city$, condition (4) only holds for $\aTup_{ex}.nList \matches \aTup.nList$. For $\aTup_{ex}'$, the definition enforces that $\matching(\nrctuple{name:Sue}, ?) > 0$ and  $\matching(\nrctuple{name:Peter}, ?) > 0$ (condition (4a)) and $\matching(\nrctuple{name:Sue}, ?) = 2$  (4b). Then, (4c) cannot hold, since the sum is
3 and $\multOf{\aTup_{ex}'}{\valPlaceholder} = 2$. Alternatively assigning each occurrence of $\nrctuple{name:Sue}$ to $\valPlaceholder$ cause a violation of (4b). 
}
\end{example}

 \ifbool{Techreport}{
\begin{example}
The following \abbrNIP $\aTup$ matches the second tuple shown in \Cref{tab:nested-data} (denoted as $\aTup'$ in the following).
\[\aTup = \nrctuple{Name: Sue, address1 : ?, address2 : \nrcbag{\nrctuple{city : ?, year : 2019}, \multPlaceholder}}\]
Indeed,
$Sue \matches Sue$, $\nrcbag{\nrctuple{\aval{city}{LA}, \aval{year}{2019}}, \nrctuple{\aval{city}{NY}, 2018}} \matches \valPlaceholder$, and
$\aTup'.address2 \matches \nrcbag{\nrctuple{\aval{city}{?}, \aval{year}{2019}}, \multPlaceholder}$, through
\begin{align*}
  \matching(\nrctuple{\aval{city}{LA}, \aval{year}{2019}}, \nrctuple{\aval{city}{?}, \aval{year}{2019}}) &= 1\\
  \matching(\nrctuple{\aval{city}{NY}, \aval{year}{2018}}, \multPlaceholder) &= 1
\end{align*}
\end{example}
}

Using \abbrNIPs, we now define why-not questions. To ensure that a why-not question asks for a tuple absent from the result, we require that none of the result tuples matches the why-not question's \abbrNIP.

\begin{definition}[Why-not questions]\label{def:whynot-q}
  Let $\query$ be a query, $\db$ a database, and $\typeOf{\qEval{Q}_\db} = \bagType{\nType}$.
  A why-not question $\whynot$ is a triple $\whynot = \aWhynot$ where why-not tuple $\aTup$ is a \abbrNIP of type $\nType$.
\end{definition}

\begin{example}\label{ex:why-not-q}
\revc{Given $\db$ and $Q$ from \Cref{fig:running-example}, and the \abbrNIP $t_{ex}$ from \Cref{ex:nip}, the example why-not question is $\whynot_{ex} = \nrctuple{Q, \db, t_{ex}}$.}
\end{example}

\subsection{Reparameterizations and Explanations}\label{sec:repar-expl}

We define query-based explanations for a given why-not question $\whynot$ as sets of operators. 
An explanation is a combination of operators that conjunctively cause tuples matching the NIP $t$ in $\whynot$ to be missing from the query result, i.e., it is possible to ``repair'' the query to return a tuple matching the NIP $t$ (the missing answer) by changing the parameters of these operators. We refer to such repairs as \textit{successful reparameterizations}.
\revm{The set of explanations produced for a why-not question should consist of sets of operators changed by successful reparameterizations. However, we do not want to return explanations that require more changes than strictly necessary. That is, we want explanations to be minimal in terms of the set of operators they include and in terms of their  ``side effects'' (changes to the original query result beyond appearance of missing answers) a reparametrization of an explanation's operators would have.}
Existing lineage-based definitions, \revm{which generally support queries in~\spcuNRAB,}  do not fulfill our desiderata:  (i) They suffer from possibly incomplete explanations (false negatives)~\cite{DBLP:conf/sigmod/ChapmanJ09,DBLP:conf/edbt/BidoitHT14,herschel:jdiq15}, i.e., changing the operator they return as an explanation may not be sufficient for returning the missing answer. This motivated alternative definitions~\cite{bidoit:cikm15,deutsch:pvldb18,diestelkaemper:tapp19}, albeit limited to conjunctive queries \revm{in~\spcNRAB}. (ii) They only reason about operators that prune data (explanations only contain
 selections and joins) and miss causes at the schema level (e.g., projecting the wrong attribute). (iii) They disregard side effects (which have been considered for instance-based and refinement based explanations~
~\cite{herschel:vldbj17}). Our formalization addresses all these drawbacks for queries \revm{in the rich algebra \abbrNRAB}.

Our formalization is based on \textbf{reparameterizations (RPs)}. A RP for an input query $\query$ is a query $\query'$ that is derived from $\query$ by altering the parameters of operators while preserving the query structure (no operators are added or removed).
 For instance,   
 changing  $\selection_{year \geq 2019}$ to $\selection_{year \geq 2018}$ in our running example is a RP, but substituting the selection with a projection is not. We made the choice to preserve query structure to avoid explanations that do not provide meaningful information about errors in the input query.

  \revm{ \Cref{tab:param} summarizes all admissible parameter changes for all \abbrNRAB operators}. They are motivated by what we consider errors commonly arising in practice. Nonetheless, our formalism also applies to alternative definitions of valid parameter changes. \revm{However, the choice of allowed parameter changes affects the compuational complexity of the problem (see \Cref{sec:complex-analy})}.

\begin{table}[t]
\scriptsize \centering
\begin{tabular}{| p{2.7cm} | p{2cm} | p{3.2cm}|} \hline
 \cellcolor{tabbg}   \textbf{Operator $op$} &  \cellcolor{tabbg}  \textbf{$param(Q,op)$} &   \cellcolor{tabbg}  \textbf{Admissible parameter changes} \\ \hline
      Selection $\selection_{\theta}(R)$, with $\theta$ including attribute references, comparison operators  ($=, >, \geq, <, \leq, \neq\}$), and constant values &  $\{ \theta \}$   &  Replacing (i) an attribute reference with another attribute from $R$ of same data type; (ii) a comparison operator by another; and (iii)~a constant with another constant of same type. \\ \hline
  Restructuring $map_f(R)$ & $\{f\}$ & Change $f$ \\ \hline
  Projection $\projection_L(R)$ & $\{A_i | A_i \in L\}$  &  Any substitution of an attribute $A_i$ with an attribute $A_j$ from $R$   \\ \hline
  Renaming $\rho_{B_1 \gets A_1, \ldots, B_n \gets A_n}(R)$ & $\{ (B_1 \gets A_1, \ldots,$ $ B_n \gets A_n) \}$ & Changing the output attributes based on a permutation of  $(B_1, \ldots, B_n)$  
 \\
       \hline
Join variants $R \somejoin_\theta S$ , where  
   $\somejoin \in \{ \join, \leftouterjoin, \rightouterjoin, \fullouterjoin \}$
   	& $\{ \theta, \typeOf{op} \}$, where  $\typeOf{op} = \somejoin$ 
	& (i) Changing the join type of $op$; (ii) replacing a reference to an attribute $A$ with a different attribute $B$ in $\theta$; (ii) modifying comparison operators in $\{=, >, \geq, <, \leq, \neq\}$ to one another. \\ 
	\hline
Flatten variants  $\flatten^{\diamond}_{A}(R)$, where $\diamond \in \{T,I, O\}$
	& $ \{ A, \typeOf{op} \}$, where $\typeOf{op} = \diamond$ distinguishes tuple flatten, relation inner flatten, and relation outer flatten
	&  (i) Replacing $A$ by an attribute $B$ in $R$ of tuple type for $\diamond = T$ or relation type otherwise, (ii) changing the flattening type from inner flatten to outer flatten or vice versa \\ \hline
Nesting variants $\nestRel{A}{C}(R)$ or $\nestTup{A}{C}(R)$ 
	& $ \{ A, C \}$  
	&   (i) Changing the attributes to be nested / grouped-on ($A$) or (ii) the name of the attribute storing the result of nesting ($C$) \\ \hline
Aggregation $\aggregation{f(A)}{B}(R)$ 
	& $\{ A, B, f \}$  &
	(i) Changing the aggregation function $f$, (ii) the attribute that we are aggregating over ($A$), or (iii) the name of the attribute storing the aggregation result ($B$) \\ \hline
\end{tabular}
Further parameter-free \abbrNRAB operators are: additive union  $ R \cup S$, difference  $ R - S $, deduplication $\epsilon(R)$, cartesian product $R \times S$, bag-destroy $\delta(R)$, and table access $R$
\caption{Admissible parameter changes of \abbrNRAB operators.
}
\label{tab:param}
\vspace{-0.8cm}
\end{table}

\begin{definition}[Valid Parameter Changes]
\revm{  Given an operator $op \in \query$ with parameters $\paramOf(\query,op)$ and a set of predefined admissible parameter changes for this operator type (~\Cref{tab:param}), a valid parameter change applies one admissible change to $\paramOf(\query,op)$.}

\end{definition}

Based on the parameter changes, we define reparameterizations.

\begin{definition}[Reparameterizations]
  Given a query~$\query$, a query $\query'$ is a reparameterization of $\query$ if it can be derived from $\query$ using a sequence of valid parameter changes. 
\end{definition}
For the ease of presentation, we assign each operator $op \in \query$ a unique identifier. Since $\query$ and $\query'$ have same structure, we further assume that an operator $op \in \query$ retains its identifier in $\query'$.
Next, we relate RPs to a why-not question. RPs are \textbf{Successful reparameterizations (SRs)} if they produce the missing answer.

\begin{definition}[Successful Reparameterizations]\label{def:srs}
  Let $\whynot = \aWhynot$ be a why-not question. Denoting $\reparams{\query}$ the set of all RPs for a query $\query$, we define $\SR{\whynot}$, the set of successful RPs for $\query$ and $\db$, as 
$$\SR{\whynot} = \{ \query' \mid \exists \aTup' \in \qEval{Q'}_\db, \aTup' \matches \aTup \wedge \query' \in \reparams{\query} \}$$
\end{definition}

\begin{example}
\Cref{fig:tree-data} shows a tree representation of nested relations (introduced here as these will become relevant later). 
The tree $T_1$ in 
\Cref{fig:tree1} corresponds to the result~$\qEval{Q}_\db$ in our example (\Cref{tab:output}).
The example why-not question asks why city NY with associated names is missing from $\qEval{Q}_\db$. One possible SR ($SR_\selection$) changes the selection predicate (e.g., to $year \geq 2018$). This SR produces the result $T_2$ in \Cref{fig:tree2}. Another SR ($SR_{\flatten\selection}$) modifies the selection and changes the flattened attribute to $address1$. It yields tree $T_3$  (\Cref{fig:tree3}). Additional SRs exist, e.g., changing the year to anything lower than 2018. However, they result in additional changes to~$\qEval{Q}_\db$.
\label{ex:SRs}
\end{example}

\begin{figure*}[t]
\centering \scriptsize
\begin{minipage}{0.72\linewidth}
\begin{minipage}{0.17\linewidth}
\centering
\includegraphics[height=2.4cm]{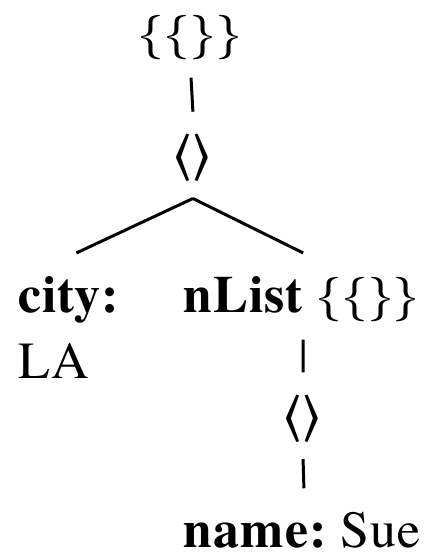}
  \subcaption{$\tree_1$}\label{fig:tree1}
\end{minipage}
\begin{minipage}{0.43\linewidth} 
\centering
  \includegraphics[height=2.4cm]{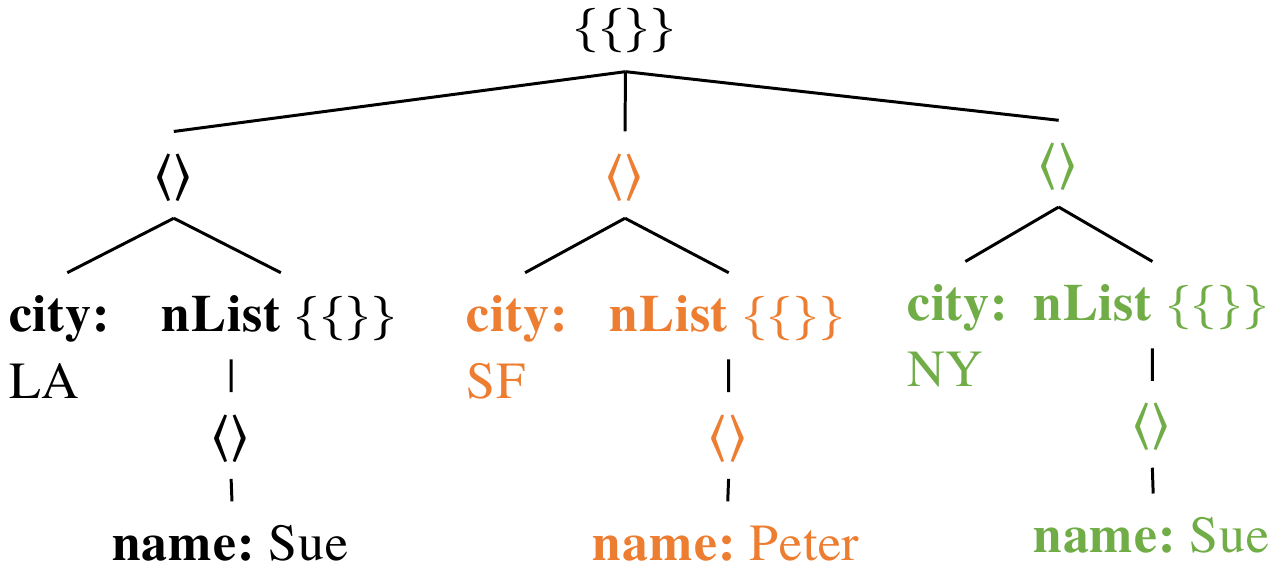}
  \subcaption{$\tree_2$,  result for  $SR_{\selection}$ }\label{fig:tree2}
\end{minipage}
\begin{minipage}{0.35\linewidth} \hspace{3mm}
\centering
  \includegraphics[height=2.4cm]{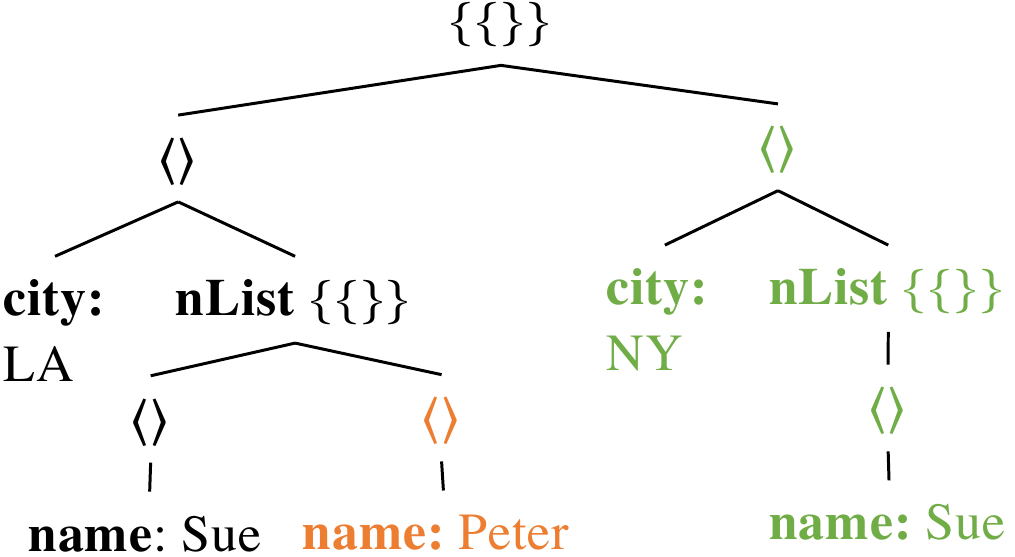}
  \subcaption{$\tree_3$, result for  $SR_{\flatten\selection}$ }\label{fig:tree3}
\end{minipage}
\trimfigspace
\caption{Tree representations of the result in \Cref{tab:output} ($\tree_1$) and results of SRs 
\label{fig:tree-data}
}
\end{minipage}
\begin{minipage}{0.25\textwidth}
\null
\removelatexerror
\begin{algorithm}[H]
\scriptsize
$ \nrctuple{\sbtMap, \overline{T} } \leftarrow$ schemaBacktracing$(\whynot)$ \\ \label{l:unrestructure}
$\aSchemaOptCollection \leftarrow$ schemaAlternatives$(\sbtMap, \overline{T} , \whynot)$ \\ \label{l:comp_alt}
$\aProvenanceAnnotationCollection \leftarrow$ dataTracing$(\aSchemaOptCollection, \whynot)$ \\ \label{l:comp_wn}
$\expl^\approx \leftarrow$ approximateMSRs$(\aProvenanceAnnotationCollection, \aSchemaOptCollection, \whynot)$ \\
\Return{$\expl^\approx$}
\label{l:comp_rewrites}
\caption{Why-Not($\whynot$)}
\label{alg:overview}
\end{algorithm}
\end{minipage}
\vspace{-0.3cm}
\end{figure*}

The above example illustrates our rationale to not consider all $\SR{\whynot}$ as explanations:. (i) some SRs may apply unnecessary changes to $\query$ (e.g., why change both selection and 
flatten operator when one is enough?) and (ii) some SRs may cause more changes to the original query result than others (e.g., the side effects caused by a less restrictive selection). \Cref{fig:tree-data} shows that these two goals (minimizing changes to operators and minimizing side effects)  may be in conflict. Green nodes indicate data matching the why-not tuple, 
orange nodes mark 
data not machting the why-not tuple. While $T_2$ has an entirely 
orange tuple $\nrctuple{city:SF, nList:\nrcbag{\nrctuple{name:Peter}}}$, $T_3$ only holds an additional name Peter in the $nList$ attribute for LA. Thus, $SR_{\selection}$ changes only a subset of $SR_{\flatten\selection}$'s operators, but $SR_{\selection}$ entails ``more significant'' changes to the data ($T_2$) than $SR_{\flatten\selection}$ ($T_3$). 
To strike a balance between changes to the query and to the data, we define a partial order $\leqMSR$ over SRs and 
\textbf{minimal successful reparameterizations~(MSRs)} as SRs that are minimal according to $\leqMSR$. We consider all MSRs as \textbf{explanations}.

\begin{definition}[MSRs] \label{def:msrs}
  Let $\whynot = \aWhynot$ be a why-not question and $\query', \query''$ be two SRs. Let $\changed{\query}{\query'}$  
denote the set of identifiers of operators whose parameters differ between $\query$ and~$\query'$, i.e.,
 $ \changed{\query}{\query'} = \{ op \mid \paramOf(\query, op) \neq \paramOf(\query', op) \}$.
Let $d$ be a distance function quantifying the distance between two nested relations.  We define a partial order $\query' \leqMSR \query''$  as follows:
\begin{align*}
&\textit{(1)} \: \changed{\query}{\query'} \subseteq \changed{\query}{\query''}\\
&\textit{(2)} \: \ted(\qEval{Q}_\db,\qEval{Q'}_\db) \leq \ted(\qEval{Q}_\db,\qEval{Q''}_\db) 
\end{align*}
  We call
  $\query' \in \SR{\whynot}$ \emph{minimal} if 
$\neg \exists \query'' \in \SR{\whynot}: \query'' \leqMSR \query'$.

\label{def:msr}
\end{definition}

\begin{definition}[Explanations]\label{def:explanation}
  Let $\whynot$ be a why-not question and $\MSR{\whynot}$ be the set of MSRs for $\whynot$. 
  We define the set of
explanations $\expl(\whynot)$ with respect to $\whynot$ as
$\expl(\whynot) = \{ \changed{\query}{\query'} \mid \query' \in \MSR{\whynot} \}$. 
\end{definition}
\begin{example}
  The 
  $SR_\selection$ and $SR_{\flatten\selection}$  from \Cref{ex:SRs} are also MSRs, because, even though $\changed{\query}{\query'_{\selection}} \subseteq \changed{\query}{\query'_{\flatten\selection}}$, we established that $\ted(\qEval{\query},\qEval{\query'_{\selection}}) > \ted(\qEval{\query},\qEval{\query'_{\flatten\selection}})$, so $\query'_{\selection} \not \leqMSR \query'_{\flatten\selection}$ (and vice versa).
We further use this example to highlight why we define query-based explanations even though refinement-based explanations are not far fetched given reparameterizations. Assuming we had $n$ address attributes (not just 2), there would be equally many refinement-based explanations involving the flatten operator, some also modifying the selection, others not. Thus, a developer would need to go through all of these and understand their similarities and differences before settling on how to fix the query. In contrast, query-based explanations identify sets of operators that need to be fixed.
\end{example}

The MSR definition leaves the choice on the distance function $d$ open. To equally support nested and flat data, a good fit is the tree edit distance for unsorted trees~\cite{BP05,MN11}. However, it is 
\nphard~\cite{KR92}. Considering an alternative \ptime distance metric $d$ will not necessarily result in an efficient algorithm for computing explanations, because, as discussed next, even for metrics computable in \ptime, computing explanations 
is in general \nphard.

\MH{old text: Note that we only identify operators as potential causes instead of actually repairing the query to limit the complexity of the problem. Furthermore, we would like to point out that, for larger instances and/or complex queries, we cannot expect the user to specify all missing answers that should be produced by a query. In this case, a repair that only returns the subset of the missing answers provided by the user is unlikely to produce all missing answers. Thus, we argue that it is more important to identify parts of the query that may be incorrect rather than returning a repair that is likely to be too restrictive (over-fitted to the missing answer specified by the user).}
\BG{The counter argument is of course that the we may in this case return only a subset of the operators that are wrong since some other operators may be needed to be changed to return the missing answers that the user has left unspecified.}

\BG{To some degree we are making contradictory arguments in motivating the definition of explanations (minimal changes) and when motivating the definition of explanation as sets of operators instead of repairs (do not overfit). Can we change the later maybe to something less contradictory?} \MH{Better with new text in last example? }

\subsection{\revm{Discussion}}\label{sec:complex-analy}

\revc{First, we demonstrate} that computing explanations for why-not questions is generally \nphard in terms of data complexity for queries in  \baseNRAB \ifnottechreport{(see~\cite{techreport})}. 
\revc{We observe that the problem is sensitive to the choice of admissible parameter changes. While it remains intractable for the parameter changes defined in \Cref{tab:param}, we identify restrictions of \Cref{tab:param} for which the problem is in \ptime.}

\begin{Theorem}\label{theo:expl-hardness}
\revc{  Given a why-not question $\whynot$ for a query $\query$, database~$\db$,  and a set $e$ of operators from the query. Testing the membership of $e$ in $\expl(\whynot)$ is \nphard in the size of $\db$ for queries consisting only of operators aggregation, map, projection, renaming, and join. \iftechreport{The problem remains hard if we disallow restructuring map or if we restrict the functions that can be used in reparameterizations for aggregation or map (but not both at the same time).} The problem is in \ptime for queries where map is restricted to be a projection and if aggregation functions are  restricted to the standard ones supported in SQL. \iftechreport{The problem is \nphard under these restrictions if we relax the set of admissible parameter changes for selection to allow the structure of the selection condition to be changed.}}
\end{Theorem}
\ifnottechreport{
  \begin{proof}[Proof Sketch]
\revc{    We first prove the hardness claim for queries including aggregation  theorem through a reduction from set cover. 
For the \ptime result, we sketch a brute force algorithm.
    For the full proof, please see~\cite{techreport}.}
  \end{proof}
}

\iftechreport{
  \begin{proof}
\revc{We first prove the hardness claim for queries including aggregation theorem through a reduction from set cover assuming the set of admissible parameter changes shown in~\Cref{tab:param}. We then show how to adapt the proof to accommodate the following restrictions: (i) map is excluded or (ii) aggregation is restricted to standard SQL aggregation functions. Afterwards, we show that the problem is \ptime if map is restricted to projection \textbf{and} aggregation is restricted to standard SQL aggregation functions by presenting a brute force \ptime algorithm.}

\myproofpar{Hardness}
\revc{
We show the claim through a reduction from the set cover problem: given a universe $\mathcal{U} = \{ 1, \ldots, n\}$ and a collection $\mathcal{S} = \{S_1, \ldots, S_m\}$ of subsets of $\mathcal{U}$ does there exists a subset $\mathcal{C} \subseteq \mathcal{S}$ of size less than or equal to $k$ such that
}
\[ \bigcup_{S \in \mathcal{C}} S = \mathcal{U} \]

\newcommand{\dbs}{\db_{\mathcal{S}}}

\revc{
For an instance of the set cover problem we construct a database $\db_{\mathcal{S}}$ and why-not questions $\whynot = (\query_{SC}, \db_{\mathcal{S}}, (n))$. Since we want to prove data complexity, the size of $\query_{SC}$ has to be in $O(1)$.
Database $\dbs$ consists of a single relation $R(S,e)$. Each tuple in $R$ encodes the membership of one element $e$ in a set $S \in \mathcal{S}$. For that we assign each set a unique identifier in $\{1, \ldots, m\}$ denoted as $id(S)$. Thus, the instance of $R$ is:}

\[ \bigcup_{S \in \mathcal{S}} \bigcup_{j \in S} \{ (id(S),j) \} \]

\revc{
The query $\query_{SC}$ is shown below. The query first applies a map operator using the identity function to the input. Without reparameterization,  this operator will return $R$ unmodified. Then all tuples are filtered where $S = -1$ and then joins the result with $R$. Afterwards, we count the number of distinct values in columns $S$ and $e$ and filter out this aggregation result if the number of distinct  values in column $S$ is larger than $k$. Finally, we project the result on attribute $els$ (the distinct number of values in column $e$).}

\begin{align*}
  \query_{SC} &\defas \projection_{els}(\selection_{sets \leq k}(\query_{cnt}))\\
  \query_{cnt} &\defas \aggregation{count(distinct\, S) \to sets, count(distinct\, e) \to els}{}(\query_{filter})\\
\query_{filter} &\defas \selection_{S \neq -1}(map_{\lambda x\, x}(R))) \join R
\end{align*}

\revc{
We claim that there exists a set cover of size $k$ or less if and only if the singleton set containing the map operators is an explanation for $\whynot$.}

\myproofpar{$\Rightarrow$}
\revc{
Assume that a set cover $\mathcal{C}$ of size less than $k$ exists, we have to show that the map operators is an explanation for the why not question. WLOG let $\mathcal{C} = \{S_1, \ldots, S_l\}$ for $l \leq k$. We can reparameterize the map function to be
}

\revc{\[
  \lambda x\, y \mathtext{{\bf for}} y =
  \begin{cases}
    (-1,x.e) &\mathbf{if} x.S \not \in \mathcal{C}\\
    x &\mathbf{otherwise}
  \end{cases}
\]}

\revc{
This function replaces value of column $S$ with $-1$ for all tuples in $R$ encoding a set that is not in $\mathcal{C}$. The selection then filters out all of these tuples. Thus, the result of $\query_{filter}$ in the reparameterization returns all tuples corresponding to the encoding of the sets in $\mathcal{C}$. Thus, the result will contain $l$ distinct value of attribute $S$ (since $\card{\mathcal{C}} = l$ and $n$ distinct values of $e$ (since $\mathcal{C}$ is a set cover). Consequently, the aggregation returns a single tuple $(l,n)$ which passes the outer selection and is projected onto $(n)$, the missing answer. Since this reparameterization returns the missing answer, the map operator is an explanation.
}

\myproofpar{$\Leftarrow$}
\revc{
Assume that the map operator is an explanation for the why-not question, we have to show that this implies that a set cover of size $k$ or less exists. First note that no matter how we reparameterize the map operator, $\query_{filter}$ always returns a subset of $R$ because the filtered result of the map is natural-joined with $R$. Since map is an explanation, that means there exists a subset of $R$ (selected through the reparameterization of the map operator) for which tuple $(n)$ is returned. From that we can follow that this subset has to contain $n$ distinct values of attribute $e$ which implies that it contains all elements of the universe $\mathcal{U}$. Furthermore, there cannot be more than $k$ distinct values of attribute $S$, because otherwise the outer selection would remove the result tuple of the aggregation. WLOG let these distinct values be $S_1$, \ldots, $S_l$ for $l \leq k$. Then $\mathcal{C} = \{S_1, \ldots, S_l\}$ is a set cover, because these sets contain all values of the universe.
}

\myproofpar{Hard restrictions}
\revc{
  We claimed above that the problem remains hard under the following restrictions: (i) map is excluded or (ii) aggregation is restricted to standard SQL aggregation functions. This claims are proven using slight modifications of the proof shown above. For (i) we replace the map with an aggregation grouping on both input columns. Since any \ptime aggregation function is allowed, we can define the aggregation function which will retrieve single values as input to simulate the lambda function shown in the proof above. The proof is otherwise analogous.}

\revc{
Note that the hardness proof above does not require us changing the function used in the aggregation. Thus, the proof holds unmodified even if we only allow standard SQL aggregation functions instead of arbitrary aggregation functions.
}

\myproofpar{\ptime Restrictions}
\revc{
If we restrict map to be projection and restrict aggregation to standard SQL aggregation functions then the problem is in \ptime. We proof this by sketching a \ptime brute force algorithm that simply enumerates a polynomial number of reparameterizations and for each reparameterization evaluates its side effects in \ptime data complexity (running a query). To show that this brute force approach is in \ptime, we first need to show that it is sufficient to enumerate a polynomial number of reparameterizations. Obviously, the number of reparameterizations is infinite over an infinite domain. However, we will argue that there is only a number of reparameterizations that is polynomial in the data size that can yield different results when evaluated over the database. Thus, if this claim holds, then for each of the polynomial number of possible results produced by reparameterizations, we only have to consider one of the infinitely many reparameterizations that produce this result (assume we can efficiently identify such reparameterizations).
}

\revc{
To proof this claim, we have to show that there are indeed only polynomially many distinguishable re-parameterizations and that we can enumerate them efficiently. We prove this by reasoning about the number of distinguishable re-parameterizations for each operator type based on their parameters. The number of distinguishable reparameterizations for an operator and query is polynomial in the number of distinct values of the database $\db$ (the database's active domain $adom(\db)$) and expoential in the size of the operatator's parameters and the query (in terms of operators). However, since we are concerned with data complexity, the query's size is constant and thus the overall number of distinguishable reparameterizations is polynomial in the data size, albeit to a possibly large, but data independent, exponent.
}

\revc{
We present the argument for selection operators here. The arguments for other operators is analogous. Consider a selection $\selection_\theta(R)$. Condition $\theta$ contains as set of attribute references $A$ from $R$ (here we consider two references to the same attributes as distinct references). Consider the case $A_1=1$ first (a single reference to a single attribute). Thus, $\theta$ is a function that takes a single value from the subset of $adom(\db)$ that exists in attribute $A_1$ and returns for each such value either true or false. There may exist infinitely many parameter changes for $\theta$ by replacing some constant in $\theta$. However, since we only allow comparison operators $\{\leq, < =, \geq, >\}$ replacing the constant $c$ in the condition $A_1\,op\,c$ containing the single attribute reference in $\theta$ we can at most get $O(adom(\db))$ different results, e.g., for a comparison $A_1 < c$ where $adom(\db) = \{1,3,5\}$ we decide which prefix of the total order of the value we include and there are linearly many such prefixes, e.g., for $c = 4$ we get $\{1,3\}$. To summarize, for conditions with a single attribute reference there are $O(adom(\db))$ many distinguishable reparameterizations. Generalizing this to a condition with $m$ attribute references, for each of the $m$ references we have at most $O(adom(\db))$ distinguishable options. Thus, the number of total distinguishable reparameterizations is $O(adom(\db)^m)$. Since the query size is constant as we are considering data complexity, $m$ is a constant.
}
\revc{
As mentioned above the arguments for other operators are analogous. From this follows that for single operator queries that number of distinguishable reparameterizations is constant. We can show by induction over the size of a query that this result implies that the same holds for queries of constant size. The argument in the induction step relies on the observation that we can treat the output of an operator as a new database with adom bound by $O(adom(\db)^m)$, because even though some operators like aggregation may produce new values not in $adom(\db)$ the number of new values they can produce is certainly bound by the number of different inputs to the operator. Thus, the active domain of the input of an operator that takes as input the result of another operator is of polynomial size  in $adom(\db)$. Thus, overall the number of distinguishable reparameterizations for a query of constant size is polynomial in $\card{D}$.
}

\myproofpar{Relaxing valid selection parameter changes}
\revc{
The last claim we need to proof is that even when map is restricted to projections and aggregation function choices are restricted to standard SQL aggregation functions, the problem is \nphard in data complexity if we relax the constraints on valid selection parameter changes by allowing any selection condition to be used (in~\Cref{tab:param} we only allow changes the preserve the structure of the selection condition by ``swapping'' attributes or changing constants). We first prove this for queries with aggregation, renaming, selection, projections, and join. Afterwards, we also proof that the problem is hard for $\RA$ (standard relational algebra which includes differences). For both classes of queries, the proof is through a reduction from 3-colorability (3C). Recall the definition of 3C. We are given a graph $G = (V,E)$ and have to decide whether it is possible to assign each vertex $v$ a color $C(v)$ from $\{r,g,b\}$ (\textbf{r}ed, \textbf{g}reen, \textbf{b}lue) such that for any two adjacent vertices $v_1$ and $v_2$ we have $C(v_1) \neq C(v_2)$.
}

\myproofpar{Queries with aggregation, renaming, selection, projection, and join}
\revc{
For a specific instance of the 3C problem we create a database $\db_G$ as follows: $V(v)$ is an unary relation storing an identifier for each vertex, $E(b,e)$ is a binary relation storing the graph's edges, and $VC(v,c)$ is a binary relation storing vertex-color pairs that contains for each vertex $v$ three tuples: $(v,r)$, $(v,g)$, and $(v,b)$. Let $n_E = \card{E}$. The query we are defining the why-not question over is:
  \begin{align*}
    \query_{choose}   & \defas \selection_{true}(VC)                                                                                                      \\
    \query_{oneColor} & \defas \projection_{a \gets n_E + 1} (                                                                                            \\
                      & \selection_{v_1 = v_2 \wedge c_1 \neq c_2}(
                        \rename_{v_1 \gets v, c_1 \gets c}(\query_{choose}) \crossprod \rename_{v_2 \gets v, c_2 \gets c}(\query_{choose})))              \\
    \query_{validE}   & \defas \projection_{a \gets 1} (\selection_{b = v_1 \wedge e = v_2 \wedge c_1 \neq c_2} (                                         \\
                      & \rename_{v_1 \gets v, c_1 \gets c}(\query_{choose}) \crossprod E \crossprod \rename_{v_2 \gets v, c_2 \gets c}(\query_{choose}))) \\
    \query_{3C}       & \defas \aggregation{sum(a)}{}(\query_{oneColor} \union \query_{validE}
  \end{align*}
  The why-not question is $\whynot = \nrctuple{\query_{3C}, \db_{G}, (n_E)}$. We claim that $G$ is 3-colorable iff $e = \{ o \}$ is in
  $\expl(\whynot)$ where $o$ is the identifier of the selection that is the root of subquery $\query_{choose}$.
}
  \myproofpar{$\Rightarrow$}:
  \revc{We have to construct an SR to show that $e$ is an explanations (since $e$ contains only a single operator there has to also exist an MSR changing only $o$). Since $G$ is 3-colorable, consider an arbitrary 3-coloring of $G$ and let $C(v)$ denote the color of a vertex in this solution. We reparametrize $o$ using the following condition:
  }

  $$\theta_{select} \defas \bigwedge_{v' \in V} v = v' \wedge c = C(v')$$

\revc{
This condition retains from $VC$ all vertex-color pairs according to the selected 3-coloring of $G$. Since every vertex only appears once in the result of the modified $\query_{choose}$ the result of $\query_{oneColor}$ is empty. Query $\query_{validE}$ returns $n_E$ copies of row $(1)$, because (i) for each edge in $E$ there will be exactly one join partner in each of the two instances of $\query_{choose}$ it is joined with and (ii) each of these tuples fulfills the additional condition $c_1 \neq c_2$, since end points of edges have different color. Thus, $\query_{3C}$ returns $(n_E)$ as required.
}
\myproofpar{$\Leftarrow$}
\revc{
Consider a reparameterization $e$ for $o$ that is an MSR. Observe that query $\query_{oneColor}$ returns a tuple $(n_E + 1)$ for every pair of tuples from $\query_{choose}$ that represent the same node, but with a different color. That is, $e$ cannot have changed the selection condition of $o$ such that the same node is returned more than once (``assigned'' more than one color), since otherwise the sum computed in the end would be larger than $n_E$. This means, that again every edge will have a unique join partner in $\query_{validE}$. Given that the reparameterized query $\query_{3C}$ returns $(n_E)$, we know that all edges have to  fulfill the condition $c_1 \neq c_2$. Thus, the result returned by the reparameterization of $\query_{choose}$ encodes a 3-coloring of $G$.
}

\myproofpar{$\RA$}
\revc{
For $\RA$ queries we introduce a new relation $T(b)$ with instance $\nrcbag{(1)}$. Furthermore, we change the schema of relation $E$ to $(id, v_1, v_2)$ where $id$ is a unique identifier for each edge. The query used in the proof is:}
\begin{align*}
  \query_{choose}     & \defas \selection_{true}(VC)                                                                                                                     \\
  \query_{multiColor} & \defas \projection_{id} (\selection_{v_1 = v_2 \wedge c_1 \neq c_2}(R \crossprod \rename_{v_1 \gets v}(V) \crossprod \rename_{v_2 \gets v}(V) )) \\
  \query_{goodE}      & \defas \projection_{id}(\selection_{c_1 \neq c_2 \wedge v_1 = b \wedge v_2 = e}(                                                                 \\
                      & \rename_{v_1 \gets v, c_1 \gets c}(\query_{choose}) \crossprod E \crossprod \rename_{v_2 \gets v, c_2 \gets c}(\query_{choose})))                \\
  \query_{3C}         & \defas T - (\projection_{1} ((\projection_{id}(E) - \query_{goodE}) \union \query_{multiColor}))
\end{align*}

\revc{
Here $\query_{multiColor}$ returns all edge identifiers once for each violation of the one-color-per-node constraint. Query $\query_{good}$ returns the identifiers of edges that do not violate the 3C condition (their endpoints have different colors). Finally, $\query_{3C}$ removes ``good'' edges from the set of all edges (making sure to include all edges if there is at least one vertex that has more than one color). The result is then projected to produce tuples $(1)$ that are removed from $T$. The net effect is that as long as either at least one node is assigned more than one color or there exists an edge whose endpoints is assigned the same color, then the result is empty. The why-not question used here is $\whynot = \nrctuple{\query_{3C}, \db_G, (1)}$. Again, let $o$ be the selection of $\query_{choose}$. We claim that  $e = \{o\}$ is an explanation for $\whynot$ iff $G$ is 3-colorable. We omit the proof of this claim since is analogous to the case of queries involving aggregation.}
\end{proof}

 }

  \revc{The algorithm we present in \Cref{sec:compute-msr} restricts aggregation and does not consider map. Thus, according to \Cref{theo:expl-hardness}, the problem is in \ptime. However, the search space is still much too large, requiring additional heuristic optimizations to scale.}

\revm{Next, we discuss that differences of reparameterization-based explanations and lineage-based explanations are not specific to our chosen algebra.
\Cref{tab:explanation-comp} summarizes the operators that both formalisms can find as part of their explanations for 
queries in \spcNRAB, \spcuNRAB, and \abbrNRAB.
Lineage-based solutions generally support $\spcuNRAB$. They only return operators that remove compatible input data. Thus, for operators overlapping with $\baseNRAB$, only selections become part of explanations. Given that the join operator can be expressed using cross product and selection,
lineage-based approaches can also find joins.
In our reparameterization-based formalism, the set of operators that can be part of an explanation is already more diverse for the least expressive query class and the minimal set of operators from $\baseNRAB$ they use. This stems from the fact that the query structure does not change when changing the function $f$ that parameterizes $map_f$. Consequently, we may return projections as causes for $\spcNRAB$ and $\spcuNRAB$. The benefits of our approach become even clearer for $\abbrNRAB$ (last row). 
}

\revm{Finally, note that the operators in an explanation depend on a query's algebraic translation and explanations may differ for equivalent translations.
For example, $\sigma_\theta (R \times S)$ may only yield the selection while $R \Join_{\theta} S$ may only yield the join. Lineage-based and reparameterization-based solutions share this property.}

\begin{table}[t]
\scriptsize \centering
\begin{tabular}{|c| p{1.6cm} | p{4.6cm} |} \hline
\textbf{Algebra} & \textbf{Lineage-based}  & \textbf{Reparameterization-based}  \\
  \hline
\spcNRAB & $\selection_\theta$*, $\Join_\theta$ & $\selection_\theta$*, $map_f$*,  $\Join_\theta$, $\projection_L$  \\ \hline
\spcuNRAB & $\selection_\theta$*, $\Join_\theta$ & $\selection_\theta$*, $map_f$*,  $\Join_\theta$, $\projection_L$  \\ \hline
\abbrNRAB &  $\selection_\theta$*, $\setminus$*,  $\Join_\theta$, $\leftouterjoin_\theta$,  $\rightouterjoin_\theta$, $\fullouterjoin_\theta$, $\flatten^{I}_{A}(R)$
	& $\selection_\theta$, $map_f$, $\Join_\theta$, $\leftouterjoin_\theta$,  $\rightouterjoin_\theta$, , $\fullouterjoin_\theta$ , $\rho_{B_1 \gets A_1, \ldots, B_n \gets A_n}$,  $\flatten^{T}_{A}(R)$, $\flatten^{I}_{A}(R)$, $\flatten^{O}_{A}(R)$, $\nestTup{A}{C}(R)$, $\nestRel{A}{C}(R)$, $\aggregation{f(A)}{B}(R)$ \\ \hline
\end{tabular}
\caption{\revm{Operators that can become part of explanations for different algebras and explanation formalisms. The $*$ marks operators in \baseNRAB; other operators are derived ones.
}}
\label{tab:explanation-comp}
\vspace{-8mm}
\end{table}

 \section{Computing Explanations}\label{sec:compute-msr}

\revb{Given our data complexity results of computing explanations, we present an algorithm that restricts admissible parameter changes to allow for \ptime computation. To also make it efficient in practice, we introduce additional novel heuristics.}  The algorithm takes a why-not question $\whynot = \aWhynot$ as input and returns a set of explanations $\expl^\approx$ that approximates  $\expl$. We first present the algorithm 
and, then, discuss in~\Cref{sec:computing-discussion} how $\expl^\approx$ relates to $\expl$ from ~\Cref{def:explanation}.

\Cref{alg:overview} shows the four main steps of our algorithm. First, given the missing tuple $\aTup$ that is defined over the output schema of $Q$, the algorithm computes a set of \abbrNIP tuples $\overline{T}$ over the schema of $Q$'s input tables in $D$ that could have contributed to the missing answer. It also computes a mapping $\sbtMap$ which associates each attribute in $\aTup$ and each attribute referenced in an operator of $Q$ with a set of attributes from the input. We refer to these input attributes as \textit{source attributes}.
In the second step, \Cref{alg:overview} determines alternatives for each source attribute in $\sbtMap$. These alternatives account for attributes that may not have been chosen appropriately when writing $Q$. They possibly require a reparameterization of the attributes referenced in $Q$'s operators, e.g., the attributes unnested by a flatten operator.
The alternative attributes allow the algorithm to enumerate a set of \emph{schema alternatives} (SAs) denoted as $\aSchemaOptCollection$. Each 
SA corresponds to a possible reparameterization of attribute references in $Q$. In the third step, $dataTracing$ traces data from the input $D$ through $Q$'s operators. It instruments each operator to include the results of its admissible reparameterizations and further annotations. They describe, e.g., whether a tuple exists under a SA. 
Based on these annotations, $approximateMSRs$ computes explanations for~$\whynot$ \revb{and returns them in the partial order of~\Cref{def:msr}}.

\vspace{-2mm}
\subsection{Step 1: Schema backtracing}
\label{sec:step1}

Taking the why-not question $\whynot = \aWhynot$ as  input, schema backtracing analyzes schema dependencies and schema transformations of the query $Q$ in a data-independent way. The goal of schema backtracing is twofold: (i) rewrite the missing answer $\aTup$ into a set of \abbrNIPs  $\overline{T}$ (\Cref{def:instances-with-place}) over the schema of $D$. $\overline{T}$ contains one \abbrNIP for each input relation in $D$ that potentially matches tuples relevant to produce $\aTup$ under some reparameterization;
(ii) identify attributes from $D$'s schema that serve as alternatives to $Q$'s source attributes.  
\BG{State whether $\overline{T}$ contains exactly one NIP (nested instance with placeholders) per input table?}
\RD{yes, I have incorporated this into the text}

\BG{I think we could be more precise in what $\overline{T}$ and $\sbtMap$ are, e.g., $\overline{T}$ is a set of NIPs one per table accessed by the query (or is it one per table access in the query?) while $\sbtMap$ associates with each attribute (in the output) with a set of attributes from the input and intermediate results.}
\RD{I changed a bit above to be more precise, $\sbtMap$ in the intro and $\overline{T}$ in the paragraph above}

To achieve the first goal, schema backtracing \revc{iterates through the query operators and analyzes each operator's parameters to trace data dependencies on the schema level. Eventually, schema backtracing returns} $\overline{T} = \{ \overline{t}_{R_1}, \ldots,  \overline{t}_{R_n}\}$, where $R_1$ through $R_n$ denote $Q$'s input relations. Each $\overline{t}_{R_i}$ is a \abbrNIP. Intuitively, the set of tuples from $R_i$ matching $\overline{t}_{R_i}$ includes all tuples that may contribute to tuples matching $\aTup$ under some schema alternative.

\begin{example} \revc{In our running example, one such NIP is $\overline{t}_{person} = \nrctuple{name: ?, address1: ?, address2: \nrcbag{\nrctuple{city: NY, year:?}}}$ computed from $\aTup = \nrctuple{city:\text{``NY"}, nList: \nrcbag{\valPlaceholder, \multPlaceholder}}$. To obtain the NIP, the algorithm traces back both dependencies for $\aTup.city$ and $\aTup.nList$. When it iterates through the operator $\nestRel{name}{nList}$, it traces the nested tuples in $nList$ back to the $name$ attribute.
    Through 
    the remaining operators, the algorithm finds the $name$'s origin in the $name$ attribute of the $person$ relation. Similarly, it traces the $city$ back to the source attribute $address2.city$, whose value has to match NY.}
\label{ex:sa}
\end{example}
$\overline{T}$  is coupled with a mapping $\sbtMap$. This mapping associates each attribute $\aTup.A$ of the why-not tuple $\aTup$ with source attributes to identify the corresponding source attributes that produce the values of $\aTup.A$.
To also identify source attributes potentially relevant for operator reparameterizations (the second goal outlined above), \revc{the backtracing algorithm} further adds associations for each attribute reference $op.A$ at operator~$op$ to $\sbtMap$ \revc{while it iterates through the query tree.}
Notationwise, we distinguish the two kinds of associations:
(i) Associations between a source attribute $\overline{t}.X$ and a missing-answer attribute $\aTup.A$ are denoted as $\frac{\color{blue}{A}}{X}$.
(ii) Associations between a source attribute $\overline{t}.X$  and an operator attribute $op.A$ are written as $\frac{\color{red}{op.A}}{X}$. In the following, we represent a pair $(\overline{t},\sbtMap)$ as a single nested tuple mirroring the nesting structure of $\overline{t}$ but using associations from $\sbtMap$ as attribute names instead. For instance, if $\frac{\color{blue}{A}}{X}$, $\frac{\color{red}{op.A}}{X}$, and  $\frac{\color{red}{op.B}}{X}$, then we substitute $X$ with $\frac{\color{blue}{A}, \color{red}{op.A}, \color{red}{op.B}}{X}$.

\begin{example}  \revc{Continuing with Example~\ref{ex:sa}, 
    $\sbtMap$ associates $t.nList$ to $\overline{t}_{person}.name$ and $t.city$ to $\overline{t}_{persons}.address2.city$. It further associates the $\selection.year$ with $\overline{t}_{person}.address2.year$.}
The associations in $\sbtMap$ coupled with  $\overline{T} = \{\overline{t}_{person}\}$ are represented as:
\begin{align*}
  \sloppy \overline{t}_{person} =& \Bigl< \sbatts{t.nList,\color{red}{\projection.name,  {\mathcal N}.nList,  {\mathcal N}.name }}{name} : \valPlaceholder, address1 : \valPlaceholder, \\
  &\sbatts{\color{red}{\flatten.address2}}{address2}: \nrcbag{\nrctuple{\sbatts{t.city, \color{red}{\projection.city}}{city}:\text{``NY''}, \sbatts{\color{red}{\selection.year}}{year}:?}}
  \Bigl>
\end{align*}

\label{ex:sb}
\end{example}

\BG{Our algorithm for schema backtracing resembles the unrenaming algorithm of~\cite{DBLP:conf/edbt/BidoitHT14}. It additionally supports nested tuples and schema manipulations on them such as nesting as opposed to flat tuples. It further tracks the ``red'' mappings that stem from query operators rather than the why-not tuple to properly cover reparameterizations of operator parameters corresponding to attribute references.}

\vspace{-3mm}
\subsection{Step 2: Schema alternatives}
\label{sec:step2}

Next, the algorithm determines schema alternatives (SAs) which have the potential to produce the missing answer, since there may exist MSR reparameterizations implementing these SAs. 
A SA 
substitutes zero or more attributes in operator parameters with alternatives. The set of all SAs 
covers all such substitutions. Thereby, the algorithm considers all possible reparameterizations that involve replacing attributes.

\noindent \textbf{Finding attribute alternatives.} The first step of identifying SAs 
is finding alternatives for attributes referenced by $Q$.  
For each $\overline{t}_{R_i} \in \overline{T}$, we identify, for each $\frac{\color{black}{A}}{X} \in \sbtMap$  
a set of alternative attributes for $X$, i.e., $\mathcal{X}' = \{X'_1, \ldots X'_k\}$ with $X'_j \in R_i$ and matching types of $X$ and $X'_j$. We restrict alternatives to attributes of the same relation, because replacing an attribute with an attribute from another relation would require more changes to the query than reparametrizations allow. We assume that the set of attribute alternatives is provided as input to our algorithm. \revb{For instance, these can be determined by hand, schema matching techniques~\cite{Do2002,Aumueller2005}, or schema-free query processors~\cite{Li2004, Li2008}. The latter may even yield entire SAs 
  rather than mere attribute alternatives.} 
\revb{These strategies} ensure that we only consider meaningful alternatives and avoid blowing up computations by considering an impractical number of alternatives.

\begin{example}
\label{ex:alternative_sets}
\sloppy
In our example, we assume the following attribute alternatives:
$ name'  {=} \{name\}$, $city' {=}  \{address2.city,$ $address1.city\}$, $year'  {=}  \{address2.year, address1.year \}$, and $address2'  {=}  \{address2, address1\}$.
\end{example}

\noindent \textbf{Enumerating and pruning SAs.} 
The attribute alternatives are used to enumerate all possible SAs, 
which requires considering alternatives for attributes  of intermediate results that appear as $\frac{\color{red}{op.A}}{X} \in \sbtMap$. 
Formally, a schema alternative $S = \nrctuple{\overline{T},  {\mathcal M} }$ is a set of \abbrNIPs $\overline{T}$ (as in schema backtracing, one tuple per table accessed by $Q$) 
and a mapping $\mathcal{M}$ (like $\sbtMap$, $\mathcal{M}$ records which input attributes are referenced by which operator and are used to derive which attribute in 
$Q'$s output).

\begin{example}
  \Cref{fig:schema-alternatives} shows how the algorithm incrementally derives all SAs 
  (ignore the dashed parts for now).
Based on the set of alternatives for $address2'$ and $year'$ from \Cref{ex:alternative_sets}, it starts evaluating options for the flatten operator's parameters. It can either use the original attribute $address2$, or the alternative attribute $address1$. For each alternative for flatten, it can then choose $address2.year$ or $address1.year$ for the selection operator.
\end{example}

\begin{figure}[t]
\trimfigspace\includegraphics[width=0.48\textwidth]{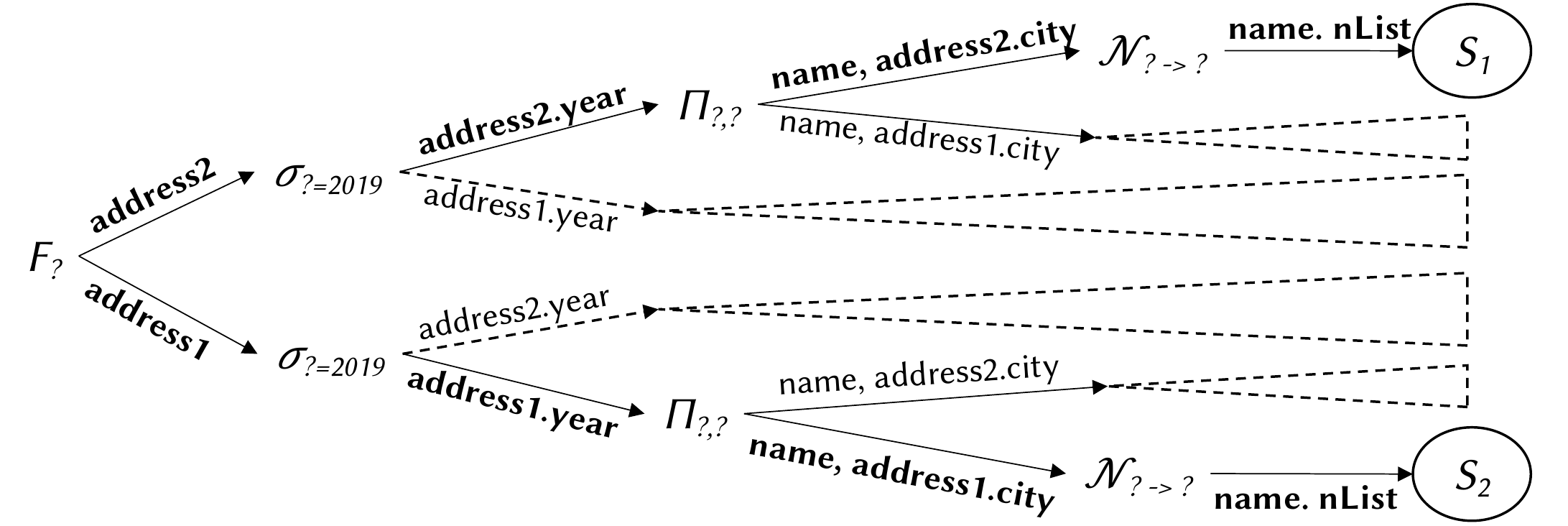}
\trimfigspace
\caption{Enumerating and pruning schema alternatives}
\label{fig:schema-alternatives}
\vspace{-3mm}
\end{figure}

SAs replace attributes of one operator independently from attributes of another operator. Thus, some 
SAs may alter \query's output schema or lead to an invalid query that references non-existing attributes in some operators.
For instance, once we flatten out $address2$, the only ``accessible'' alternative for $year$ is $address2.year$ in the selection. We further prune alternatives that alter the output schema (that is fixed by definition). For instance, assuming the source data included $address1.city1$ instead of $address1.city$, flattening $address1$ changes $Q$'s output schema to $\nrcbag{\nrctuple{city1, nList}}$, which is not allowed.
\begin{example}
  In our example, all dashed subtrees in \Cref{fig:schema-alternatives} are pruned. Only two SAs 
  remain, denoted as $S_1$ and $S_2$. The SA $S_1 = \nrctuple{\{\overline{t}_1\},  {\mathcal M}_1 }$, with $\overline{t}_{1}$ being equal to $\overline{t}_{person}$ shown in \Cref{ex:sb}, and $S_2 = \nrctuple{\{\overline{t}_2 \},  {\mathcal M}_2 }$ with $\overline{t}_{2}$ ``swapping'' the address attribute, i.e.,
  \begin{align*}
    \overline{t}_{2} =&   \langle  \sbatts{t.nList,\color{red}{\projection.name,  {\mathcal N}.nList,  {\mathcal N}.name }}{name}: \valPlaceholder,  address2: \valPlaceholder, \\
    &\sbatts{\color{red}{\flatten.address1}}{address1}: \nrcbag{\nrctuple{\sbatts{t.city, \color{red}{\projection.city}}{city}:\text{``NY''}, \sbatts{\color{red}{\selection.year}}{year}:?}} \rangle
  \end{align*}
\MH{For conciseness, I am thinking of combining equations with figure.}

\label{ex:schema-alternatives}
\end{example}
\MH{Special case of derived attributes not covered here. }

\subsection{Step 3: Data tracing}
\label{sec:step3}

At this point, the algorithm has identified the source attributes to consider for reparameterizations (``blue numerators'' identified during schema backtracing) and has determined the reparameterizations to consider for attributes (through SAs). Next, it identifies and traces data that may yield the missing answer through reparameterizations of query operators. It instruments operators to compactly keep track of possible reparameterizations and their results.

We define individual tracing procedures for each operator. The procedures commonly take the operator $op$, an annotated relation $\aRel^A$, and schema alternatives $\aSchemaOptCollection$ as input. Their output consists of an annotated relation $\aRel^{A'}$ and updated schema alternatives $\aSchemaOptCollection'$. In general, the algorithm extends genuine operator semantics to further collect result tuples producible by possible reparameterizations as well as annotation columns for each schema alternative. It encodes the operator results of each alternative in $\aSchemaOptCollection$ into $\aRel^{A'}$.

\MH{Possibly for the implementation section: While we present schema alternatives as a separate step, it can be implemented together with tracing. }

We distinguish four annotation types that introduce additional attributes to the output tuples $\aTup'$ in the operator's output $\aRel^{A'}$.
\begin{itemize}
\item $id$: Each top-level tuple is assigned a unique identifier. 
\item $\valid S_i$: For each schema alternative $S_i$, this boolean annotation describes whether  $\aTup'$ is part of the operator output under schema alternative $S_i$. Our algorithm leverages it to determine which $\aTup' \in \aRel^{A'}$ correspond to which $S_i$.
\item $\compatible S_i$: For each schema alternative $S_i$, this boolean annotation identifies if  a tuple $\aTup'$ is consistent with the why-not question. $\aTup'$ is consistent if it potentially contributes to the missing answer. This annotation stores the result of re-validating compatibles as hinted at in the introduction.
\item $\survivor S_i$ indicates if $\aTup'$ is an output tuple of the original query except for attribute changes given by $S_i$ (true) or if it can result from other operator reparameterizations~(false), e.g., by changing constants in a selection condition.
\end{itemize}

In the following, we describe the tracing algorithms for the operators used in our running example, omitting projection since it simply propagates consistent and valid annotations of its input. 

\noindent \textbf{Table access.} 
The tracing procedure for the table access operator iterates over each tuple $\aTup$ in the input relation $\aRel$ and extends $\aTup$ with annotation attributes. It adds the $id$ attribute and a $\compatible S_i$ attribute for each SA $S_i$. The value $v$ of this attribute is only true if $\aTup$ matches the tuple $\overline{\aTup}_R$ in the set of tuples $\overline{T}_i$ of  $S_i$.
To add correctly named annotations in function of $S_i$, we use the $annotate$ function (Algorithm~\ref{alg:ann-ta}), e.g., we call $annotate(t', [(\compatible, v)], S_i, \anOpDS)$. 
The table access operator does not change the structure of its input, so input SAs are simply propagated to its output.
\RD{if we need space we could inline the annotate function}
\RD{if we need even more space, maybe we could drop the entire table access description and continue with the example below?}

\begin{example}
Applying the table access procedure to our running example yields the annotated relation in \Cref{fig:ann-ta}. Schema alternative $S_1$ is associated to $\overline{t}_1$ shown in \Cref{ex:schema-alternatives} and considers $address2.city$, while $S_2$ comprises $\overline{t}_2$ using $address1.city$. The first tuple in \Cref{fig:ann-ta} has $\compatible S1\_1 = 0$ because it has no value in $address2.city$ that matches $\overline{t}_1$'s constraint $city = $``NY'', while $\compatible S2\_1 = 1$ because $address1.city$ nests $\nrctuple{city: \text{``NY''}, 2010}$.
\end{example}

\begin{figure}[t]
\begin{minipage}{0.2\textwidth}
\null
\removelatexerror
\begin{algorithm}[H]
\scriptsize
 \SetKwFunction{Annotation}{$annotate$}
  \SetKwProg{Pn}{Function}{:}{}
  \Pn{
  \Annotation{$t$, $avMap$, $S_i$, $\anOpDS$}}{
	\ForEach{ $(a, v) \in avMap$}{
		$label \leftarrow a+$``S''$+i+$“\_”$+op.getID()$\\
	 	$t \leftarrow t \circ \nrctuple{label: v}$
	}
	\Return{t}
  }
\caption{$annotate$}
\label{alg:ann-ta}
\end{algorithm}
\end{minipage}
\hspace{0.5cm}
\begin{minipage}{0.24\textwidth}
\centering
\includegraphics[width = 1.15  \textwidth]{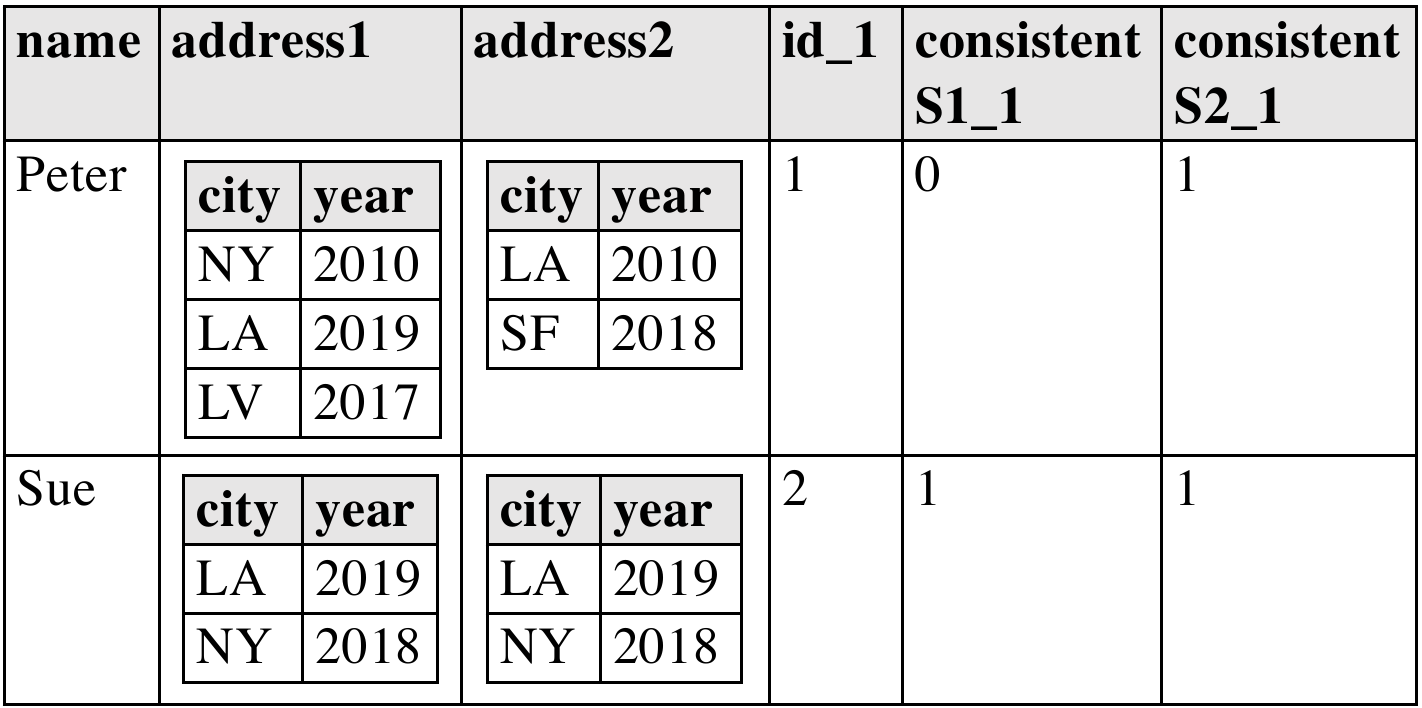}
\trimfigspace
\vspace{-3mm}
\caption{Example of annotations after table access}
\label{fig:ann-ta}
\end{minipage}
\vspace{-4mm}
\end{figure}

\noindent \textbf{Flatten.} 
The tracing procedure for the flatten operator (\Cref{alg:fl}) computes the results of the operator under all schema alternatives \revc{utilizing the concepts of zip and flatMap functions in functional programming.}
It obtains the result $O_i$ of the outer flatten for each SA $S_i$. It uses an outer flatten for two reasons. First, changing an inner flatten to an outer flatten is a valid parameter change. Second it has to track tuples that the inner flatten filters because the flattened attribute is null or the empty set.
Next, the algorithm updates the SAs to reflect the restructuring of the tuples. It then combines all $O_i$ as follows. Lines~\ref{ln:flatten-begin}--\ref{ln:flatten-end} process $O_1$, i.e., the result of the outer flatten parameterized as given by $S_1$. For each tuple $t$ in $O_1$, it evaluates boolean conditions to determine the values $c$ and $r$ for the $\compatible$ and $\survivor$ flags.
  The algorithm sets the $\valid$ annotation to $1$. To process the remaining SAs (lines~\ref{ln:f-alt-begin}--\ref{ln:f-alt-end}), it uses the $merge$ function. Intuitively, $merge$ concatenates tuples with the same $id$ across the outer flatten results of all SAs, ensuring not to replicate columns that remain the same across all SAs. Since the number of tuples with a given $id$ may vary across the different results (due to nested relations of varying cardinality), it pads missing ``concatenation partners'' with null values ($\bot$). The algorithm creates annotations for each SA. Thus, it sets annotations corresponding to null-padded (non-existent) alternatives to $0$. The annotations of tuples in each $O_i$ are set analogously to the ones for $S_1$. Each tuple produced in the output also receives a fresh unique~$id$.

{
\begin{figure}[t]
\begin{minipage}{\columnwidth}
\null
\removelatexerror
\begin{algorithm}[H]
  \scriptsize
  \SetKwFunction{Flatten}{$Flatten$}
  \SetKwProg{Fn}{Function}{:}{}
  \Fn{
  \Flatten{$\anOpDS$, $\aRel$, $\aSchemaOptCollection$}}{
     \addtolength{\hsize}{10mm}
     \hspace{-1mm}$\squeezespaces{0.5} \forall S_i \in  \aSchemaOptCollection$, let $\squeezespaces{0.5}O_i$ be the result of executing $op$ wrt $S_i$ and generalized to an outer flatten\\
     \hspace{-1mm}$\squeezespaces{0.5} \forall S_i \in  \aSchemaOptCollection$, let $\squeezespaces{0.5} S_i' = \nrctuple{\overline{T}_i', {\mathcal{M}}_i'}$ be the schema alternative reflecting the flattening wrt $S_i$ \\
     \hspace{-1mm}$O_{merged} \leftarrow \emptyset$\\
     \hspace{-1mm}\ForEach{$t \in O_1$}{ \nllabel{ln:flatten-begin}
	$r \leftarrow$ $t$ is in the result of original flatten wrt $S_1$\\
	$c \leftarrow t \matches \overline{t}_R'$, where $\overline{t}_R' \in \overline{T}_1'$\\
	$avMap \leftarrow [(\valid,1), (\survivor, r), (\compatible, c)]$\\
	$O_{merged} \leftarrow O_{merged} \cup \{annotate(t, avMap, S_i, \anOpDS) \}$ \nllabel{ln:flatten-end}
     }
     \hspace{-1mm}\ForEach{$O_i, 1 < i \leq |\aSchemaOptCollection|$ }{\nllabel{ln:f-alt-begin}
     	$ \overline{t} \leftarrow \overline{t}_R' \in \overline{T}_i'$\\
     	$O_{merged} \leftarrow merge(O_{merged}, O_i, S_i, \anOpDS, \overline{t})$ \nllabel{ln:f-alt-end}
     }
     \hspace{-1mm}\Return{$\nrctuple{O_{merged},  \bigcup_{S_i \in \aSchemaOptCollection } S_i'}$}
     }
     \caption{Function $Flatten(\anOpDS, \aRel, \aSchemaOptCollection)$}
\label{alg:fl}
\end{algorithm}
\vspace{-0.5cm}
\end{minipage}
\end{figure}
}

\begin{example}
\vspace{-2mm}
Given the annotated relation in \Cref{fig:ann-ta} and the SAs in \Cref{ex:schema-alternatives}, the inner flatten produces the annotated relation shown in \Cref{fig:ann-fl} and updates $S_1$ with $\overline{T}'_1 =  \{\nrctuple{ \sbatts{nList}{name}: \valPlaceholder, {\sbatts{city}{cityS1}:\text{``NY''}, yearS1:?}}  \}$ and $S_2$ analogously. It combines both SAs, as both $address1$ and $address2$ are flattened. The column marked with $\ldots$ summarizes all annotation columns of the input. They are treated as ``regular'' input columns when executing the outer flatten. Focusing on the new annotations, we see in the column $consistentS1\_2$ that only the last tuple is consistent with $\overline{T}'_1$, because it is the only tuple that features ``NY'' in $cityS1$. Further, the $1$ values in $validS1\_2$ indicate that the flatten  produces 4 tuples under $S_1$. The third tuple is not valid under $S_1$, being an artifact of unnesting $address1$ for SA $S_2$. The other tuples all have the $retainedS1\_2 = 1$. Thus, no tuple is lost due to the more restrictive inner flatten type.  
\vspace{-2mm}
\end{example}

\begin{figure}[t]
\centering 
\includegraphics[width = 0.43 \textwidth]{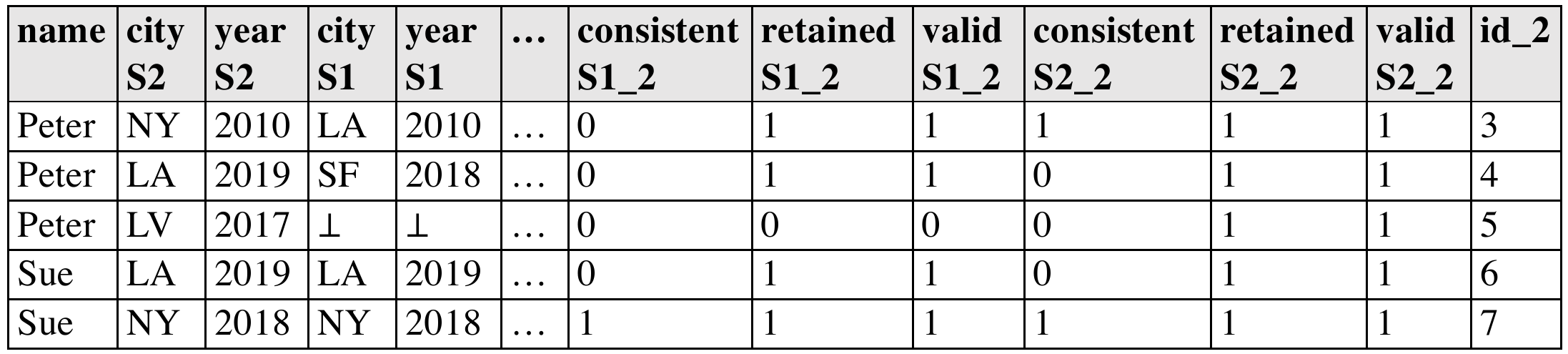}
\trimfigspace
\caption{Example of annotations after flatten}
\label{fig:ann-fl}
\end{figure}

\noindent \textbf{Selection.}
The tracing procedure for the selection operator returns all input tuples with additional annotation columns. It propagates the $\compatible$, $\valid$ and $id$  attributes of the previous operator, since it neither manipulates the schema nor the identity of top-level tuples. However, the procedure adds a new $\survivor$ attribute for each $S_i$. The value of the $\survivor$ attributes is 1 if a tuple from the input under $S_i$ satisfies the selection condition $\theta$, and 0 otherwise.

\begin{figure}[b]
\centering
\includegraphics[width = 0.43 \textwidth]{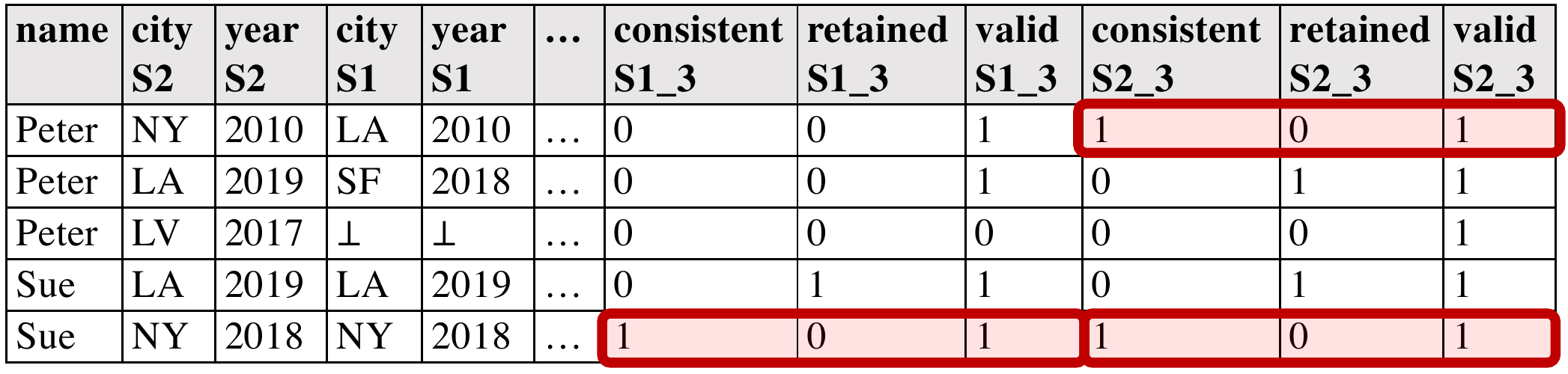}
\trimfigspace
\caption{Example of annotations after selection}
\label{fig:ann-fi}
\end{figure}

\begin{example}
Ignoring the red highlighting for now, \Cref{fig:ann-fi} shows the tracing output after the selection checking if $year \geq 2019$.  For instance, the last tuple has $year = 2018$ under $S_1$, so $\survivor S1\_3~=~0$.
\RD{Do we really need two examples for reading the table? In order to shorten the text, I commented the sentence out}
\end{example}

\noindent \textbf{Relation nesting.}
Due to space constraints, we explain the algorithm for relation nesting  only based on our running example.
Since the nesting changes the structure of the input relation, our algorithm first updates the set of SAs. For each $S_i$, it  derives an alternative $S_i'$ from $S_i$ that reflects nesting. This, for instance, yields $S_1'$ with $\overline{T}'_1  = \{\nrctuple{ \sbatts{nList}{nListS1}: \nrcbag{\nrctuple{name: \valPlaceholder}, *}}, \sbatts{city}{cityS1}: \text{``NY''}  \}$ . 
Then, the algorithm computes the result of relation nesting considering the schema alternatives and annotates the result tuples as shown in
\Cref{fig:ann-rn}. First, for each $S_i$, it computes $R_i$ by ``isolating'' all columns involved in schema alternative $S_i$ and retaining valid tuples only. Similarly,  $S_i$  yields $R_i^{prov}$ by projecting on all annotation columns related to $S_i$ and selecting valid tuples. \Cref{fig:ann-rn}~\textcircled{1} shows the result of tracing the preceding projection operator. It highlights data of $R_1$ in yellow and $R_2$ in cyan, while data of $R_1^{prov}$ and $R_2^{prov}$ are highlighted in orange and dark blue, respectively. In step \textcircled{2}, the algorithm nests $R_i$ and $R_i^{prov}$.
Processing $S_1$ results in the top row of tables for step \textcircled{2}, while $S_2$ yields the two bottom relations. Annotations are added to tuples of $R_i$ in step \textcircled{3}, resulting in $R_i^A$. For all tuples, the valid annotation is set to 1, whereas the consistent annotation is set to 1 only if $t \in R_i$ matches $\overline{t}_R' \in \overline{T}_i'$. For instance, in \Cref{fig:ann-rn}\textcircled{3}, the third tuple of $R_1$ (left) is flagged as consistent, because it matches the constraints defined by $\overline{T}_1'$. Finally, in step \textcircled{4}, all relations $R_i^A$ and $R_i^{prov}$ of all schema alternatives are combined using a function similar to a full outer join. Instead of padding values with nulls when no join partner exists, the algorithm pads the nested relations with $\emptyset$ and the annotations with $0$. That allows the algorithm to compose operators extended with our tracing procedure. It also collapses joined columns (the non-nested attributes) from the different schema alternatives (e.g., $cityS1$ and $cityS2$), by coalescing their values. The final result of step~\textcircled{4} is shown at the bottom of \Cref{fig:ann-rn} (ignore red highlighted boxes for now).

\begin{figure}[t]
  \includegraphics[width = 0.5 \textwidth]{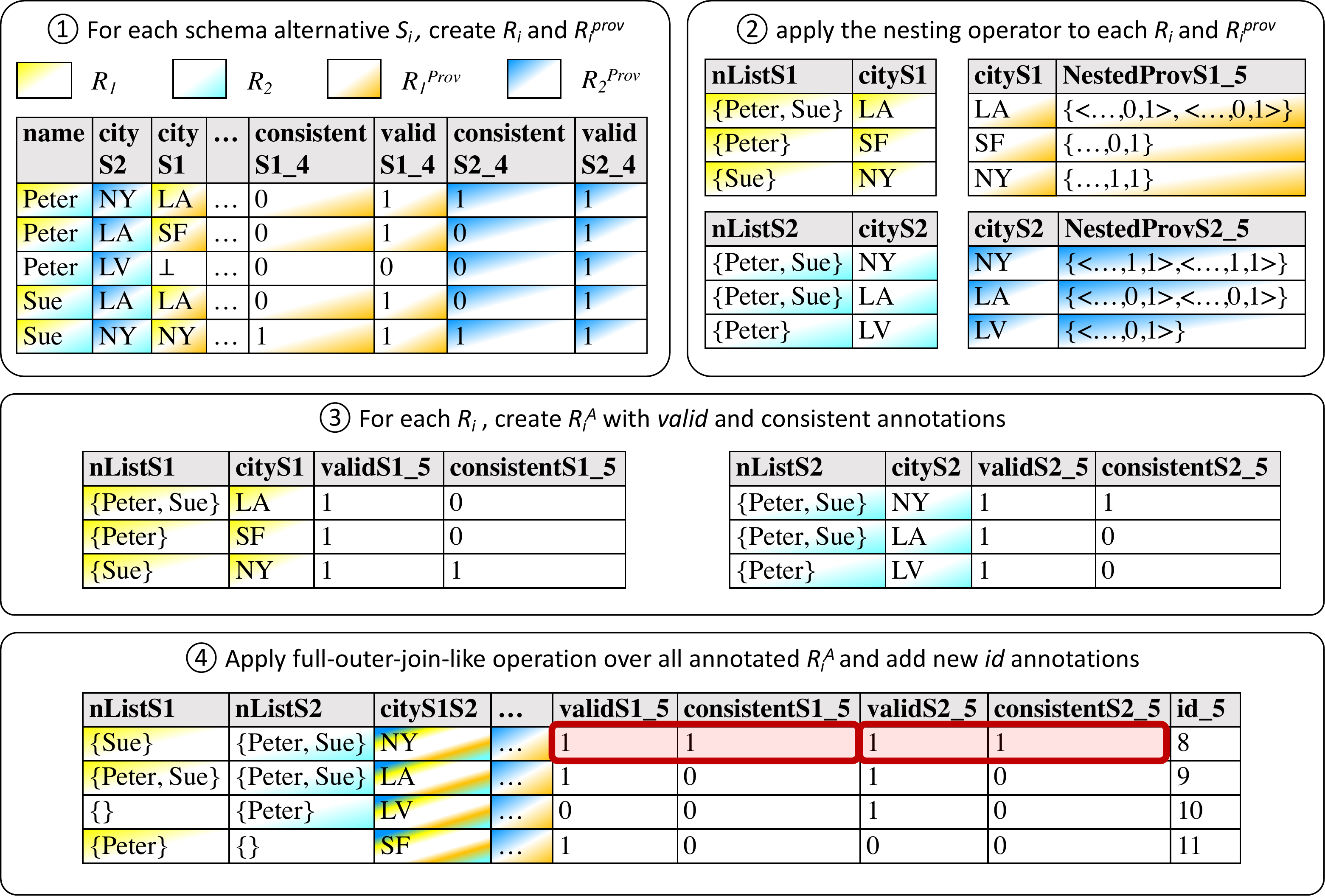}
\vspace{-7mm}
\caption{Example of annotations after relation nesting}
\label{fig:ann-rn}
\end{figure}

\subsection{Step 4: Computing MSRs}
\label{sec:step4}

The result of the data tracing step is a nested relation that extends the original query result with (i) data that could belong to the result under some reparameterization and (ii) annotations needed to identify the operators that require a reparameterization to obtain the missing data. \Cref{alg:msr} approximates the set of MSRs formally defined in \Cref{sec:repar-expl}. It first initializes a queue of partial SRs, each associated with an operator to consider to extend the SRs. The initial partial SRs are based on operator reparameterizations imposed by schema alternatives. For instance, in \Cref{fig:schema-alternatives}, $S_1$ does not involve any change in the attributes referenced by query operators (and thus, $SR_1 = \emptyset$), whereas $S_2$ involves changing the attribute referenced by the flatten operator (and thus $SR_2 = \{\flatten\}$). The algorithm then retrieves each operator $op_j$ of the query (top-down) and their associated partial SRs $SR_i$ from $queue$ to check if $op_j$ needs to be added to an $SR_i$. More precisely, $op_j$ extends $SR_i$ when the annotations relative to $op_j$ and the schema alternative $S_i$ contain at least one valid tuple that is consistent with the why-not question, not retained, and in the lineage of a consistent tuple of the final result.
The algorithm further adds $op_j$'s predecessor $op_{j-1}$ with unchanged $SR_i$ to the queue when it is possible that explanations without $op_j$ but with some of its predecessors can be found (i.e., all annotations are set to 1 for $op_j$). 
 When no further operators can be added, it adds $SR_i$ to the set $\expl^\approx$.

\begin{algorithm}[t]
\scriptsize
Let $SR_i$ be the SR prefix determined for each $S_i$ \\
$queue \leftarrow $ add all pairs $(\anOpDS_{|\aSchemaOptCollection|}, SR_i)_{S_i}$ in context $S_i$ \\
$SR \leftarrow \emptyset$\\
\While{$queue \neq \emptyset$}{
	$(op_j, SR_i)_{S_i} \leftarrow queue.removeFirst() $\\
	$R_{ij}^A \leftarrow $ annotations relative to $op_j$ and $S_i$ \\
	$extendWithOp \leftarrow false$ \\
	\If{ $R_{ij}^A$ contains a valid tuple $t$ where $retainedSi\_j = 0$ and $consistentSi\_j = 1$ and $t$ in the lineage of a consistent output tuple}{
		$extendWithOp \leftarrow true$
	}
	\If{$j > 1$}{
		\If{ $extendwithOp$} {
			$queue.append(op_{j-1}, SR_i \cup \{op_j\})$
		}
		\If{ $R_{ij}^A$ contains a valid tuple with all its annotations being set to 1}{
			$queue.append(op_{j-1}, SR_i)$
		}
	}
	\Else{
		\If{ $extendwithOp$} {
			$SR \leftarrow SR \cup \{ SR_i \cup \{(op_{j}\}\}$
		}
		\If {$R_{ij}^A$ contains a valid tuple with all its annotations being set to 1}{
			$SR \leftarrow SR \cup SR_i$, if $SR_i \neq \emptyset$}
	}
}
Prune $SR$ based on upper and lower bound of side effects for each explanation in $SR$ \revb{and sort them according to the partial oder defined by \Cref{def:msr}.}\\
\Return{$SR$}
\caption{$approximateMSRs(\anOpDS, \aRel^A, \aSchemaOptCollection)$}
\label{alg:msr}
\end{algorithm}

\MH{This algorithm needs to be verified.}

\begin{example}
$\expl^\approx_{example}$ holds two SRs: $\squeezespaces{0.5} SR_1 = \{\selection\}$ and $\squeezespaces{0.5} SR_2 = \{\flatten \selection\}$ computed from the annotations in red boxes in \Cref{fig:ann-fi,fig:ann-rn}.
\label{ex:msr-alg}
\end{example}

Currently, we only compute loose upper and lower bounds (UB and LB) for side effects. \revc{Obtaining the exact number of side effects would require comparing the original query result to the result of any possible actual reparameterization for each operator. For example, $year \geq 2018$ and $year \neq 2019$ are both possible actual reparameterizations of the selection operator in $SR_1$ and $SR_2$, but may yield a different number of side effects.}

\revc{
We compute $LB =  LB(\Delta+) + LB(\Delta^-)$ and $\squeezespaces{0.5} UB = UB(\Delta^+) + UB(\Delta^-)$ based on estimates on the maximum (for $UB$) and minimum (for $LB$) number of top-level tuples \emph{any} operator reparameterization in an explanation adds ($\Delta^+$) or removes ($\Delta^-$) from the original query result $\qEval{\query}_{\db}$. 
For explanations within the original schema alternative, which we will consistently denote as $S_1$, $UB(\Delta^+)$ equals the number of valid top-level tuples in the result
that have at least one retained flag set to 0 for one of the explanation's operators.
For instance, in~\Cref{fig:ann-rn}, tuples 9 and 11 satisfy this condition for explanation $SR_1$. 
For explanations linked to a SA $S_i, i \neq 1$ that does not represent the original query $Q$, the upper bound is the number of 
valid top-level tuples with values under $S_i$ different from tuples under $S_1$ having all their retained and valid flags set to 1,
e.g., tuple 9 and tuple 10. $UB(\Delta^-)$ equals $|\qEval{\query}_{\db}|$ minus the number of valid top-level tuples under the considered SA that match an original tuple (with only true valid and retained flags) under $S_1$.
In our example, all result tuples not matching the why-not question have at least one nested value with a false retained flag, so we get $UB(\Delta^-) = 1$ for both explanations.
For explanations involving a selection or join, the lower bound is always set to 0, because we do not know if a reparametrization different from the ``full relaxation'' of the operator that our tracing algorithms model may avoid the side effects. In all other cases, we estimate $\squeezespaces{0.5}LB(\Delta^+) = max( \text{number of valid and retained tuples} - |\qEval{\query}_{\db}|, 0)$ and $\squeezespaces{0.5} LB(\Delta^-) = max( |\qEval{\query}_{\db}| - \text{number of valid and retained tuples}, 0)$. 
}
We leave algorithms that compute tighter bounds to future work. \revb{Finally, the explanations are ordered following the partial order defined in Definition~\ref{def:msr}, ranking $SR_1$ higher than $SR_2$.}

\subsection{Discussion}\label{sec:computing-discussion}

\revc{First, we observe that our algorithm guarantees that any returned explanation is a correct SR. However, given our loose bounds on side effects, we cannot guarantee that they are all MSRs. Furthermore, we may miss some operators / explanations due to the algorithm's heuristic nature. Essentially, the proposed algorithm cuts the following corners for efficiency, causing certain cases not to be accurately covered: (i) It considers only equi-joins  and does not model a reparameterization to theta-joins. This avoids cross products that enumerate all possible outputs of join reparameterizations. If such a reparameterization was an explanation, our algorithm misses it. (ii) The tracing procedures for selection, join, and flatten faithfully cover reparameterizations yielding more tuples, compared to the original query operator. So we miss explanations where a more restrictive selection condition, join type, or flatten type would yield a missing answer. (iii) Finally, for aggregations, we generally do not trace the result for different subsets of their input data, which is particularly problematic when selections precede it (for changing equi-join types and flatten types, this is manageable). Also, we do not consider changing the aggregation function.}

 \section{Implementation and Evaluation}
\label{sec:experi}

\begin{figure*}[t]
$\,$\\[-3mm]\hspace{-1mm}
\begin{minipage}{0.27\linewidth}
 \centering$\,$\\[-4mm]
     \includegraphics[width=\columnwidth, trim=20 20 20 20]{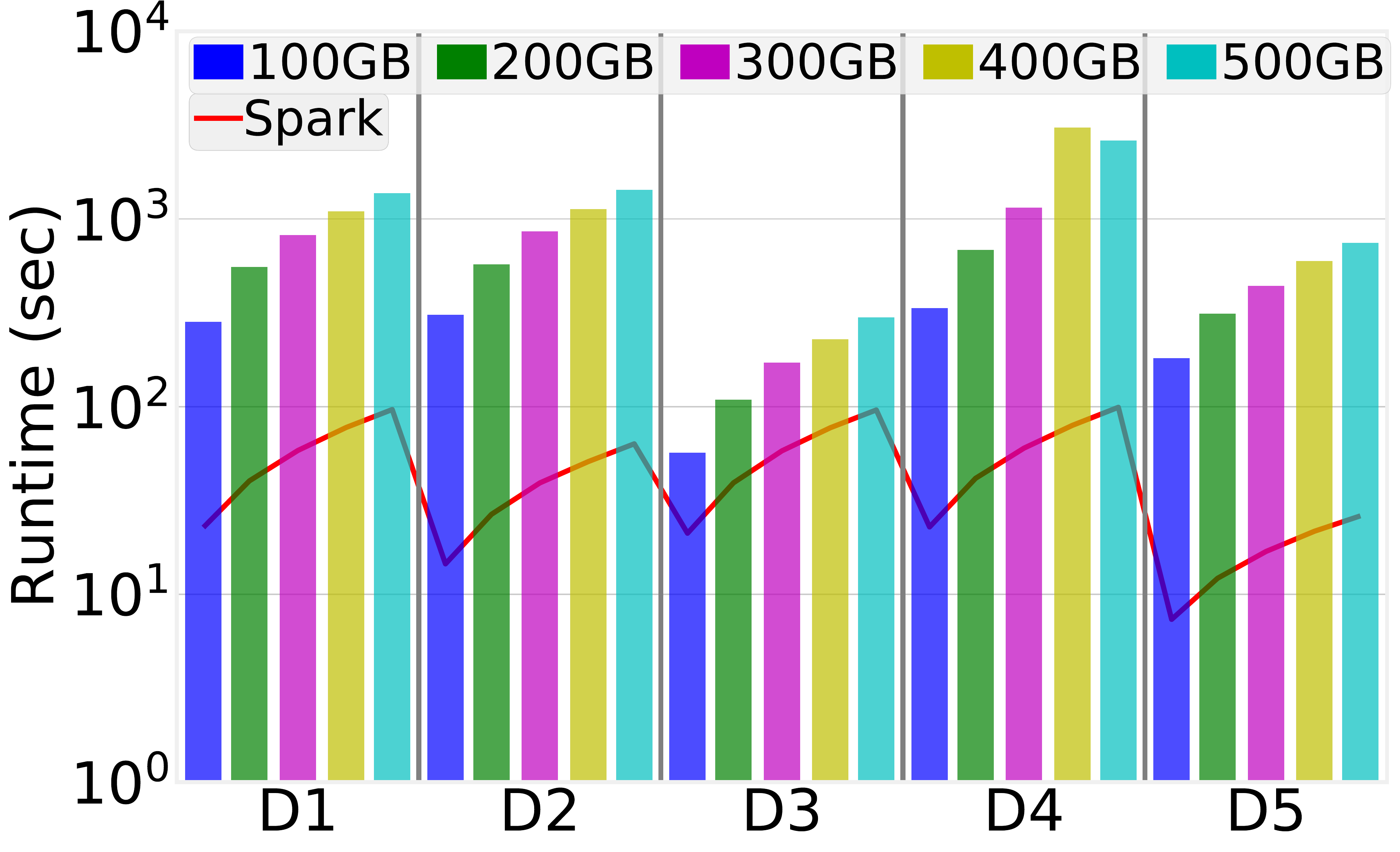}
\\[-3mm]
\caption{Runtime for DBLP} 
\label{fig:perf-dblp}
\end{minipage}
\hspace{0.1cm}
\begin{minipage}{0.27\linewidth}
 \centering $\,$\\[-4mm]
     \includegraphics[width=\columnwidth, trim=20 20 20 20]{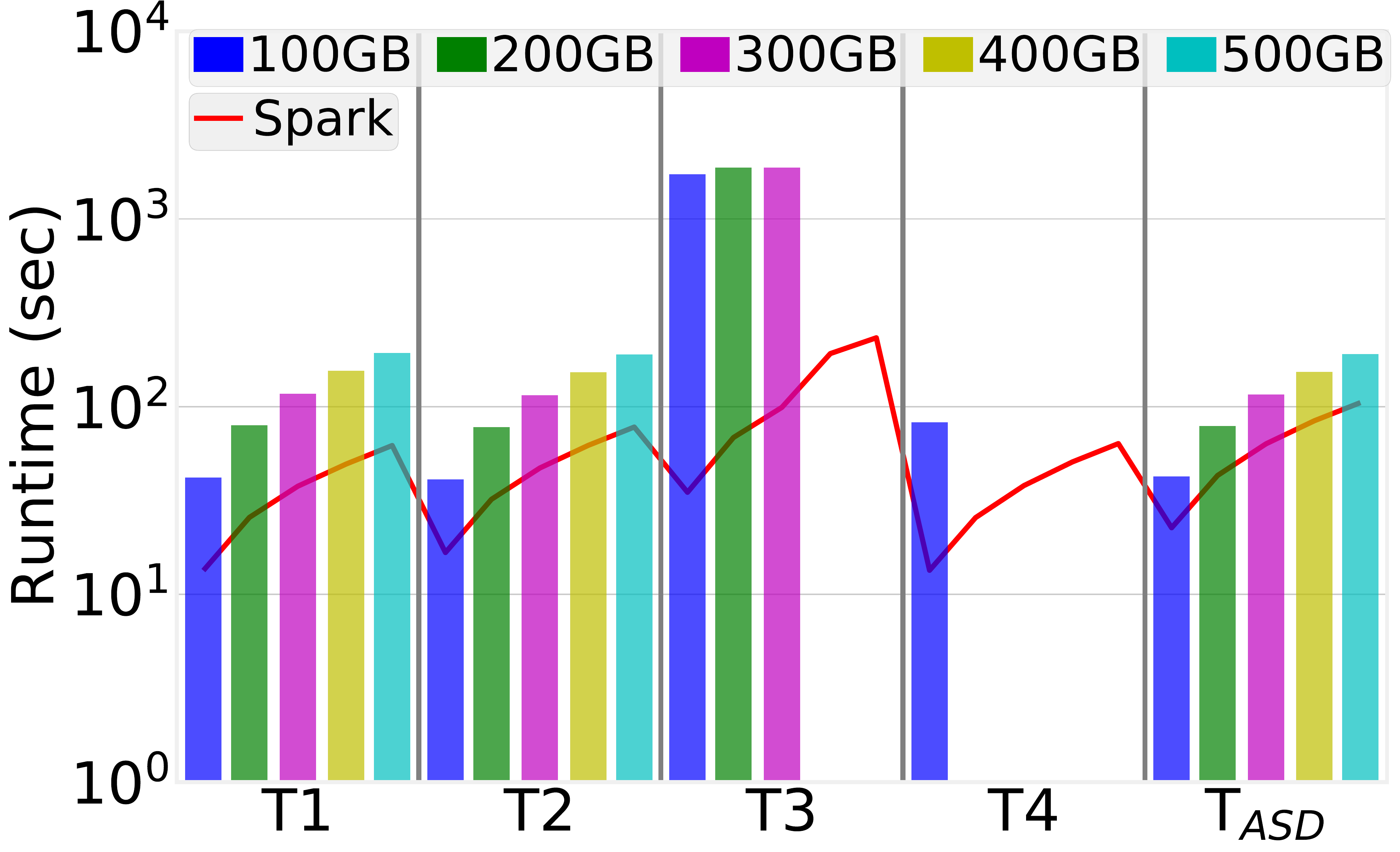}
\\[-3mm]
\caption{Runtime for Twitter} 
\label{fig:perf-tw}
\end{minipage}
\hspace{0.3cm}
\begin{minipage}{0.18\linewidth}
 \centering $\,$\\[2.5mm]
     \includegraphics[width=\columnwidth, trim=20 20 20 20]{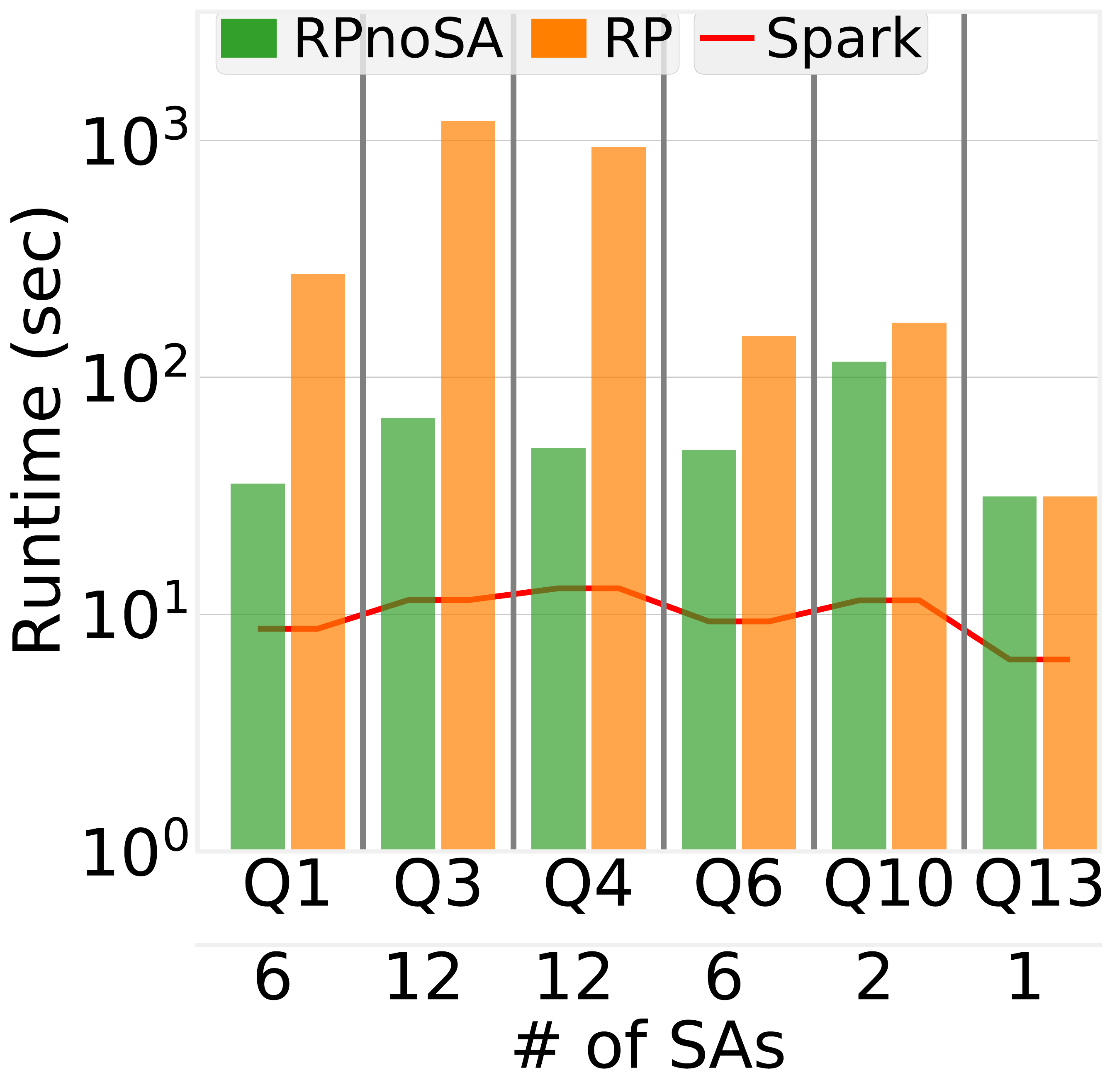}
\\[-4mm]
\caption{Runtime for TPC-H}
\label{fig:perf-tpch}
\end{minipage}
\hspace{0.2cm}
\begin{minipage}{0.19\linewidth}
 \centering $\,$\\[1mm]
     \includegraphics[width=\columnwidth, trim=20 20 20 20]{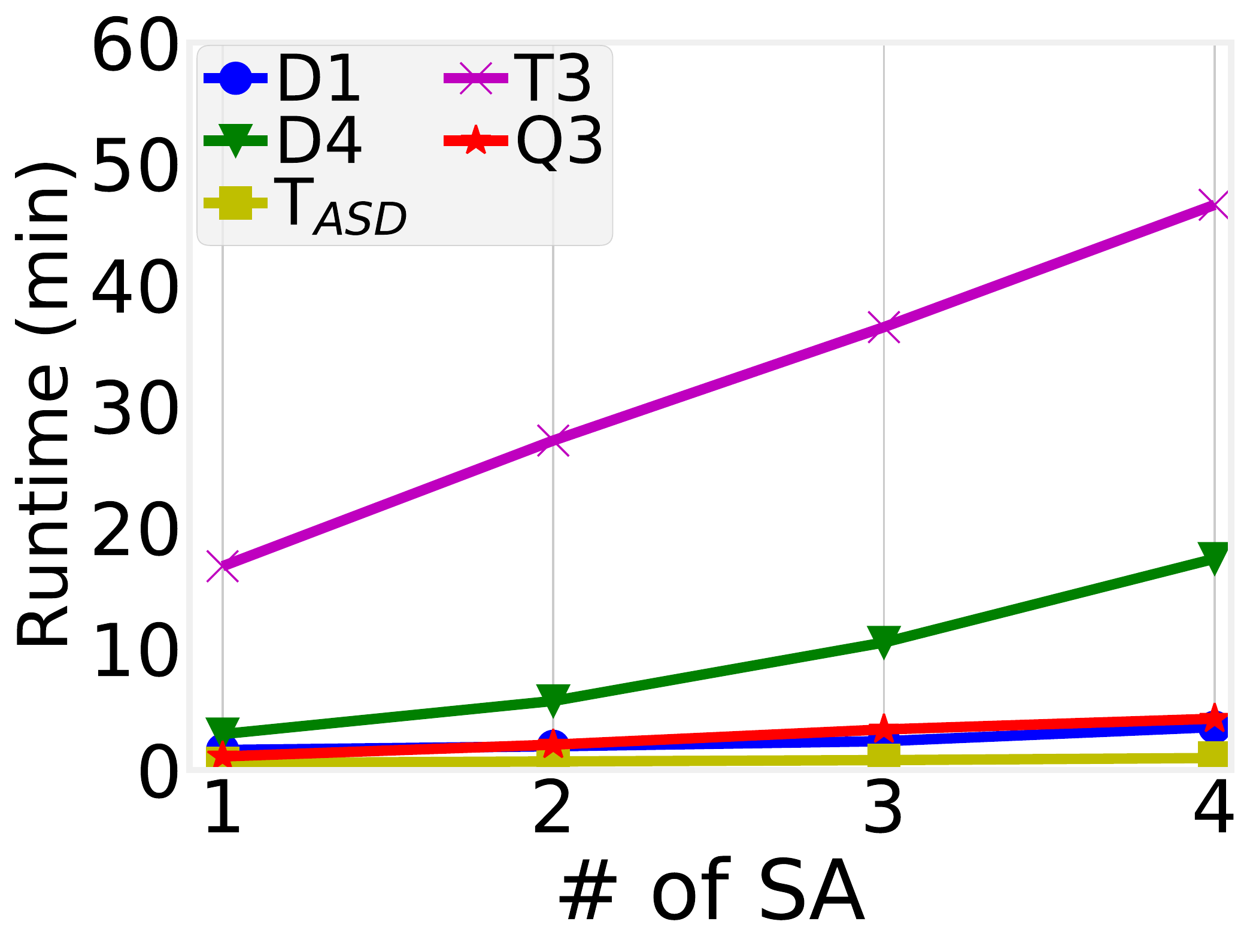}
\\[-2mm]
\caption{Runtime varying schema alternatives (SA)}
\label{fig:perf-sa}
\end{minipage}
\end{figure*}

\iftechreport{

 \begin{table*}[t]
  \centering \scriptsize$\,$\\[-3mm]
\begin{minipage}{0.49\linewidth}
\begin{tabular}{| c p{2.7cm}  p{1.9cm} |} \hline
 \cellcolor{tabbg} \textbf{Set of used operators} & \cellcolor{tabbg} \textbf{Why-not questions} & \cellcolor{tabbg} \textbf{Schema alternatives}\\ \hline
   \multicolumn{3}{| c |}{ \cellcolor{tabbg!40} \textbf{\color{blue}D1}: Computes all authors and titles of papers that are published in SIGMOD proceedings} \\ 
   $\projection, \selection, \Join, \flatten^{I}, \flatten^{T}$ & 
     Why is a paper with a certain title missing? 
     & P.title 
     
     $\to$ P.booktitle 
     \\  
   \multicolumn{3}{| c |}{ \cellcolor{tabbg!40} \textbf{\color{blue}D2}: Computes the number of articles for authors who do not have "Dey" in their name} \\ 
    $\projection, \selection, \flatten^{I}, \flatten^{T}, \nestRel{A}{B}, \aggregation{f}{A} $ 
    & Why is a certain author with a minimum of 5 articles missing? 
    & I.title.bibtex 
    
    $\to$ I.title.text \\  
  \multicolumn{3}{| c |}{  \cellcolor{tabbg!40} \textbf{\color{blue}D3}: Lists all author-paper-pairs per booktitle and year } \\ 
   $\projection, \flatten^{T}, \nestTup{A}{B}, \nestRel{A}{B} $ 
   & Why is a an expect author missing for a fixed booktitle and year? 
   & A.author 
   
   $\to$ A.editor \\  
  \multicolumn{3}{| c |}{  \cellcolor{tabbg!40} \textbf{\color{blue}D4}: Yields a collection of papers per author who have published through ACM after 2010} \\ 
   $\projection, \selection, \flatten^{I}, \flatten^{T}, \Join, \nestRel{A}{B}, \aggregation{f}{A}$ 
   & Why is an expected author missing? 
   & I.publisher 
   
   $\to$ I.series 
   \MH{\&   I.year 
    $\to$ I.modif\_date 
:we said we only use one, to fix} 
\\  
  \multicolumn{3}{| c |}{  \cellcolor{tabbg!40} \textbf{\color{blue}D5}: Computes a list of (hompage) urls for each author} \\ 
   $\projection, \flatten^{I}, \flatten^{T}, \nestRel{A}{B} $ 
   & Why is an author known to have a homepage missing? 
   & U.url
   
    $\to$ U.note \\ \hline
\end{tabular}
\caption{Summary of DBLP scenarios D1 -- D5}
\label{tab:scenarios-dblp}
\end{minipage}
\hspace{0.2cm}
\begin{minipage}{0.49\linewidth}
  \centering \scriptsize
\begin{tabular}{| c  p{2.4cm}  p{2.2cm} |} \hline
 \cellcolor{tabbg} \textbf{Set of used operators} & \cellcolor{tabbg} \textbf{Why-not questions} & \cellcolor{tabbg} \textbf{Schema alternatives}\\ \hline
  \multicolumn{3}{| c |}{  \cellcolor{tabbg!40} \textbf{\color{blue}T1}: Returns tweets providing media urls about a basketball player} \\ 
  $\projection, \selection, \flatten^{I}, \flatten^{T}$ 
  & Why is a certain tweet missing in the result? 
  & T.entities.media
  
  $\to$ T.entities.urls \\ 
  \multicolumn{3}{| c |}{  \cellcolor{tabbg!40} \textbf{\color{blue}T2}: Computes all users who tweeted about BTS in the US} \\
   $\projection, \selection, \flatten^{T}$
   & Why is known fan from the US missing? 
   & T.place.country 
   
   $\to$ T.user.location \\ 
  \multicolumn{3}{| c |}{  \cellcolor{tabbg!40} \textbf{\color{blue}T3} : Yields hashtags and medias for users that are mentioned in other tweets } \\ 
  $\projection, \selection, \flatten^{I}, \flatten^{T}, \Join $ 
  & Why is a user mentioned in a tweet with a certain hashtag missing? 
  & T.entities.media 
  
  $\to$ T.entity.urls \\ 
  \multicolumn{3}{| c |}{  \cellcolor{tabbg!40} \textbf{\color{blue}T4}: Computes a nested list of countries for each hashtag, if the tweeted text contains "UEFA" } \\
  $\projection, \selection, \flatten^{I}, \flatten^{T}, \nestRel{A}{B}, \aggregation{f}{A} $ 
  & Why is a soccer club from England missing? 
  & T.place.country
  
   $\to$ T.user.location \\ 
  \multicolumn{3}{| c |}{  \cellcolor{tabbg!40} \textbf{\color{blue}T$_{ASD}$}: Extracts a flat relation of retweeted tweets }\\
  $\projection, \selection, \flatten^{T}$ 
  & Why is a famous tweet missing?
   & T.retweet\_status

    $\to$ T.quoted\_status \\ \hline
\end{tabular}
\caption{Summary of Twitter scenarios T1 -- T4 and T$_{ASD}$}
\label{tab:scenarios-twitter}
\end{minipage}
\begin{minipage}{\linewidth}
\centering \scriptsize
\trimfigspace
\begin{tabular}{| c | p{10.3cm} | p{3.7cm} | } \hline
  \cellcolor{tabbg} \textbf{Scenario} & \cellcolor{tabbg} \textbf{Descriptions} & \cellcolor{tabbg} \textbf{Why-not questions} \\\hline
C1 & $\projection_{name, type} \left(C \Join_{sector}  \left(W  \Join^2_{name=witness} \left(S \Join_{hair, clothes}  \selection^1_{hair=\text{``blue''}}\left(P\right) \right) \right) \right)$ 
& $\nrctuple{name: \text{``Roger''}, type: ?}$\\
C2 & $\projection_{P.name} \left( P  \Join_{hair, clothes} \left( S \Join_{W.name=witness} \left(C \Join_{sector} \left( \selection^4_{name = \text{``Susan''}} \left( \selection^3_{sector > 90}\left( W \right) \right) \right) \right) \right) \right)$ 
& $\nrctuple{P.name: \text{``Conedera''}}$ \\
C3 & $\projection^6_{name, desc \leftarrow hair} \left(S \Join^5_{name=witness}  \left(W \Join_{sector} \left(C\right) \right) \right)$  
& $\nrctuple{name: \text{``Ashishbakshi''}, desc: \text{``snow''}}$ \\
\hline
\end{tabular}
\caption{Crime scenarios C1 -- C3}
\label{tab:cs}
\end{minipage}

\end{table*}

 }

We implement the algorithm of \Cref{sec:compute-msr} as summarized in \Cref{subsec:impl}. We describe the test setup in \Cref{subsec:setup}. \Cref{subsec:eval-quantitative} covers our quantitative evaluation on scalability, while \Cref{subsec:eval-qualitative} discusses the quality of returned explanations.

\subsection{Implementation}
\label{subsec:impl}
While the concepts apply to DISC systems in general, we implement them in Spark's DataFrames, 
which are tuple collections matching our data model from \Cref{sec:prelim}. They are modified by transformations matching the algebra in \Cref{subsec:nra}. To express and process the why-not questions (\Cref{def:whynot-q}), we leverage tree-patterns~\cite{Lu2011}.

Our prototype integrates into Spark's query planning and execution phases. 
The schema backtracing (\Cref{sec:step1}) and schema alternatives computation (\Cref{sec:step2}) integrate into the query planning phase. 
Data tracing (\Cref{sec:step3}) and computing approximate MSRs (\Cref{sec:step4}) span across both phases. Similar to~\cite{mueller:vldb18},  our prototype rewrites the query plan to directly obtain the MSRs from provenance annotations added for data tracing. 

\RD{While these implementation decisions leverage the distributed computing power of DISC systems and the capabilities available in the query planner, a straightforward implementation of plan rewriting is insufficient to generate efficient plans.}
Furthermore, a straightforward implementation of plan rewriting does not generate efficient plans.
We incorporate multiple optimizations to avoid operator blow-ups in the plan and cross products over the data.
These careful design choices make our algorithm scale to dataset sizes several orders of magnitude larger than those any other state-of-the-art solution can handle. At the same time, it can produce explanations that lineage-based approaches miss.

\subsection{Test Setup}
\label{subsec:setup}

\RD{Our evaluation runs on a Spark 2.4.0 cluster with Hadoop 3.2.0. The cluster hosts 50 executors of 16GB main memory each.}

We test on a Spark 2.4 cluster with  50 executors of 16GB RAM each.
We define \revm{16 scenarios on three nested datasets: T1 to T4 and $\text{T}_{ASD}$~(the latter adapted from \cite{Spoth2017}) on Twitter data, D1 to D5 on DBLP data, and 6 scenarios on a nested version of TPCH that nests lineitems into orders~\cite{Pirzadeh2017} with queries corresponding mostly (as explained later) to the benchmark queries Q1, Q3, Q4, Q6, Q10, and Q13 without the unsupported sorting and top-k selection.} 
\iftechreport{
\revm{
We also implement the TPCH queries on the relational data denoted as Q1F, Q3F, Q4F, Q6F, Q10F, and Q13F to compare the explanations in the nested scenarios with the explanations in the flat data. 
}
}
The Twitter dataset consists of tweets with roughly 1000 mostly nested attributes~\cite{Wang2017}. 
\ifnottechreport{DBLP contains records of different types, such as article, author, etc.}
\iftechreport{
The DBLP dataset contains records of one of ten types, such as \textbf{A}rticle, \textbf{P}roceeding, \textbf{I}nproceeding, or a\textbf{U}thor~\cite{Ley2009}.
} 
\revm{ \Cref{tab:scenarios} summarizes our scenarios. For each scenario, it provides a short description and highlights its query operators (ignore the rest for now).
By default, each Twitter and DBLP scenario has 2 schema alternatives (SAs), i.e., the unmodified SA plus one SA using \ifnottechreport{an} \iftechreport{the specified} attribute alternative.} 
\revm{For the TPCH scenarios, we identify three sets of attribute alternatives:  
  (i) $\nrcset{l\_discount, l\_tax}$,
  (ii) $\nrcset{l\_shipdate, l\_commitdate, l\_receiptdate}$, and
  (iii) $\nrcset{o\_orderpriority, o\_shippriority}$.
  This can yield up to $12$ SAs, 
depending on the attributes used in each query.
}
\ifnottechreport{\cite{techreport} provides more details, e.g., 
  the queries in \abbrNRAB or why-not questions.}
\iftechreport{The scenarios in complete \abbrNRAB are available in \Cref{tab:tpch-scen} and \Cref{tab:NRAB}. Blue colors indicate the introduced errors for the scenarios with a gold standard. \Cref{tab:scenarios-dblp} and \Cref{tab:scenarios-twitter} further describe the DBLP and Twitter scenarios in more detail.}

\revm{When not mentioned otherwise, we apply a scale factor of 10 for TPCH and consider 100GB of DBLP or Twitter data. } To evaluate runtime and scalability, we vary the DBLP and Twitter dataset size between 100GB and 500GB. \revm{To assess explanation quality, we deliberately modified operators in the $\text{T}_{ASD}$ and TPCH queries. The unmodified queries serve as gold standard, such that the explanations precisely containing the modified operators are the correct ones. We study the explanations returned by our reparameterization-based algorithm with (\textbf{\rp}) and without (\textbf{\rpnosa}) multiple schema alternatives. We further compare these to the explanations of a lineage-based approach \textbf{WN++}. To this end, we extended Why-Not \cite{DBLP:conf/sigmod/ChapmanJ09} to scale to big data and to support nested data.}

\vspace{-3mm}
\subsection{Performance Evaluation}
\label{subsec:eval-quantitative}

\mypar{Varying dataset size} The bars in Figures~\ref{fig:perf-dblp} and \ref{fig:perf-tw} report 
\rp's runtime for DBLP and Twitter scenarios for varying dataset sizes given a 2 hours time-out. The line reports the original query runtime.

First, we note linear scalability with the input size. Second, our implementation exceeds the runtime of the original query by a factor between $2.4$ and $78.2$, depending on the scenario. This overhead is in line with the overhead of state-of-the art solutions on relational data. A closer analysis reveals that the overhead is particularly low in queries with a low number of operators, such as D3, T2, and \revb{$\text{T}_{ASD}$}. The overhead increases when the queries become more complex (D4, D5, T3, T4). For such queries, our annotations grow in size, causing additional runtime overhead and even exceeding our time-out limit for larger input sizes of T3. Furthermore, joins are expensive. Spark rewrites the joins in D4 and T3 from Hash-Joins to much slower Sort-Merge-Joins, since it does not support outer Hash-Joins. However, we require the outer joins to accurately trace tuples without a join partner. Moreover, high runtime overhead occurs when the output is based on a small subset of the input tuples. For example, in D5, two inner flatten operators on nested relations that are empty for most tuples yield much fewer output tuples than input tuples. In contrast, our tracing algorithm retains at least one output tuple for each input tuple. Finally, for T4, our results limit to 100GB input data, because we hit a Spark limitation for larger sizes. It is related to a reported bug in Spark's grouping set implementation, which we use in the aggregation tracing procedure, and Spark's current item limit in nested collections ($2^{31}$).

\revm{
  In the TPCH scenarios (\Cref{fig:perf-tpch}), the overhead has a factor of
  $3.9$ and $10.1$ 
  for \rpnosa, and up to 
  $105.2$ 
  for \rp. It is larger for two reasons. First, all TPCH queries use aggregations. Thus, their result size is insignificant compared to the number of traced tuples (analogous to D5). Second, the higher numbers of SAs cause higher overhead (up to 12 compared to 2 for DBLP and Twitter). }

 \MH{Is this still accurate? Or is this better: It is related to a reported bug in Spark's grouping set implementation, which we use in the aggregation tracing procedure, and Spark's current size limit of nested collections ($2^{31}$).}

\iftechreport{
	\switchon
}
\ifnottechreport{
	\switchoff
}

\begin{table*}[t]
\centering \small
\begin{tabular}{| c | p{8.3cm} | p{1.0cm}|p{0.2cm}|p{0.2cm}|p{1.0cm}|p{0.2cm}|p{0.2cm}| c | c | c | } \hline
Scen. & Query & \multicolumn{6}{ c }{Operators} & \multicolumn{3}{|c|}{\# explanations } \\ 
	&  & $\selection$ & $\projection$  &  $\Join$ & $\flatten$ & $\mathcal{N}$ & $\gamma$ & WN++ & RPnoSA & RP  \\ \hline 
D1 & All authors and titles of papers that are published at SIGMOD & \cellcolor{tabbg!40} \pieempty & \cellcolor{tabbg!40} \piefull & \cellcolor{tabbg!40} &  \cellcolor{tabbg!40} &  \cellcolor{tabbg!40}   & & 1 & 1 & 2  \\ \hline
D2 & Number of articles for authors who do not have "Dey" in their name &  \cellcolor{tabbg!40} &  \cellcolor{tabbg!40} &  &  \cellcolor{tabbg!40} \piefull &  \cellcolor{tabbg!40} &  \cellcolor{tabbg!40}  & 0& 0& 1  \\ \hline
D3 & Lists all author-paper-pairs per booktitle and year  & &   \cellcolor{tabbg!40} &  &  & \cellcolor{tabbg!40} \piefull &  & 0& 0& 1 \\ \hline
D4 & Collection of papers per author having published through ACM after 2010 &  \cellcolor{tabbg!40}\pieempty  \piehalf &  \cellcolor{tabbg!40} &   \cellcolor{tabbg!40} &  \cellcolor{tabbg!40} \piefull &  \cellcolor{tabbg!40} &  \cellcolor{tabbg!40}  & 1&2& 4 \\ \hline
D5 & List of (hompage) urls for each author & & \cellcolor{tabbg!40} \piefull& &  \cellcolor{tabbg!40} \pieempty & \cellcolor{tabbg!40} &  & 1& 1& 2  \\ \hline
T1 &  List of tweets providing media urls about a basketball player& \cellcolor{tabbg!40} \piehalf & \cellcolor{tabbg!40} &  & \cellcolor{tabbg!40} \piehalfred \piehalf \piefull & &  & 1& 1& 2 \\ \hline
T2 & All users who tweeted about BTS in the US & \cellcolor{tabbg!40} \pieempty \piehalf  &\cellcolor{tabbg!40}  &  & \cellcolor{tabbg!40}  \piefull & &  & 1& 2& 4 \\ \hline
T3 & Hashtags and medias for users that are mentioned in other tweets  &  & \cellcolor{tabbg!40}  & \cellcolor{tabbg!40} & \cellcolor{tabbg!40} \pieempty \piefull & &  & 1& 1& 2 \\ \hline
T4 & Nested list of countries for each hashtag, if tweet contains  ``UEFA''  &  \cellcolor{tabbg!40} \piehalfred \piehalf &  \cellcolor{tabbg!40} &  &  \cellcolor{tabbg!40} \piefull &  \cellcolor{tabbg!40}  & \cellcolor{tabbg!40}   & 1& 1& 3  \\ \hline
\revb{$\text{T}_{ASD}$} &  \revb{ASD example~\cite{Spoth2017}: flatten, filter, project quoted tweets (2 modifications)}  &  \cellcolor{tabbg!40} \piefull &  \cellcolor{tabbg!40} &  &  \cellcolor{tabbg!40} \piefull & &  & 0 (-)& 0 (-)& 2 (2)  \\ \hline
\revm{Q1} & \revm{TPCH query 1 with one modified aggregation} &  \cellcolor{tabbg!40} \pieempty &  & & \cellcolor{tabbg!40} & & \cellcolor{tabbg!40} \piefull  & 1 (-)& 1 (-)& 3 (2)  \\ \hline
\revm{Q3} & \revm{TPCH query 3 with two modified selections} &  \cellcolor{tabbg!40} \piehalfred \piehalf \piefull &  \cellcolor{tabbg!40} &   \cellcolor{tabbg!40} &  \cellcolor{tabbg!40}& &  \cellcolor{tabbg!40} \piefull & 1 (-)& 1 (1)& 2 (1) \\ \hline
\revm{Q4} &  \revm{TPCH query 4 with a modified selection and aggregation} &  \cellcolor{tabbg!40} \piefull & & &  \cellcolor{tabbg!40}& &   \cellcolor{tabbg!40} \piefull & 0 (-)& 0 (-) & 4 (3)   \\ \hline
\revm{Q6} & \revm{TPCH query 6 with one modified selection} &  \cellcolor{tabbg!40} \pieempty \piehalf \piefull&  \cellcolor{tabbg!40}&  &  \cellcolor{tabbg!40} & &   \cellcolor{tabbg!40}& 1 (-)&  7 (2)& 11 (2)\\ \hline
\revm{Q10} & \revm{TPCH query 10 with two modified selections and a modified projection} &  \cellcolor{tabbg!40} \piehalf \piefull &  \cellcolor{tabbg!40} \piefull  &  \cellcolor{tabbg!40} \piered&  \cellcolor{tabbg!40}& &  \cellcolor{tabbg!40} & 1(-)& 2(-)& 4 (4) \\ \hline
\revm{Q13} & \revm{TPCH query 13 with one modified join}  &  \cellcolor{tabbg!40}& & \cellcolor{tabbg!40} \pieempty  & &  &  \cellcolor{tabbg!40} & 1 (1)& 1(1)& 1 (1)  \\  
\rowswitch \hline \revm{Q1F} & \revm{TPCH query 1 with one modified aggregation} &  \cellcolor{tabbg!40} \pieempty &  & & & & \cellcolor{tabbg!40} \piefull  & 1 (-)& 1 (-)& 3 (2)   \\ 
\rowswitch \hline \revm{Q3F} & \revm{TPCH query 3 with two modified selections} &  \cellcolor{tabbg!40} \piehalfred \piehalf \piefull &  \cellcolor{tabbg!40} &   \cellcolor{tabbg!40} & & &  \cellcolor{tabbg!40} \piefull & 1 (-)& 1 (1)& 2 (1) \\ 
\rowswitch \hline \revm{Q4F} &  \revm{TPCH query 4 with a modified selection and aggregation} &  \cellcolor{tabbg!40} \piefull & & \cellcolor{tabbg!40} & & &   \cellcolor{tabbg!40} \piefull & 0 (-)& 0 (-) & 4 (3)   \\ 
\rowswitch \hline \revm{Q6F} & \revm{TPCH query 6 with one modified selection} &  \cellcolor{tabbg!40} \pieempty \piehalf \piefull& &  & & &   \cellcolor{tabbg!40}& 1 (-)&  7 (2)& 11 (2)\\ 
\rowswitch \hline \revm{Q10F} & \revm{TPCH query 10 with two modified selections and a modified projection} &  \cellcolor{tabbg!40} \piehalf \piefull &  \cellcolor{tabbg!40} \piefull  &  \cellcolor{tabbg!40} \piered& & &  \cellcolor{tabbg!40} & 1(-)& 2(-)& 4 (4) \\ 
\rowswitch \hline \revm{Q13F} & \revm{TPCH query 13 with one modified join}  &  \cellcolor{tabbg!40}& & \cellcolor{tabbg!40} \pieempty  & &  &  \cellcolor{tabbg!40} & 1 (1)& 1(1)& 1 (1)  \\  
\hline
\end{tabular}

\pieempty: Found by all algorithms, \piehalf: found only by RPnoSA and RP,  \piefull: found only by RP, \piehalfred: WN++ is incomplete \piered WN++ is incorrect
\caption{Summary of explanations returned for the lineage-based approach WN++, our reparameterization-based approach without SAs (RPnoSA) and our fully fledged approach RP.  Shaded fields indicate that a scenario's query uses one or more operators of this type, and the shaded circles indicate operators found by the different approaches (see legend).}
\label{tab:scenarios}
\vspace{-7mm}
\end{table*}
 
\mypar{Varying the number of SAs} 
To study the runtime impact of SAs, 
we consider between 1 and 4 SAs.
We report results for a simple scenario
with only a few operators and insignificant changes in intermediate result sizes (\revb{T$_{ASD}$}),
 two scenarios of intermediate difficulty with relation flatten and join operators (D1, T3), and \revm{two} difficult scenarios featuring flatten, join, nesting, and aggregation (D4, \revm{Q3}).
 Figure~\ref{fig:perf-sa} shows the results.
 For all but the most difficult scenarios, the runtime increases by a constant factor of \revb{0.15 (T$_{ASD}$),} 0.5 (D1), or 0.8 (T3) per added SA. 
 Since the factor is below 1, adding an SA to the rewritten query is faster than executing seperate queries for each SA.
 In D4, adding SAs causes some deceleration. While the factor is 0.96 for adding the first alternative, it is 1.47 for adding the last alternative. \reva{Similarly, for Q3, the deceleration occurs when going up to 12 SAs with a factor of 4.76 from 4 SAs and of 17.92 from one SA.} 
 The reason is twofold. First, with each added SA, each tuple's size increases. Second, the grouping set implementation in Spark used for our aggregations duplicate each input tuple for each alternative. Thus, both the tuple width and the tuple number increase with each SA, explaining the growing factor.

\vspace{-0.2cm}
\subsection{Explanation Quality}
\label{subsec:eval-qualitative}

\revm{We summarize the explanations returned by WN++, \rpnosa, and \rp\ for all scenarios in~\Cref{tab:scenarios}. We flag \pieempty if all algorithms find an explanation involving the same operators of this type. The flag \piehalf indicates that WN++ misses an explanation involving an operator of this type, which both \rpnosa\ and \rp\ find. Operators appearing only in \rp's explanations are marked \piefull. \piehalfred marks WN++'s incomplete explanations, i.e., explanations that involve the marked operator, but require modifying another operator that WN++ misses. \piered describes incorrect explanations only found by WN++. Note that a query may contain multiple operators (with different id) of the same type, which explains that some cells have multiple flags.}
\revm{As shown in the three rightmost columns in~\Cref{tab:scenarios}, WN++ finds 12, \rpnosa\ detects 21, and \rp\ yields 48 explanations. The numbers in brackets behind the number of explanations indicate the positions of the correct explanations in the scenarios with a gold standard.}
\iftechreport{\Cref{tab:answers} lists all explanations. Incorrect explanations found by WN++ are highlighted in red. The correct explanations regarding the gold standard are shown in blue. The operators in these explanations refer to operators in \Cref{tab:tpch-scen} and \Cref{tab:NRAB} with the same superscript index.}

\revm{Next, we describe scenarios Q3, Q10, and $\text{T}_{ASD}$ in detail, before we conclude with generalized observations. We use $op^{id}$ notation to distinguish multiple operators of the same type
 \ifnottechreport{(see \cite{techreport} for the queries in \abbrNRAB)}.}
\revm{Scenario Q3 computes unshipped orders. We have introduced a typo in the constant commitdate in $\selection^{27}$ and replaced the marketsegment in $\selection^{26}$ as errors and miss a certain order in the output. WN++ finds only $\selection^{27}$ as an explanation, because it removes the order entirely from its output. Unlike our solution, WN++ misses that $\selection^{26}$ on the marketsegment would also remove the missing order. Thus, \rpnosa, and \rp\ return $\{\selection^{26}, \selection^{27}\}$ as their first and correct explanation. \rp\ further yields $\{\selection^{26}, \selection^{27}, \gamma^{25}\}$ as explanation. It is based on schema alternatives reflecting the tax as alternative to the discount.  The last explanation yields the missing order since the order appears in the result regardless of the SA. This scenario shows that our solution already outperforms WN++ without SAs.}
\iftechreport{
\revm{
To show that WN++ also struggles to find the complete explanation on flat data, we repeat this scenario on relational data (Q3F). We introduce the same errors and ask the same why-not question. On flat data, WN++ returns only the manipulated selection on the marketsegment since it removes all compatible customers before they are joined with the lineitems. It misses the manipulated selection on the commitdate. \rpnosa and \rp return both selections as their first explanation. Once again \rp\ further yields the two selections together with the aggregation based on the according SA. Thus, we conclude that our solution provides explanations that WN++ misses on relational data without schema alternatives.
}
}

\revm{Scenario Q10 reports returned items and the associated revenue loss. We introduce three errors in the query. First, we replace the constant value in the selection $\selection^{35}$ on the returnflag. Second, we replace the constants in the selection $\selection^{36}$ on the orderdate.
Third, we replace the discount with the tax in the projection $\projection^{37}$ that computes the discount, which is considered to compute the correct, non-zero revenue. We expect a missing customer in the result who generates noticable revenue. WN++ finds the $\join^{38}$ on customer and order as explanation because it removes the expected customer from the result. This explanation is misleading or even incorrect, since it makes the customer appear in the result, but cannot provide a non-zero revenue, which we ask for. \rpnosa\ and \rp\ first point to $\selection^{35}$.
It removes all potential join partners for the expected customer, because the customer has no lineitems with the modified returnflag. Next, \rpnosa\ and \rp\ find both selections $\{\selection^{35}, \selection^{36}\}$, because $\selection^{36}$ also removes tuples that join with the expected customer.  \rp\ further returns $\{\selection^{35}, \projection^{37}\}$ and $\{\selection^{35}, \selection^{36}, \projection^{37}\}$
, which add $\projection^{37}$ to the discussed explanations. 
The last explanation is based on SAs and precisely points at all our modifications. It is ranked last, since it modifies the most operators. However, note that one would have obtained the correct solution iteratively when observing the provided selections before the projection. Our solution does not return $\join^{38}$, because it cannot yield the non-zero revenue.}

\revm{We finally describe the \revb{T$_{ASD}$} scenario from~\cite{Spoth2017} in detail. An adaptive schema database (ASD) extracts and refines relational schemata from semi-structured or unstructured data.
In~\cite{Spoth2017}, the ASD extracts one relation each for the nested retweeted tweets, and the nested quoted tweets. To extract the retweeted tweets the ASD (i) flattens them with $\flatten^{21}$, (ii) filters a non-null retweet count in $\selection^{22}$, and (iii) projects only the attributes from the retweet. We run this query on our Twitter data after adding two errors to it (reflecting an ambiguity between retweets and quotes): We flatten the quoted tweets and filter on the quote count. The missing answer is a certain retweet. As finding these errors requires schema alternatives, \rp\ is the only algorithm to find explanations, i.e., 
 $\{\flatten^{21}\}$ and $\{\flatten^{21}, \selection^{22}\}$. 
 The example shows that \rp\ adds two key features to ASDs. It helps resolving schema ambiguities through SAs and provides means to find missing, but expected data in the flat output relations.
}

\iftechreport{
\revm{
\mypar{DBLP Explanations} We describe the explanations of each DBLP scenario in detail. In D1, WN++ and our solution identify the selection on the $ptitle$ as explanation to the missing paper title. The explanation is misleading, since the missing paper is published on the $SIGMOD$ conference. However, the projection $\projection^1$ projects on the $title$ instead of the $booktitle$ onto the $ptitle$. While the $booktitle$ contains the string $SIGMOD$, the $title$ holds the written out version the proceedings name. Leveraging the schema alternatives, our solution identifies the faulty projection $\projection^1$ as the second explanation. Thus, in this scenario, WN++ and our solution yield a misleading explanation. However, our solution also pinpoints $\projection^1$ as an explanation.
In D2, we miss a certain author in the result of whom we are sure to have more than five publications. In the output, it has zero publications. While WN++ does not provide an explanation, our solution provides the tuple flatten $\flatten^{T^3}$ as an explanation. $\flatten^{T^3}$ flattens the attribute $title.bibtex$. It holds null values for more than 99\% of the titles in the data in general and for all instances of the author under observation. Since Spark only nests non-null values, the query generates an empty $ctitle$ relation. A count on this relation yields zero. Thus, choosing the non-null $title.text$ in $\flatten^{T^3}$ instead of $title.bibtex$ is the only and correct explanation to provide the author with the correct paper count.
Scenario D3 lists all author-paper-pairs per $booktitle$ and $year$. However, the listing misses a certain person with a certain $booktitle$ and $year$. The missing combination can only appear if the query uses the $editor$ instead of the $author$. While WN++ does not find any explanation, our solution yields the explanation as first and only explanation.
D4 yields a collection of papers per author who have published through ACM in 2010. We wonder why an expected author is missing. WN++ yields one explanation, our solution four. Both solutions explain that $\selection^6$ requires reparameterization. Our solution further explains to reparameterize $\selection^6$ and $\selection^7$. Both explanations potentially yield the missing author, but require the constants in the filter conditions to be changed. We are sure that the $\selection^6$ correctly filters on $ACM$. However, $\selection^7$ filters on year 2015 rather than 2010, which was our initial intention. Thus, we have a closer look at the other two explanations containing $\selection^7$. Both require reparameterization of the tuple flatten $\flatten^{T^5}$. Flattening the $series$ instead of the $publisher$, yields the value $ACM$ in the attribute that $\selection^6$ filters on. Thus, changing $\flatten^{T^5}$ and the $\selection^7$ yields the result, we have expected. Additionally changing $\selection^6$, as the last explanation suggests, is not needed. Thus, our solution provides the expected explanation as the third result, whereas WN++ only provides a misleading explanation.
In D5, the query computes a list of (hompage) urls for each author but lacks a certain author known to have a homepage. WN++ and our solution yield the inner flatten $\flatten^{I^9}$ as only or first explanation, respectively. While changing the inner to an outer flatten makes the author appear in the result the homepage url is still missing, for it is stored in the $note$ attribute instead of the $url$ attribute, which is common in the DBLP dataset. Our solution points at $\projection^8$ in its second explanation. Replacing the $url$ with the $note$ in this projection lets the author appear with the correct homepage url. It does not require changing the inner flatten to an outer flatten.
}
}

\iftechreport{
\revm{
\mypar{Further Twitter Explanations} The Twitter scenarios show additional advantages of our solution compared to WN++.
Scenario T1 provides a relation of all tweets with media URLs about a basketball player. The output relation misses a famous tweet about player, because the query has two errors. First, the tweet is about \textit{LeBron James} and not about \textit{Michael Jordan}, as we thought when writing the query. Second, the schemas records the URLs in the $url$ attribute rather then the used $medias$ attributes. The $medias$ attribute is empty for the missing answer. WN++ yields the flatten operator $\flatten^{I^{11}}$ as its only explanation. Changing the inner flatten to an outer flatten does not suffice to make the missing tweet appear, since $\selection^{12}$ on \textit{Michael Jordan} would filter the missing tweet. Thus, WN++ provides an imcomplete explanation. In contrast, our solution yields the set containing $\flatten^{I^{11}}$ and $\selection^{12}$ as a complete explanation. It is the first of two explanations. The second explanation is computed from the schema alternative. It suggests to modify the $\selection^{12}$ and $\flatten^{T^{10}}$, which yields the query we intended to write.
In T2, which computes all users who tweeted about \textit{BTS} in the US is a known US-based fan missing. While WN++ provides one explanation, our solution yields four explanations. Both suggest to reparameterize $\selection^{15}$, which filters on the country $US$. Changing the filter makes the user appear in the result, but we are shure that they tweeted from the $US$. Thus, we have a closer look at the remaining three explantions our solution provides. The second explanation suggests to modify the tuple flatten $\flatten^{T^{12}}$. Indeed, replacing the $place.country$ attribute with the $user.location$ attribute in $\flatten^{T^{12}}$ yields the intended query. However, our solution provides two further explanations. As third explanation, it suggests to reparameterize $\selection^{15}$ and $\selection^{14}$. The latter selection intentionally filters on the text \textit{BTS}. The last explanation suggests to modify $\flatten^{T^{12}}$ in addition to the two selections. Therefore, the last two explanations do not yield the intended query, but they are ranked lower than the applied explanation.
T3 computes hashtags and media for users that are mentioned in other tweets. An expected user is missing. Our solution applies the same schema alternative as in T1. It yields two explanations, whereas WN++ yields only one. Both suggest to make the inner flatten on $entities.media$ $\flatten^{I^{17}}$ an outer flatten as first or only explanation, respectively. This explanation makes the user appear, but lacks the proper media urls. Our solution's second explanation suggests to reparameterize $\flatten^{T^{16}}$ based on the schema alternative. This reparameterizatian yields the user and the media urls.
T4 computes a nested relation of countries for each hashtag used in a tweet about \textit{UEFA}. It counts the number of countries per hashtag and removes all tuples, whose count is zero. We expect an English soccer club to appear in at least one of the nested hashtags. WN++ suggests to modify $\selection^{19}$ on the \textit{UEFA} as only explanation. This explanation is incomplete since modifying that filter will not make the missing tuple appear in the result. Our solution provides three explanations. First, it suggests to reparameterize the tuple flatten $\flatten^{T^{12}}$ based on the applied schema alternative. It yields the missing result and the intended query. Second, it suggests to modify $\selection^{19}$ and $\selection^{20}$ that filters on the count. Unlike WN++'s explanation, this explanation also yields the missing result, but requires the modification of two operators. Thus, it is ranked second. As third explanation, our solution suggests to modify all three mentioned operators. While it yields the missing tuple, it requires the most reparameterizations to the query and is ranked last.
}
}

\iftechreport{
\revm{
\mypar{Further TPCH Explanations}
In Q1, we replace the $l\_tax$ with the $l\_discount$ in the aggregation and expect the average discount to be smaller than the value in the result. Both, WN++ and our solution without SA find filter $\selection^{24}$ as an explanation. While careful filtering may yield a smaller average discount, the correct solution is to reparameterize the aggregation $\gamma^{23}$. Our solution finds this explanation as second result when applying schema alternatives. As final result, it returns both mentioned operators as explanation, which also potentially yields the expected result.
Query Q4 counts the orders by order priority. As first error, we replace the commitdate with the shipdate in one of the query's selections. As second error, we replace the orderpritority with the shippriority in the aggregation. We expect the value \emph{3-MEDIUM} as priority value. Further, we expect a different count. WN++ and our solution without SAs fail to yield an explanation, because the values in the shippriority attribute lack the \emph{MEDIUM} string. Consequently, they cannot produce the missing result. When our solution leverages schema alternatives, it finds four explanations. All these explanations contain the aggregation $\gamma^{30}$ which is one of our introduced errors and the cause for the absence of the \emph{3-MEDIUM} value. In addition to the explanation, that only holds the aggregation, it finds three explanations that also contain the selections in the query: $\{\gamma^{30}, \selection^{29}\}$, \textcolor{blue}{$\{\gamma^{30}, \selection^{28}\}$}, $\{\gamma^{30}, \selection^{29}, \selection^{28}\}$. It finds all filter combinations, because each tuple that ends up in the final aggregation influences the aggregated order count. Further, each selection removes tuples from their input. Thus, reparameterizing them potentially impacts the order count. In this scenario, the
third explanation correctly pinpoints introduced errors. Even though it is ranked quite low, keep the following two aspects in mind: (i) the explanation extends the first explanation that points at the aggregation, (ii) existing solutions would not have found any explanations.
Q6 computes a single revenue value. We replace the discount with the tax attribute in one of its three selections as error. Further, we expect less revenue than we get after introducing the error. WN++ yields the last selection $\selection^{32}$ as explanation. Our solution yields the powerset of the three selections, because it marks all input tuples as compatibles. All of them can contribute to the aggregated value. Further, each of the filters remove tuples from the input data in such a way that all selection combinations occur in the explanations. With the help of schema alternatives, our solution further finds combinations that replace the tax with the discount.
In this example, the correct solution is ranked second. If we introduced more errors, the correct solution moves further back in the ranking, showing the limitations of our solution.
In scenario Q13, we are missing the count of customers, who have not placed any order yet. Instead of using the left outer join, we initially applied an inner join. Both solutions identify the inner join as the root cause for the missing customer count as the only explanation. We rerun the scenario on a schema that has the orders relation nested into customers relation and apply an inner flatten instead of the join for further processing. In this scenario, our solution correctly pinpoints the inner flatten as explanation, which is the anologous explanation to the join operator only on the more deeply nested customers relation.
}
}

\iftechreport{
\revm{
Since the TPCH scenarios run on nested and on flat data, we have also run them on flat data to compare the explanation. In~\Cref{tab:scenarios}, the flat scenarios are indicated by a trailing F, e.g. Q3F refers to TPCH query 3 on flat data. \Cref{tab:scenarios} further reveals that the flat scenarios do not contain the flattening operator unlike the according nested scenarios. However, the explanations in the TPCH scenarios only contain the selection, projection, join, and aggregation in the explanation. Unlike flattening and nesting, are not specific to nested data. Thus, our solution finds the same explanations on the nested and the flat data. Only in Q3/Q3F the order of the selections in the queries is different. Thus, WN++ finds different selections in this scenario, as explained above.
}
}

\revm{In general, even \rpnosa\ finds explanations that WN++ misses (T1, T4, Q3, Q6, Q10) because \rpnosa\ traces through the entire query. While WN++ runs on nested data here, it suffers the same problem on relational data 
\ifnottechreport{
  as further described in~\cite{techreport}.
}
\iftechreport{
  as described in detail in scenario Q3.
} 
Furthermore, \rp\ may find explanations based on schema alternatives that \rpnosa\ and WN++ miss (happens in all scenarios except Q13). In fact, the schema alternatives may be the only means to obtain an explanation at all (D2, D3, \revb{T$_{ASD}$}, Q4). When multiple operators need reparameterizations, our solution provides the correct explanation, but it is generally ranked lower, like in Q10 and \revb{T$_{ASD}$}. However, the operators of higher ranked explanations typically intersect with the operators in the correct explanation. Thus, starting investigations with the higher-ranked explanations seems a viable option to incrementally correct a query.
}

\iftechreport{
\begin{table}[t]
\vspace{-2mm}
\centering \scriptsize
\begin{tabular}{| c | p{0.65cm} | p{1.7cm} | p{4.3cm} |} \hline
\cellcolor{tabbg} \textbf{Scen.}                                                                                                                                                                                                                                       & \cellcolor{tabbg} \textbf{WN++}          & \cellcolor{tabbg} \textbf{without SA}                         & \cellcolor{tabbg} \textbf{with SA} \\\hline
D1                                                                                                                                                                                                                                                                     & $\{\selection^2\}$                       & $\{\selection^2\}$                                            & $\{\selection^2\}$, $\{\projection^1\}$  \\
D2                                                                                                                                                                                                                                                                     & $\emptyset$                              & $\emptyset$                                                   & $\{\flatten^{T^3}\}$ \\
D3                                                                                                                                                                                                                                                                     & $\emptyset$                              & $\emptyset$                                                   & $\{{\nestTupSimp}^4\}$\\
D4                                                                                                                                                                                                                                                                     & $\{\selection^6\}$                       & $\{\selection^6\}$, $\{\selection^6, \selection^7\}$          & $\{\selection^6\}$, $\{\selection^6, \selection^7\}$, $\{\flatten^{T^5}, \selection^7\}$, $\{\flatten^{T^5}, \selection^6, \selection^7\}$\\
  D5                                                                                                                                                                                                                                                                   & $\{\flatten^{I^9}\}$                     & $\{\flatten^{I^9}\}$                                          & $\{\flatten^{I^9}\}$, $\{\projection^8\}$\\
  \hline
T1                                                                                                                                                                                                                                                                     & \textcolor{red}{$\{\flatten^{I^{11}}\}$} & $\{\flatten^{I^{11}}, \selection^{12}\}$                      & $\{\flatten^{I^{11}}, \selection^{12}\}$, $\{\flatten^{T^{10}}, \selection^{12}\}$ \\
T2                                                                                                                                                                                                                                                                     & $\{\selection^{15}\}$                    & $\{\selection^{15}\}$, $\{\selection^{14}, \selection^{15}\}$ & $\{\selection^{15}\}$, $\{\flatten^{T^{13}}\}$, $\{\selection^{14}, \selection^{15}\}$,  $\{\flatten^{T^{13}}, \selection^{14}, \selection^{15}\}$\\
T3                                                                                                                                                                                                                                                                     & $\{\flatten^{I^{17}}\}$                  & $\{\flatten^{I^{17}}\}$                                       & $\{\flatten^{I^{17}}\}$, $\{\flatten^{T^{16}}\}$ \\
T4                                                                                                                                                                                                                                                                     & \textcolor{red}{$\{\selection^{19}\}$}   & $\{\selection^{19}, \selection^{20}\}$                        & $\{\flatten^{T^{18}}\}$, $\{\selection^{19}, \selection^{20}\}$, $\{\flatten^{T^{18}}, \selection^{19}, \selection^{20}\}$ \\
T$_{ASD}$                                                                                                                                                                                                                                                                   & $\emptyset$                    & $\emptyset$                                         & $\{\flatten^{T^{21}}\}$, \textcolor{blue}{$\{\selection^{22}, \flatten^{T^{21}}\}$}  \\
\hline
\hline
Q1                                                                                                                                                                                                                                                                     & $\{\selection^{24}\}$                    & $\{\selection^{24}\}$                                         & $\{\selection^{24}\}$, \textcolor{blue}{$\{\gamma^{23}\}$}, $\{\gamma^{23}, \selection^{24}\}$  \\
Q3                                                                                                                                                                                                                                                                     & \textcolor{red}{$\{\selection^{27}\}$}   & \textcolor{blue}{$\{\selection^{26}, \selection^{27}\}$}      & \textcolor{blue}{$\{\selection^{26}, \selection^{27}\}$}, $\{\selection^{26}, \selection^{27}, \gamma^{25}\}$ \\
Q4                                                                                                                                                                                                                                                                     & $\emptyset$                              & $\emptyset$                                                   & $\{\gamma^{30}\}$, $\{\gamma^{30}, \selection^{29}\}$, \textcolor{blue}{$\{\gamma^{30}, \selection^{28}\}$}, $\{\gamma^{30}, \selection^{29}, \selection^{28}\}$ \\
Q6                                                                                                                                                                                                                                                                     & $\{\selection^{32}\}$                    &
$\{\selection^{32}\}$, \textcolor{blue}{$\{\selection^{33}\}$}, $\{\selection^{34}\}$, $\{\selection^{32}, \selection^{33}\}$, $\{\selection^{32}, \selection^{34}\}$, $\{\selection^{33}, \selection^{34}\}$, $\{\selection^{32}, \selection^{33}, \selection^{34}\}$ &
$\{\selection^{32}\}$, \textcolor{blue}{$\{\selection^{33}\}$}, $\{\selection^{34}\}$, $\{\selection^{32}, \selection^{33}\}$, $\{\selection^{32}, \selection^{34}\}$,  $\{\selection^{33}, \selection^{34}\}$, $\{\projection^{31}, \selection^{33}\}$, $\{\selection^{32}, \selection^{33}, \selection^{34}\}$, $\{\projection^{31}, \selection^{32}, \selection^{33}\}$, $\{\projection^{31}, \selection^{33}, \selection^{34}\}$, $\{\projection^{31}, \selection^{32}, \selection^{33}, \selection^{34}\}$  \\
Q10 & \textcolor{red}{$\{\join^{38}\}$} & $\{\selection^{35}\}$, $\{\selection^{35}, \selection^{36}\}$ & $\{\selection^{35}\}$, $\{\selection^{35}, \selection^{36}\}$, $\{\selection^{35}, \projection^{37}\}$, \textcolor{blue}{$\{\selection^{35}, \selection^{36}, \projection^{37}\}$}  \\
Q13 & $\color{blue}\{\join^{39}\}$ & $\color{blue}\{\join^{39}\}$ & $\color{blue}\{\join^{39}\}$  \\
\hline
\end{tabular}
\caption{Explanations produced by our system with schema alternatives (\textbf{with SA}), without schema alternatives (\textbf{without SA}), and by our implementation of Whynot (\textbf{WN++}) for the DBLP (D), Twitter (T), and TPC-H (Q) scenarios.}
\label{tab:answers}
\end{table}

 }
\iftechreport{
   \begin{table*}[t]
 \centering
   \scriptsize 
\begin{minipage}{\linewidth}
 \centering
  \scriptsize
\trimfigspace
\begin{tabular}{| c | p{11.5cm} | p{4.3cm} | } \hline
  \cellcolor{tabbg} \textbf{Scenario} & \cellcolor{tabbg} \textbf{Descriptions} & \cellcolor{tabbg} \textbf{Why-not questions} \\\hline
  Q1 &
  \begin{minipage}{0.7\linewidth}
\begin{dmath*}
  {\aggregation{sum({\color{blue}l\_tax})}{avgDisc}}^{23} \left(\selection^{24}_{l\_shipdate \leq 1998-09-02} \left(\flatten^{I}_{o\_lineitems} \left( nestedOrders \right)\right)\right)
     \end{dmath*}
   \end{minipage}
 &
$\nrctuple{avgDisc: > 0.45, ?}$
  \\ \hline
  Q3 &
\begin{minipage}{0.7\linewidth}
\begin{dmath*}
  {\aggregation{o\_orderkey, o\_orderdate, o\_shippriority, sum(disc\_price)}{revenue}}^{25} \left( \\ \projection_{nestedOrders, disc\_price \leftarrow (l\_extendedprice \times (1 \:-\: {\color{black}l\_discount}))} \left( \selection^{26}_{{\color{blue} c\_mktsegment = ``HOUSEHOLD``}} \left( \\ \selection_{o\_orderdate < 1995-03-15}  \left( \selection^{27}_{{\color{blue}l\_commitdate > 1995-03-25}} \left(  customer \join \left( \flatten^{I}_{o\_lineitems} \left( nestedOrders \right) \right)\right)\right)\right)\right)\right)
\end{dmath*}
\end{minipage}
&
 \begin{minipage}{0.5\linewidth}
\begin{dmath*}
  \nrctuple{l\_orderkey = 4986467, o\_orderdate: ?, \\ {\color{black}o\_shippriority}: ?, revenue: ?}
\end{dmath*}
\end{minipage}
  \\ \hline
  Q4 &
\begin{minipage}{0.95\linewidth}
\begin{dmath*}
  distOrd \Leftarrow \aggregation{l\_orderkey, count(*)}{cnt} \left(\selection^{28}_{{\color{blue}l\_shipdate < l\_receiptdate}} \left( \flatten^{I}_{o\_lineitems} \left( nestedOrders \right) \right) \right)
\end{dmath*}
  \begin{dmath*}
  filterOrd \Leftarrow \selection^{29}_{1993-07-01 \:\le\: o\_orderdate \:\le\: 1993-09-30} \left( nestedOrders \right)
\end{dmath*}
  \begin{dmath*}
  {\aggregation{{\color{blue}o\_shippriority}, count(o\_orderkey)}{order\_count}}^{30} \left( filterOrd \join  distOrd \right)
\end{dmath*} 
\end{minipage}
 &
\begin{minipage}{0.5\linewidth}     
  \begin{dmath*}
  \nrctuple{o\_shippriority: \text{``3-MEDIUM''}, \\ order\_count: \: < 11000}
\end{dmath*}
\end{minipage}
  \\ \hline
  Q6 &
\begin{minipage}{0.8\linewidth}     
\flushleft
  \begin{dmath*}
    \aggregation{sum(disc\_price)}{revenue} \left( \projection^{31}_{nestedOrders, disc\_price \leftarrow (l\_extendedprice \times l\_discount)} \left( \\ \selection^{32}_{1994-01-01 \:\le\: l\_shipdate \:\le\: 1994-12-31} \left(\selection^{33}_{{\color{blue} 0.05 \le l\_tax \le 0.07}} \left(\selection^{34}_{l\_quantity < 24} \left( \flatten^{I}_{o\_lineitems} \left(nestedOrders\right) \right) \right) \right) \right) \right)
\end{dmath*}
\end{minipage}
                                     & $\nrctuple{revenue: \: < 1.24 \times 10^8}$
  \\\hline
  Q10 &
\begin{minipage}{0.95\linewidth}
  \begin{dmath*}
    flatOrd \Leftarrow \selection^{35}_{\color{blue}l\_returnflag = ``A``} \left( \selection^{36}_{\color{blue}1997-10-01 \:\le\: o\_orderdate \:\le\: 1997-12-31} \left(\flatten^{I}_{o\_lineitems} (nestedOrders) \right)\right)
\end{dmath*}
  \begin{dmath*}
  \aggregation{c\_custkey,c\_name,c\_acctbal,c\_phone,n\_name,c\_address,c\_comment,sum(disc\_price)}{revenue} \left( \\ \projection^{37}_{customer,flatOrd,nation, disc\_price \leftarrow (l\_extendedprice \times (1 - {\color{blue}l\_tax}))} \left( customer \join^{38} flatOrd \join nation \right) \right)
\end{dmath*}
\end{minipage}        
                                      &
\begin{minipage}{0.6\linewidth}
\begin{dmath*}
  \nrctuple{c\_custkey: \: 61402, c\_name: ?,c\_acctbal: ?,\\c\_phone: ?,n\_name: ?,c\_address: ?, \\ c\_comment: ?, revenue: \: > 0}
\end{dmath*}
\end{minipage}
  \\ \hline
  Q13 &
\begin{minipage}{0.7\linewidth}     
\flushleft
  \begin{dmath*}
   \aggregation{c\_count, count(c\_custkey)}{custdist} \left( \aggregation{c\_custkey, c\_count(o\_orderkey)}{c\_count}  \left( \\ \selection_{\text{ ``special''} \not\in o\_comment \:\&\: \text{``requests''} \not\in o\_comment} \left( customer {\color{blue}\join^{39}} nestedOrders \right)\right) \right)
 \end{dmath*}
 \end{minipage}
                                                                                & $\nrctuple{c\_count: 0, custdist: ?}$\\
\hline
\end{tabular}
\caption{TPC-H scenarios in \abbrNRAB}
\label{tab:tpch-scen}
\end{minipage}
\end{table*} }

\iftechreport{
\mypar{Comparison to other approaches with crime dataset}
We further validate the added value of query-based explanations using reparametrization by comparing the explanations of our algorithm to the ones of the baseline algorithms Why-Not~\cite{DBLP:conf/sigmod/ChapmanJ09} and Conseil~\cite{herschel:jdiq15} on scenarios C1 to C3 in \Cref{tab:cs}.  Note that the parentheses in their descriptions indicate the order in which the operators are processed during lineage tracing.
 \MH{moved to setup: As these work on relational data only, we resort to a relational dataset and workload. It is based on floppy-disc-sized 
relations of \textbf{C}rimes, \textbf{W}itnesses, \textbf{S}awpersons, and \textbf{P}ersons, already used for benchmarking in previous work~(including \cite{DBLP:conf/sigmod/ChapmanJ09,herschel:jdiq15}). We reimplemented the following approaches as baselines: Why-Not~\cite{DBLP:conf/sigmod/ChapmanJ09} as a representative of a purely lineage-based approach and the query-based part of Conseil~\cite{herschel:jdiq15}, which injects fictive tuples into the tracing procedure to continue tracing and potentially identifying further problematic operators to mitigate the problem of ``drying out'' during tracing. This resembles the use of our retained flag. It ``pretends'' intermediate result tuples succeed an operator to trace them through consecutive operators. We discuss three scenarios considering operators supported by all approaches, summarized in \Cref{tab:cs}. Note that the parentheses in the scenario description indicate the order in which operators are processed.}

In scenario C1, Why-Not returns the selection $\selection^1$, because the person Roger exists, but only without blue hair. Thus, any compatible tuple featuring Roger in \textbf{P} is pruned by  $\selection^1$, at which point Why-Not terminates as there is nothing more to trace. It thereby misses the fact that even if some Roger passed the selection, no such tuple would satisfy the join condition  $\join^2$. Both Conseil and our approach find the combined explanation $\{\selection^2, \join^1\}$. In C2, both Why-Not and Conseil return $\selection^4$ as explanation, because there are witnesses in \textbf{W} satisfying $\selection^3$. However, their names do not match the constraint of $\selection^4$. If such a tuple existed in \textbf{W}, all subsequent joins would succeed, so Conseil does not return any further operator. Opposed to Conseil, our approach returns two explanations, i.e., $\{ \{ \selection^4\}, \{\selection^3, \selection^4\} \}$, because rewriting both filters is an alternative option to obtain the missing answers with potentially less side effects than just changing $\selection^4$. Finally, in C3, Why-Not and Conseil both return $\join^5$, since no witness named ``Ashishbakshi'' finds a join partner that yields the expected answer. On the contrary, our approach does \emph{not} return this join as an explanation at all. The only way to fix the join is to apply a cross product, which we do not consider a valid reparametrization. In this example, our approach suggests to revise $\projection^6$. This result is based on a schema alternative, as the description related to ``hair'' could be replaced by ``clothes''.

Clearly, our solution outperforms existing solutions in finding complex operator combinations as explanation. With the addition of SAs, it is also the first to include operators beyond tuple filtering operators (typically join and selection) in explanations.
}

 \vspace{-0.3cm}
\section{Conclusions}\label{sec:conclusions}

We presented a novel approach for query-based explanations for missing answers that is the first to (i) support nested data, (ii)~consider changes to the query that affect the schema of intermediate results, and (iii)~scale to big data (100s of GBs). 
Even for queries over flat data
, which prior work is limited to,
it produces explanations that existing systems miss. One avenue for future research is to define and efficiently compute tighter bounds for side~effects.

\iftechreport{\begin{landscape}
\begin{table}[t]
\resizebox{21.5cm}{!}{
\begin{minipage}[t]{28.5cm}
\centering 
\footnotesize
\begin{tabular}{| c | p{28cm} | } \hline
\cellcolor{tabbg} \textbf{Scen.} & \cellcolor{tabbg} \textbf{Descriptions} \\\hline
D1 & $\selection^2_{ptitle = "SIGMOD"}\left(\left(\projection_{author, ititle, ptitle}\left({\flatten^{T}_{author \leftarrow iauthor.\_VALUE}}\left({\flatten^{T}_{ititle \leftarrow title.\_VALUE}}\left(\flatten^{I}_{iauthor \leftarrow author}\left(\flatten^{I}_{crf \leftarrow crossref}(I)\right)\right)\right)\right)\right)  \join_{\_key = crf} \left(\projection^1_{\_key, ptitle \leftarrow title}(P)\right)\right)$ \\
D2 & $\aggregation{count(ctitle)}{cnt}\left(\nestRel{title}{ctitle}\left(\selection_{\text{``Dey''}\ \in author}\left(\projection_{author, title}\left({\flatten^{T}_{author \leftarrow aauthor.\_VALUE}}\left({\flatten^{T^3}_{title \leftarrow title.\_bibtex}}\left(\flatten^{I}_{aauthor \leftarrow author}(A)\right)\right)\right)\right)\right)\right)$ \\
D3 & $\nestRel{authorPaper}{aplist}\left(\projection_{booktitle, year, authorPaper}\left({\nestTup{author, title}{authorPaper}}^4(I)\right)\right)$  \\
D4 & $\aggregation{count(tlist)}{cnt}\left(\nestRel{title}{tlist}\left(\projection_{author, title}\left(\selection^7_{year = 2015}\left(\selection^6_{ppublisher = \text{``ACM''}}\left(\projection_{\_key, year, ppublisher}\left({\flatten^{T^5}_{ppublisher \leftarrow publisher.\_VALUE}}(P)\right)\right)\right)\right) \join_{\_key = crf} \left(\projection_{crf, author, title}\left(\flatten^{I}_{iauthor \leftarrow author}\left(\flatten^{I}_{crf \leftarrow crossref}(I)\right)\right)\right)\right)\right)  $ \\
D5 & $\nestRel{url}{lurl}\left(\projection_{name, url}\left({\flatten^{T}_{name \leftarrow author.\_VALUE}}\left({\flatten^{T}_{url \leftarrow urls.\_VALUE}}\left({\flatten^{I^9}_{urls \leftarrow url}}\left(\flatten^{I}_{authors \leftarrow author}\left(\projection^8_{author, url}(U)\right)\right)\right)\right)\right)\right)$ \\ \hline
T1 & $ \selection^{12}_{\text{``Michael Jordan''} \in text}\left({\flatten^{I^{11}}_{medias \leftarrow media}}\left(\projection_{text, id, media} \left({\flatten^{T^{10}}_{media \leftarrow entities.media}}(T)\right)\right)\right) $ \\
T2 & $ \selection^{15}_{\text{``United States''} \in country} \left(\selection^{14}_{\text{``BTS''} \in text}\left(\projection_{text, country, sht \leftarrow size(ht), uLang, uLoc, uName, fCnt} \left({\flatten^{T^{13}}_{country \leftarrow place.country}}\left({\flatten^{T}_{ht \leftarrow entities.hashtags}}\left({\flatten^{T}_{uLang \leftarrow user.lang}}\left({\flatten^{T}_{uLoc \leftarrow user.location}}\left({\flatten^{T}_{uName \leftarrow user.name}}\left({\flatten^{T}_{fCnt \leftarrow user.followers\_count}}(T)\right)\right)\right)\right)\right)\right)\right)\right)$ \\
T3 & $ \projection_{uName, ht, medias}\left(\left(\flatten^{T}_{uName \leftarrow user.name}\left(\flatten^{T}_{uid \leftarrow user.id}(T)\right)\right) \join_{uid = mid} \left({\flatten^{I^{17}}_{medias \leftarrow media}}\left({\flatten^{T^{16}}_{media \leftarrow entities.media}}\left(\flatten^{T}_{ht \leftarrow entities.hashtags}\left(\flatten^{T}_{mName \leftarrow muser.name}\left(\flatten^{T}_{mid \leftarrow muser.id}\left(\flatten^{I}_{muser \leftarrow musers}\left(\flatten^{T}_{musers \leftarrow entities.
mentioned\_user}(T)\right)\right)\right)\right)\right)\right)\right)\right)$  \\ 
T4 & $\selection^{20}_{cnt > 0}\left(\aggregation{count(lcountry)}{cnt}\left(\nestRel{country}{lcountry}\left(\projection_{country, htText}\left(\selection^{19}_{\text{``Uefa''} \in text}\left({\flatten^{T}_{htText \leftarrow fht.text}}\left(\flatten^{I}_{fht \leftarrow ht}\left({\flatten^{T}_{ht \leftarrow entities.hashtags}}\left({\flatten^{T^{18}}_{country \leftarrow place.country}}(T)\right)\right)\right)\right)\right)\right)\right)\right)$  \\
T$_{ASD}$ & $ \projection_{id, id\_str, text, ...}\left(\selection^{22}_{\color{blue}{quote_count > 0}}\left({\flatten^{T}_{\color{blue}{quoted\_status}}}^{21}(T)\right)\right)$ \\
\hline
\end{tabular}

\end{minipage}
}
\caption{DBLP and Twitter scenarios in \abbrNRAB}
\label{tab:NRAB}
\end{table}
\end{landscape}
 }

{\small
\bibliographystyle{abbrv}

	 \end{document}